\documentclass[runningheads,10pt]{llncs}

\usepackage[T1]{fontenc}
\usepackage[utf8]{inputenc}
\usepackage[dvipsnames]{xcolor}
\definecolor{darkred}{rgb}{0.64, 0.0, 0.0} 

\usepackage{xr}
\usepackage{
  alphalph,
  amsmath,
  amsfonts,
  amssymb,
  booktabs,
  comment,
  colortbl,
  ifthen,
  enumerate,
  listings,
  nicefrac,
  makecell,
  mathtools,
  microtype,
  multirow,
  paralist,
  rotating,
  setspace,
  semantic,
  stmaryrd,
  subcaption,
  tabularx,
  wrapfig,
  xargs,
  xspace
}

\usepackage{pifont}
\newcommand{\cmark}{\ding{51}}%

\usepackage[inline]{enumitem}
\usepackage[scale=.77]{noto-mono}
\usepackage[scaled=.84]{helvet}

\usepackage[
  pdftex,
  colorlinks = true,
  bookmarks = false,
  linkcolor = blue,
  citecolor = blue,
  urlcolor = blue,
  bookmarks
]{hyperref}

\usepackage[capitalize]{cleveref}

\usepackage[color=black!20,textsize=scriptsize]{todonotes} 
\crefname{section}{Sect.}{sections}
\Crefname{section}{Section}{Sections}
\crefname{table}{Tbl.}{tables}
\Crefname{table}{Table}{Tables}

\definecolor{darkgoldenrod}{rgb}{0.72, 0.53, 0.04}
\definecolor{goldenrod}{rgb}{0.85, 0.65, 0.13}
\definecolor{darkpastelpurple}{rgb}{0.59, 0.44, 0.84}
\definecolor{darkblue}{rgb}{0.43, .5, 0.9}
\definecolor{darkgreengray}{rgb}{0.33, .4, 0.53}

\renewcommand{\phi}{\varphi}
\renewcommand{\arraystretch}{1.1}
\newcommand{\code}[1]{\text{\lstinline[basicstyle={\normalsize\ttfamily}]{#1}}}
\newcommand{\codescriptsize}[1]{\text{\lstinline[basicstyle=\scriptsize\ttfamily]{#1}}}

\newcommand{\textcmtt}[1]{\fontfamily{lmtt}\selectfont#1} 

\renewcommand{\textcmtt}[1]{\texttt{#1}}
\newcommand{\tildecustom}{\raisebox{0.5ex}{\texttildelow}}

\newcommand{\Dist}{\ensuremath{\mathcal{D}}}

\newcommand{\Env}{\textbf{Env}}
\newcommand{\Var}{\textbf{Var}}
\newcommand{\Con}{\textbf{Con}}
\newcommand{\Xpr}{\textbf{Exp}}

\newcommand{\Val}{\textbf{Val}}

\newcommand{\Binomial}{\ensuremath{\mathcal{B}}}
\newcommand{\Normal}{\ensuremath{\mathcal{N}}}
\newcommand{\Uniform}{\ensuremath{\mathcal{U}}}
\newcommand{\Constant}{\ensuremath{\mathcal{C}}}
\newcommand{\Laplace}{\ensuremath{\mathcal{L}}}

\renewcommand{\Pr}{\ensuremath{\mathit{P}}} 
\newcommand{\Expected}[1]{\ensuremath{\mathrm{E}}[#1]}

\newcommand{\alicerow}{\ensuremath{a}}

\newcommand{\Lift}{\ensuremath{\mathbf{lift}\,}}

\newcommand{\Bind}{\text{\textcmtt{\,>\hspace{-0.22em}>\hspace{-.08em}=\,}}}
\newcommand{\Return}{\text{\textcmtt{return}}}
\newcommand{\monad}[1]{\ensuremath{\textit{Monad}\ #1}}

\newcommand{\var}{\ensuremath{\mathit{x}}}

\newcommand{\env}{\ensuremath{\varepsilon}}
\newcommand{\xpr}{\ensuremath{\mathit{e}}}
\newcommand{\val}{\ensuremath{\mathit{v}}}
\newcommand{\cse}[3]{\ensuremath{\text{\textcmtt{case}}\ #1\ #2\ #3}}
\newcommandx{\con}[4][2=,3=,4=]{\ensuremath{%
\textsf{#1}%
\ifthenelse{\equal{#2}{}}{}{\,#2}%
\ifthenelse{\equal{#3}{}}{}{\,#3}%
\ifthenelse{\equal{#4}{}}{}{\,#4}%
}}
\newcommand{\step}[3]{\ensuremath{#2 \to_{#1} #3}}
\newcommand{\fun}[2]{\ensuremath{#1(#2)}}

\newcommand{\lam}[2]{\ensuremath{\lambda#1\,\text{\textcmtt{.}}\,#2}}
\newcommand{\app}[2]{\ensuremath{#1\ #2}}

\newcommand{\sub}[3]{\ensuremath{#3{[}#1\mapsto#2{]}}}

\newcommand{\Dst}[1]{\ensuremath{\fun{\Dist{}}{#1}}}

\newcommand{\eaf}[2]{\ensuremath{\llbracket#2\rrbracket_#1}}

\newcommand{\lstof}[1]{\ensuremath{\overline{#1}}}

\newcommand{\ie}{\textit{i.e.}}
\newcommand{\eg}{\textit{e.g.}}
\newcommand{\etal}{\textit{et al.}\xspace}

\newcommand{\wsem}[2]{\ensuremath{{\llbracket #2 \rrbracket_{#1}}}}
\newcommand{\yes}{\cmark}
\newcommand{\no}{~}

\newcommand{\privug}{{\sc Privug}\xspace}
\newcommand{\privugbold}{{\bfseries\scshape Privug}\xspace}

\newcommand{\kal}{%
  \ensuremath{%
    {
      \mathrm{k}_{%
        \mathrm{al}
      }%
    }%
  }%
  \xspace}
\newcommand{\kab}{%
  \ensuremath{%
    {
      \mathrm{k}_{%
        \mathrm{ab}
      }%
    }%
  }%
  \xspace}
\newcommand{\doublemid}{\ensuremath{\mid\mid}\xspace}

\newcommand{\dkl}{\ensuremath{\mathrm{D}_\mathrm{KL}}}
\newcommand{\entropy}{\ensuremath{\mathrm{H}}}
\newcommand{\mi}{\ensuremath{\mathrm{I}}}

\makeatletter
\newcommand\footnoteref[1]{\protected@xdef\@thefnmark{\ref{#1}}\@footnotemark}
\makeatother

\renewcommand{\subparagraph}[1]{\vspace{6.5pt plus 2.1pt minus 1pt}{\noindent\sffamily\itshape #1.}}

\newcommand{\smath}[1]{{\small #1}}

\newcommand{\senv}[1]{%
  \begin{center}%
    \smath{%
      \begin{minipage}{1.13\linewidth}%
        \vspace{-.2em plus .2em minus .2em}%
        #1%
        \vspace{-.5em plus .2em minus .2em}%
      \end{minipage}%
    }%
  \end{center}%
}

\renewcommand{\smath}[1]{#1}
\renewcommand{\senv}[1]{#1} 

\newcommand{\FOURTH}{0.25}
\newcommand{\SIXTH}{0.166666666666666666666666666666666666666666666666666666666666666666666666666666666666666666666666666666666666666666666666666}

\newcommand{\PLOTSW}{\FOURTH\linewidth}
\newcommand{\PLOTSH}{21.9mm}
\newcommand{\PLOTTW}{\SIXTH\linewidth}
\newcommand{\PLOTTH}{13.16mm}

\newcommandx{\labelS}[2][1=7.8mm]{%
  \phantomsubcaption\label{#2}
  \llap{%
    \scriptsize%
    \raisebox{#1}[0mm][0mm]{%
      (\subref*{#2})                
    }\hspace{.2mm}%
  }}%

\newcommand{\labelT}[1]{%
  \phantomsubcaption\label{#1}
  \llap{%
    \scriptsize%
    \hspace{18mm}\raisebox{4.4mm}[0mm][0mm]{%
      (\subref*{#1})                
    }%
  }}%

\begin{document}

\author{
  {Ra\'ul Pardo}\inst{1} \and
  {Willard Rafnsson}\inst{1} \and
  {Christian W.~Probst}\inst{2} \and
  {Andrzej W\k{a}sowski}\inst{1}
}
\authorrunning{R. Pardo et al.}

\institute{
  IT University of Copenhagen, Denmark\\\email{\{raup,wilr,wasowski\}@itu.dk} \and
  Unitec Institute of Technology, New Zealand\\\email{cprobst@unitec.ac.nz}
}

\title{Privug: Using Probabilistic Programming for Quantifying Leakage
  in Privacy Risk Analysis\thanks{Work partially supported by the
    Danish Villum Foundation through Villum Experiment project
    No.~00023028 and New Zealand Ministry of Business, Innovation and
    Employment -- H\:\!\={\i}\:\!kina Whakatutuki through Smart Ideas
    project No.~UNIT1902.}}

\titlerunning{Privug: Using Probabilistic Programming for Privacy Risk Analysis}

\maketitle

\begin{abstract}

  Disclosure of data analytics results has important scientific and commercial justifications.  However, no data shall be disclosed without a diligent investigation of risks for privacy of subjects. \privug\ is a tool-supported method to explore information leakage properties of data analytics and anonymization programs.  In \privug, we reinterpret a program probabilistically, using off-the-shelf tools for Bayesian inference to perform information-theoretic analysis of the information flow.  For privacy researchers, \privug\ provides a fast, lightweight way to experiment with privacy protection measures and mechanisms.  We show that \privug{} is accurate, scalable, and applicable to a range of leakage analysis scenarios.
  \looseness=-1

\end{abstract}

\section{Introduction}
\label{sec:introduction}

\noindent
However high the value of data becomes, we cannot ignore the risks that data disclosure presents to personal privacy.   Consequently, general privacy protection methods like differential privacy\,\cite{DBLP:conf/tcc/DworkMNS06},  comprehensibility and communication of privacy issues\,\cite{DBLP:conf/smsociety/NadonFJS18}, industrial processes for data management\,\cite{DBLP:journals/pacmhci/HargitaiSW18}, and debugging and analyzing privacy risk problems in program code\,\cite{chothia.leakest.2013,chothia.leakwatch.2014,cherubin.fbleau.2019} have become intensive areas of research.  This paper falls into this last group; we present tools for data scientists who create data analysis programs and would like to disclose the results of the computation.  Our primary goal is to create a method that supports a \emph{privacy debugging process}, \ie\ assessing effectiveness of such algorithms, and indeed of any calculations on the data, for concrete programs and datasets, in the style of debuggers. We want to help \emph{identifying} and \emph{explaining} the leakage risks, as the first step towards eliminating them.

As an example, consider the following Scala program that, given a list of names and ages, computes the mean age of the persons in a map-reduce style:
\begin{center}
\begin{tabular}{c}
\begin{lstlisting}
def agg (^records^: List[(String,Double)]): Double =
  records.map    { (n, a) => (a, 1) }
         .reduce { (x, y) => (x._1 + y._1, x._2 + y._2) }
         .map    { (sum, count) => sum / count }
\end{lstlisting}\label{scala:agg}
\end{tabular}
\end{center}

\noindent
Let the age of each individual be the sensitive secret in this example.  One attack could be that underage individuals can be identified. An analyst would like to ask: How much of sensitive information leaks when the mean is disclosed? In what situations is this leak not ignorable?  What kind of attackers may discover the secret by observing the mean?
\looseness=-1

\privug\ is an analysis method for privacy risks in data processing.  A data analyst using \privug\ models an attacker's knowledge about the secret as a probability distribution.  \privug\ re-interprets the program as an information transformer that operates on distributions instead of concrete inputs.  The analyst analyzes the attacker's confidence about the secret, using a combination of probability queries, standard information-theoretic measures, and visualizations. She explores and assesses the information leakage to the result of the program  by varying attacker knowledge, the queries and the leakage measures. For our example, the analyst may learn that the leakage is ignorable if the subjects are drawn from general population, but if the attacker knows that they come from a homogeneous group, she could, for example, conclude that a specific individual is under age.
\looseness=-1

Since \privug\ is based on probabilistic reasoning, it can be facilitated by \emph{probabilistic programming}, a lively field in data science, with many tools available.  \privug is not tied to any particular probabilistic programming framework.  In this paper, we implement queries, measures, and visualization in Figaro\,\cite{pfeffer2016practical} and PyMC3\,\cite{pymc3}.  For programs seen as functions, a probabilistic programming framework can automatically build a Bayesian model which represents the information transform. This transform supplemented by a model of attacker can be used to explore re-identification risks\,\cite{clarkson.belief.2005}.

\privug\ offers three distinct advantages over state of the art tools for privacy risk analysis:
\begin{enumerate*}[label = (\roman*)]%

  \item It focuses on the analysis of programs not data, which means that a what-if  analysis can be performed before data is available, or without authorizing access to a sensitive database.

  \item It is largely automatable using off-the-shelf systematic Monte-Carlo inference tools already used by data analysts, but which have not been used for this purpose before.

  \item \privug\ is easy to extend with new estimators of leakage thus  serves as a good test-bed for privacy mechanism research.  To the best of our knowledge, \privug as a method and probabilistic programming as a platform are the only basis that can offer such versatility at this point.
\end{enumerate*}
Our contributions include:

\begin{enumerate}%

  \item A widely applicable and extensible method, \privug, to analyze privacy risks. The first such method based on probabilistic programming frameworks.
  \looseness=-1

  \item An implementation of \privug{} in Figaro and PyMC3, the first versatile tool supporting such a wide range of measures over continuous and discrete inputs and outputs.
    \looseness=-1

  \item An empirical evaluation of the accuracy, scalability, and applicability of \privug\ for analyzing systems of different size and complexity, showing that probabilistic programming is an excellent base for implementing leakage analysis tools.
    \looseness=-1

\end{enumerate}

\noindent
We evaluate applicability, accuracy, and scalability of \privug, using well known privacy mechanisms (differential privacy, $k$-anonymity, naive anonymization) and synthetic cases that can be scaled up for higher dimensionality and using. Our experiments demonstrate \privug's versatility to realize many analysis scenarios,  and its interoperability with existing tools (by integrating external estimators). The source code and experiment data is available at \url{https://bitbucket.org/itu-square/privug-experiments}. The repository contains additional experiments showing the use of \privug\ in a realistic case study: an experiment using the differential privacy library OpenDP (\url{https://opendp.org/}).
\looseness=-1

\begin{figure*}[!t]
  \centering
  \renewcommand{\arraystretch}{1.2}
  \hspace*{-2mm}
  \begin{tabular}{
      >{\small}l
      >{\hspace{-0.0mm}\small}l
      >{\hspace{-0.0mm}\small}l
      >{\hspace{-0.0mm}\small}c
      >{\hspace{-0.0mm}\small}l
    }

    \textbf{name}
    & \textbf{zip}
    & \textbf{birthday}
    & \textbf{sex}
    & \textbf{diag.}
    \\
    \midrule

    Alice
    & 2300
    & 15.06.1954
    & F
    & ill
    \\

    Bob
    & 2305
    & 15.06.1954
    & M
    & healthy
    \\

    Carol
    & 2300
    & 09.10.1925
    & F
    & healthy
    \\

    Dave
    & 2310
    & 01.01.2000
    & M
    & ill
    \\
\end{tabular}
  \hfill$\xrightarrow{\code{ano}}$\hfill
  \begin{tabular}{
      >{\small}l
      >{\hspace{-0.0mm}\small}l
      >{\hspace{-0.0mm}\small}c
      >{\hspace{-0.0mm}\small}l
    }

    \textbf{zip}
  &
  \textbf{birthday}
  &
  \textbf{sex}
  &
  \textbf{diag.}
  \\
  \midrule
  \textbf{2300}
  &
  \textbf{15.06.1954}
  &
  \textbf{F}
  &
  ill
  \\
  2305
  &
  15.06.1954
  &
  M
  &
  healthy
  \\
  2300
  &
  09.10.1925
  &
  F
  &
  healthy
  \\
  2310
  &
  01.01.2000
  &
  M
  &
  ill
  \\

  \end{tabular}%

  \medskip

  \begin{tabular}{
    >{\small}l
    >{\hspace{-0.0mm}\small}l
    >{\hspace{-0.0mm}\small}c
    >{\hspace{-0.0mm}\small}l
  }

    \textbf{zip}
  &
  \textbf{birthday}
  &
  \textbf{sex}
  &
  \textbf{diag.}
  \\
  \midrule
  \textbf{2300}
  &
  \textbf{15.06.1954}
  &
  \textbf{F}
  &
  ill
  \\
  2305
  &
  15.06.1954
  &
  M
  &
  healthy
  \\
  2300
  &
  09.10.1925
  &
  F
  &
  healthy
  \\
  2310
  &
  01.01.2000
  &
  M
  &
  ill
  \\

  \end{tabular}%
  \hfill{\large$\oplus$}\hfill
  \begin{tabular}{
      >{\small}l
      >{\hspace{-0.0mm}\small}l
      >{\hspace{-0.0mm}\small}l
      >{\hspace{-0.0mm}\small}l
    }

    \textbf{name}
    & \textbf{zip}
    & \textbf{birthday}
    & \textbf{sex}
    \\
    \midrule

    Mark
    & 2450
    & 30.09.1977
    & M
    \\

    Rose
    & 2870
    & 24.12.1985
    & F
    \\

    Alice
    & \textbf{2300}
    & \textbf{15.06.1954}
    & \textbf{F}
    \\

    Dave
    & 2310
    & 01.01.2000
    & M
    \\

  \end{tabular}
  \hfill{\large$\rightsquigarrow$}\hfill%
  \parbox{1.8em}{\small\centering Alice is ill}\hspace{-.0em}

  \caption{Privacy violation: The data is anonymized (\lstinline{ano}), then  the diagnosis of Alice is recovered by an attacker who links the result with another data set ($\oplus$).}%
  \label{fig:example1}


\end{figure*}

\begin{figure}[!t]

  \renewcommand{\arraystretch}{0.95}
  \centering

  \begin{tabular}{
    >{\small}l
    >{\hspace{-0.0mm}\small}l
  }

    \textbf{name}
    &
    \textbf{age}
    \\
    \midrule
    Alice
    &
    42
    \\
    Bob
    &
    25
    \\
    Carol
    &
    25
    \\
    Dave
    &
    25\\

  \end{tabular}
  \hfill
  $\xrightarrow{\code{agg}\,}$
  \hfill
  \begin{minipage}{2em}
    {\small\fbox{\textsf{29.25}}}\vspace{.2em}
  \end{minipage}
  \hfill
  {\large $\oplus$}
  \hspace{1em}
  \begin{tabular}{
    >{\small}l
    >{\hspace{-0.0mm}\small}l
  }

    \textbf{name} & \textbf{age} \\ \midrule
    Alice & {\small\textsf{$\mathcal{N}$}(40,3)} \\
    Bob & {\small\textsf{$\mathcal{N}$}(23,10)} \\
    Carol & {\small\textsf{$\mathcal{N}$}(23,10)} \\
    Dave & {\small\textsf{$\mathcal{N}$}(23,10)}\\

  \end{tabular}%
  \hfill
  $\rightsquigarrow$%
  \hfill
  \parbox{4em}{\small\centering%
    Alice's age is\\\textsf{$\mathcal{N}$(41,2.5)}}%

    \caption{Privacy violation: The program computes a mean of ages, the mean is released, but an attacker with prior knowledge can reduce the uncertainty regarding Alice's age.}
    \label{fig:example2}

    \vspace{-2.5mm plus 0.5mm minus 0.5mm}

  \end{figure}

\section{Overview}%
\label{sec:overview}

We consider data disclosure programs seen as functions that transform an input dataset to an output. The output is then disclosed to a third party, called an attacker.   The aggregation example \code{agg} from the introduction translates a database with two columns: name (\smath{\code{String}}) and age (\smath{\code{Double}}) to a number representing mean age which is then published. The second running example, \code{ano},  \emph{anonymizes} medical records in a dataset.  The input data has five columns: name, zip code, birthday, sex, and diagnosis.  The program simply drops the name column, before the data is released to an attacker:
\begin{center}
\begin{tabular}{c}
\begin{lstlisting}
def ano (records: List[(Name,Zip,Day,Sex,Diag)]): List[(Zip,Day,Sex,Diag)] =
  ^records^.map { (n, z, b, s, d) => (z, b, s, d) }
\end{lstlisting}\label{scala:ano}
\end{tabular}
\end{center}

\noindent
Suppose that in the anonymization example, subjects have not consented to disclosure of their diagnosis.  Despite anonymization, the diagnosis of individuals may be revealed by a \emph{linking attack} (see \cref{fig:example1}).  If an attacker has access to a dataset with zip codes, birthdays, sex, and, crucially, names, a simple join could reveal the names of the individuals from the disclosed medical records.  Zip code, birthday, and sex form a quasi-identifier in both datasets.  (Sweeney famously joined medical records disclosed by the Group Insurance Commission with a voter registration list to reveal the health record data of the then-governor of Massachusetts\,\cite{sweeney.kanonymity.2002}.)  Similarly, suppose that in the aggregation example (\code{agg}) users have not consented that their age is disclosed.  Despite the disclosed data being an aggregate, it carries some information about individual ages.  If you knew that Alice is around 40, as modeled by a Normal distribution in \cref{fig:example2}, then after learning the average, your uncertainty decreases: the final mean age raises, and the standard deviation decreases, making extreme values of Alice's age less likely.
\looseness=-1

\privug aims to help data scientists investigate the information revealed to an attacker from the output of a program. We frame this scenario as an adversarial problem. We assume a \emph{threat model} in which an information theoretical attacker has some prior knowledge about the input, has access to the program code, and observes the output. There are no bounds on the computational resources available to the attacker when analyzing the posterior knowledge to learn information about the secret input, e.g., probability of an outcome or event.
\looseness=-1

We model this scenario using probabilistic programming.  First, we build a probabilistic model of the \emph{prior knowledge} of the attacker.  Intuitively, the prior of the attacker captures what she knows, with (un)certainty, about the input of the program before observing the output.  We then express the attacker's view of the program, by transforming the program to operate on probabilistic models of datasets instead of actual data.  We do this by lifting the algorithm into the probability monad\,\cite{DBLP:conf/popl/RamseyP02}.  Next, we introduce \emph{observations} modeling the concrete output of the program that the attacker sees.  Observations constrain the prior of the attacker and produce the \emph{posterior knowledge} of the attacker, \ie, what the attacker knows about the input.  We use Bayesian inference to estimate the posterior,  Figaro for Scala\,\cite{pfeffer2016practical} and PyMC3~\cite{pymc3} for Python, but many other probabilistic frameworks can be used (Pyro~\cite{pyro}, Tensorflow Probability~\cite{tensorflow-probability}, Anglican~\cite{anglican}, etc.). Finally, we \emph{analyze} the posterior to quantify how much the attacker learns by observing the output.  This lets us determine whether specific attackers are capable of learning specific things, to assess the risk of disclosing the output of the program.
\looseness=-1


\section{\privugbold}
\label{sec:model}

We present each step of the \privug\ method in detail.  We model disclosure problems probabilistically and express models directly in a probabilistic programming language to enable automatic analysis. Let $D(X)$ denote a distribution over a set $X$. We write \smath{$x \sim D(X)$} to denote that random variable \smath{$x$} is distributed according to \smath{$D(X)$}; thus \mbox{\smath{$x \sim \Uniform(0,10)$}} means that \smath{$x$} is uniformly distributed from \smath{$0$} to \smath{$10$}.  In a programming language, this corresponds to \mbox{\smath{\code{x = Uniform(0,10)}}}. We also use composition operators of the language to define $y$ in terms of $x$, define a distribution over datasets, and so on.
\looseness=-1

\paragraph{Step 1: Attacker Knowledge (Prior).}

We model the prior knowledge of an attacker as a probability distribution over the possible input values of the program. In the \code{agg} program the input ranges over an array of pairs (name,age). Therefore, attacker prior knowledge is defined as a distribution over lists of pairs, $D(\code{List[(String,Double)]})$. Consider the following two examples of attackers:
\looseness=-1

\begin{enumerate}

  \item[\llap\kal] The \emph{knows-a-lot} attacker knows that the input dataset contains exactly four rows and that the age of all individuals, except Alice, is \smath{$55.2$}.  This is modeled by distributions \Constant(4) for the size and \Constant(55.2) for the age column.

  \item[\kab] The \emph{knows-just-a-bit} attacker knows that the input has approximately hundred entries (\mbox{\smath{$|\textit{records}| \sim \Binomial(300,\nicefrac13)$}}, a binomial distribution), and that the average age of an individual in the list is 55 (distributed with \smath{$\Normal(55.2,3.5)$}).

\end{enumerate}

\noindent
Both attackers know that Alice's record is in the dataset, and that no other record in the dataset has that name.  They do not know anything about Alice's age upfront:  all ages from 0 to 100 are equally likely, a uniform distribution \smath{$\Uniform(0,100)$}.  Implementations of \kal and \kab are shown below,  the differences highlighted in bold.  Here, \smath{\code{Element[T]}} denotes a distribution over \smath{\code{T}}, and \smath{\code{FixedSizeArrayElement[T]}} denotes a distribution over fixed-but-unknown-size arrays of \smath{\code{T}}s.  When sampled, \kab yields an array of random size, containing \smath{\code{(String,Double)}} pairs. The first pair represents Alice.
\looseness=-1

\begin{lstlisting}
def prior_kal: FixedSizeArrayElement[(String,Double)] =
   VariableSizeArray (^\bfseries Constant (4)^, i => for
      n <- if i==0 then Constant ("Alice") else Uniform (names: _*)
      a <- if i==0 then Uniform (0,100) else ^\bfseries Constant (55.2)^
   yield (n, a))^\vspace{-2.3mm}^

def prior_kab: FixedSizeArrayElement[(String,Double)] =
   VariableSizeArray (^\bfseries Binomial (300, 0.3)^, i => for
      n <- if i==0 then Constant ("Alice") else Uniform (names: _*)
      a <- if i==0 then Uniform (0,100) else ^\bfseries Normal (55.2,3.5)^
   yield (n, a))
\end{lstlisting}

\paragraph{Step 2: Attacker Prediction (Program).}%
\label{sec:experiment}

We obtain the attacker's prediction of the output of running a program by transforming---\ie\, lifting---the program to operate on distributions instead of concrete datasets, and applying it to the attacker model.  Let \smath{$\Dst{X}$} denote the set of distributions on set \smath{$X$}.  Since distributions form a monad\,\cite{DBLP:conf/popl/RamseyP02}, several useful functions are well defined on distributions,  including $\Lift$ that, here, has type
\looseness=-1
\begin{equation*}
  {\Lift{}} : \big(A \to B\big) \to \big(\Dst{A} \to \Dst{B}\big).
\end{equation*}
%
A function from $A$ to $B$ becomes a function from distributions over $A$ to distributions over $B$.  Recall that the type of \code{agg} is \code{List[(String, Double)]}\,$\to$\,\code{Double}.  The lifting of \smath{\code{agg}} has type: \Dst{\code{List[(String, Double)]}}\,$\to$\,\Dst{\code{Double}}. In Figaro:
\looseness=-1

\begin{lstlisting}
def agg_p (records: FixedSizeArrayElement[(String,Double)]): Element[Double] =
  records.map    { (n, a) => (a, 1) }
         .reduce { (x, y) => (x._1 + y._1, x._2 + y._2) }
         .map    { (sum, count) => sum / count }
\end{lstlisting}

\noindent
Note that only types change; \Dist{} is \code{Element} in Figaro, and \code{FixedSizeArrayElement[T]} is an efficient implementation of \Dst{\code{List[T]}}.  For a distribution over input datasets, \code{agg_p} yields a distribution over average ages (\code{Double}).  Running \code{agg_p} on a prior modeling the attacker's knowledge yields the attacker's prediction of the average age. Formally, the distribution of the output (the attacker prediction) is defined as $\Pr(o) = \int_x \Pr(o|x)\Pr(x)dx$. \Cref{sec:language} describes the semantic details of computing $\Pr(o)$.
\looseness=-1

\paragraph{Step 3: Attacker Observation.}
\label{sec:observation}

We use \emph{observations} to condition the attacker's prediction of the output. Since the prediction depends on the prior, conditioning it conditions the prior, and updates the attacker's knowledge about the input. We write $\Pr(x\,|\,E)$ to denote the conditional distribution of $x$ given evidence $E$. Let \mbox{\smath{$x \sim D(X)$}}, the evidence $E$ is a predicate over $X$. For instance, we write \mbox{\smath{$\Pr(x\,|\,4<x<8)$}} to denote the conditional distribution where only the outcomes $x$ in the interval $(4;8)$ are possible. We use conditions to model attacker observations of the output. For our aggregation example, to assert that the attacker observes 55.3, we define the predicate $E$ as \smath{\code{(x : Double) => (55.295<=x && x<55.305)}} as evidence on the prediction. The observation is typically known as \emph{likelihood function}~\cite{kruschke2014doing}, and it is modeled as a distribution (denoted as \mbox{\smath{$\Pr(E|x)$}}) assigning high probability to the values satisfying $E$. For instance, for the observation above we define $\Pr(E|x)$ equals $1/0.005$ for $55.295 \leq x \leq 55.305$ and $0$ otherwise. In Scala, the predicate $E$ is written as an anonymous function stating that the output is within $0.005$ from $55.3$. We cannot assert that the output is exactly $55.3$, since the output space is continuous; each individual outcome occurs with probability zero. In Figaro, we set $E$ on prediction \smath{\code{o}} with \smath{\code{o.setCondition(}$E$\code{)}}.
\looseness=-1

\paragraph{Step 4: Attacker Posterior.}
\label{sec:posterior}
We use Bayesian inference to compute the updated attacker's knowledge upon the observation. We put together the elements of our model using the Bayes rule as follows:
\looseness=-1
\vspace*{-4mm}
\begin{equation}
  \underbrace{\Pr(x,o\,|\,E)}_{\mathit{posterior}} = {\underbrace{\Pr(E\,|\,x,o)}_{\mathit{observation}} \cdot \overbrace{P(o|x) \cdot \underbrace{\Pr(x)}_{\mathit{prior}}}^{\mathit{prediction}} \cdot \Pr(E)^{-1}}
  \vspace*{-0mm}
\end{equation}
Our goal is to use the attacker prior $\Pr(x)$ (step 1), attacker prediction $\Pr(x,o)=\Pr(o|x)\Pr(x)$ (step 2), and observation $\Pr(E|x,o)$ (step 3) to compute the posterior knowledge $\Pr(x,o|E)$. Note that the equation above is expressed in terms of the joint distribution of the random variables for input $x$ and output $o$. The marginal distributions can be obtained by integrating out the corresponding variables.
\looseness=-1

We use Markov Chain Monte Carlo (MCMC) methods~\cite{mcmc} to estimate the posterior distribution by generating samples from a probabilistic program. MCMC algorithms are simulation methods that efficiently generate samples from the high density intervals of the target distribution, in our case $P(x,o|E)$. We refer the interested readers to Robert and Casella~\cite{mcmc} for details. We consider only terminating programs; as no samples can be generated from non-terminating programs using these methods. We remark that MCMC methods do not require computing or specifying the normalization factor $\Pr(E)^{-1}$. Their convergence conditions are well-known~\cite{BDA.gelman.2013}, but the number of samples determines their accuracy. In \cref{sec:evaluation}, we evaluate the accuracy and efficiency of several MCMC methods for this application. In Figaro, we use the MCMC algorithm \emph{importance sampling}\,\cite{pfeffer2016practical}. Let $a$ and $o$ denote Alice's age and the outcome in the aggregation example. If we define the evidence \mbox{\smath{$E = \text{``}o \approx 55.3\text{''}$}} on the prediction as above, \mbox{\smath{\code{Importance(10000, a)}}} produces $10000$ samples that estimate the distribution \mbox{\smath{$\Pr(a\,|\,o \approx 55.3)$}}.
\looseness=-1

\paragraph{Step 5: Leakage (Posterior Analysis).}

\begin{table}[t]

  \renewcommand{\arraystretch}{1.09}

  \scalebox{1}{
  \hspace*{-2mm}
  \begin{tabular}{
    >{\scriptsize}l
    >{\hspace*{-4mm}\scriptsize}c
    >{\scriptsize}r
    >{\scriptsize}r
    >{\scriptsize}c
    >{\scriptsize}l
    >{\hspace*{-2mm}\scriptsize}c
    >{\scriptsize}r
    >{\scriptsize}r
    }
    &
    & \kal
    & \kab
    & \hspace*{3mm}
    &
    &
    & \kal
    & \kab
    \\
    \cmidrule{1-4}\cmidrule{6-9}
    Expectation
    & $\Expected{\alicerow | o \approx 55.3}$
    & $55.60$
    & $64.000$
    &
    &
    Standard deviation
    & $\sigma[\alicerow | o \approx 55.3]$
    & $0.01$
    & $14.00$
    \\
    Probability query
    & $\Pr(\alicerow | o \approx 55.3)$
    & $0.00$
    & $0.004$
    &
    &
    Shannon Entropy
    & $\entropy(\alicerow | o \approx 55.3)$ & $-3.08$
    & $5.83$
    \\
    KL-divergence
    & $\dkl(\alicerow | o \approx 55.3 \doublemid a)$
    & $5.64$
    & $0.770$
    &
    &
    Mutual Information
    & $\mi(\alicerow; o)$
    & $9.37$
    & $0.60$
    \\
  \end{tabular}}

  \medskip

  \caption{Posterior analysis summary. Results of each measure for the two attackers.}%
  \label{tab:summary-analyses}

  \vspace*{-5mm plus 0.8mm minus 0.5mm}

\end{table}

\noindent
We analyze the posterior distribution to investigate what the attacker learns. \Cref{tab:summary-analyses} shows an overview of analyses for the \code{agg} example. Using multiple measures gives a multi-perspective analysis for complex problems.

To query the probability \mbox{$\Pr(x\!\mid\! \phi)$} of a random variable $x$ satisfying a predicate $\phi$, in Figaro we write \mbox{\smath{\code{alg.probability(}$x$\kern-.2mm\code{,}\kern-.2mm$\phi$\code{)}}} where \code{alg} is the inference algorithm.  Other available queries estimate the histogram of the attacker's posterior, its expectation, and variance.  The probability query allows to estimate whether an attacker learns a fact, effectively encoding a knowledge-based security policy check (\cref{sec:related}).  The strengths of an attack checking if Alice is underage in the \code{agg} example is captured by the query: \mbox{\smath{$\Pr\left(\alicerow \!<\! 18 \mid o \!\approx\! 55.3\right)$}}. The prior probability of $\alicerow \!<\! 18$ is \smath{$0.17$}. It reduces to \smath{$0.004$} for \kab and to \smath{0} for \kal in the posterior. Both attackers can conclude that Alice is an adult.  To visualize information gain, we plot the \emph{kernel density estimates}\,\cite{silverman.kde.1986}.  \Cref{fig:alice-age_CONSTANT} plots the age of Alice in the prior \mbox{\smath{$P(a)$}} and the posterior \mbox{\smath{$\Pr(a \mid o \! \approx \! 55.3)$}} for \kal. \Cref{fig:alice-age-with-observation} shows the same for \smath{\code{o = agg_p(prior_kab)}}. The plots confirm that \kal can make stronger conclusions than \kab; the posterior of the former is taller and narrower than the one of the latter; note the y-axis scale.  The uniform prior has expected value \mbox{\smath{$\Expected{\alicerow} \! \approx \! 50$}} and standard deviation \mbox{\smath{$\sigma_\alicerow \! \approx \! 29$}}.  As listed in \cref{tab:summary-analyses}, the posterior expectation increases to 55.60 for \kal with standard deviation 0.01: \kal effectively learns $a$ from the output.  For \kab the posterior has larger standard deviation (14), so \kab's uncertainty about the age of Alice is greatly reduced, yet remains high.
\looseness=-1

Moving beyond measuring and visualizing probability, we quantify attacker's learning using quantitative information flow measures: entropy, KL-divergence, mutual information, and Bayes risk. These and other measures are added to \privug\ as libraries, which estimate the corresponding measure using the samples of the MCMC algorithm of Step 4.
\looseness=-1

Shannon's \emph{entropy} quantifies the uncertainty about the value of a random variable (\eg,\,\cite{malacaria.entropy.2007,kpof.basin.adaptive.2007}). A decrease in entropy from prior to posterior signifies an increase in knowledge. Entropy (in bits) is defined as \mbox{\smath{$\entropy(x)=\sum_{x \in X} \Pr(x)\log_2\Pr(x)$}} for discrete random variables. Since \privug\ works with an inferred distribution, we estimate the entropy using the classic algorithm~\cite{ibrahim.pi.entropy.estimation.1976}, which is known to be accurate and easy to implement.  In the \code{agg} example, the entropy of $\alicerow$ in the prior is \mbox{\smath{$\entropy(\alicerow) = 6.67$}}\,bits.  At the same time, the conditional entropy of $\alicerow$ in the posterior for \kab\ is \mbox{\smath{$\entropy(\alicerow \,|\, o \approx 55.3) = 5.84$}}\,bits.  The attacker gained \smath{$0.83$}\,bits of information about the age of Alice.  For \kal, the posterior entropy is \mbox{$\entropy(\alicerow \,|\, o \approx 55.3) = -3.08$}. Here the difference is \smath{$9.75$}\,bits, twelve times more than what \kal\ learned. (The entropy of a continuous variable (replace $\sum$ with $\int$ above), \emph{differential entropy}, can be negative\,\cite{ibrahim.pi.entropy.estimation.1976}.) Clearly, \kal is an example of an attacker able to amplify the disclosed information thanks to its additional pre-existing knowledge---a situation often referred to as a \emph{linking attack}. The ability of \kab in this respect is much weaker.  \privug allows experimenting with the attacker space in this way, to let the data controller understand what attacks are successful, and then assess whether they are of concern.
\looseness=-1

Relative entropy~\cite{kullback.leibler.kldivergence.1951} or \emph{KL-divergence} measures how much two distributions differ.  In Bayesian inference, the KL-divergence of a posterior \smath{$P(x)$} and a prior \smath{$Q(x)$}, defined as \mbox{\smath{$\dkl(P \doublemid Q) = \sum P(x)\log_2(P(x)/Q(x))$}}, expresses the amount of \emph{``information lost when \smath{$Q$} is used to approximate \smath{$P$}''}~\cite[page 51]{burnham.anderson.klinformationgain.2002}.  Thus, KL-divergence is a measure of information gained by revising one's knowledge of the prior to the posterior.  As with entropy, since we are working with an inferred distribution, we can estimate KL-divergence from samples.  We use the algorithm by Wang \etal\,\cite{wang.sanjeev.verdu.kldivergence.2005}.  For the aggregation example, the KL-divergence between the posterior and prior of $\alicerow$ is a measure of the amount of information that the attacker gained about Alice's age by observing the output of the program.  For \kab, \mbox{\smath{$\dkl(\alicerow \,|\, o \approx 55.3 \doublemid \alicerow) = 0.77$}}.  For \kal, on the other hand, \mbox{\smath{$\dkl(\alicerow \,|\, o \approx 55.3 \doublemid \alicerow) = 5.64$}}.  These results indicate that the observation yields an information gain of \smath{$0.77$} bits for \kab and \smath{$5.64$} bits for \kal. More important is the difference; \kal gains over \smath{$7$} times more information than \kab. In \cref{subsec:dp} we show how KL-divergence can be used to measure utility when programs add noise to their output.
\looseness=-1

\emph{Mutual information} between two random variables $x$, $y$, defined as $\mi(x;y)=\sum_{y\in Y}\sum_{x \in X}\Pr(x,y)\log_2(\Pr(x,y)/\Pr(x)\Pr(y))$, measures the reduction of the uncertainty of $x$ by the knowledge of $y$ \cite{elementsofinformationtheory.2006}.  We estimate $\mi(i;o)$ where $i$ is a secret input and $o$ a public output (the attacker's prediction) to quantify how much information $o$ shares with~$i$.  Mutual information is well studied as a quantitative information flow measure (cf.~\cref{sec:related}).  A privacy protection mechanism typically aims at minimizing $\mi(i;o)$.  In \privug, we use the mutual information estimator\,\cite{PhysRevE.69.066138} provided by SKlearn\,\cite{scikit-learn} for continuous variables and LeakiEst\,\cite{chothia.leakest.2013} for discrete variables. In our example, we have $\mi(a;o) = 9.37$ bits for $\kal$ and $\mi(a;o) = 0.60$ bits for $\kab$.  This is consistent with our intuition: when the attacker knows everything about the input except for Alice's age, observing the output greatly reduces their uncertainty.
\looseness=-1

\privug can incorporate estimators of other measures.  In \cref{sec:evaluation} we show that other tools can be incorporated on the example of F-BLEAU\,\cite{cherubin.fbleau.2019}, to estimate \emph{Bayes risk}---the expected probability of an attacker guessing a secret ($s$) by observing the output of the program ($o$); formally: \mbox{\smath{$1 - \sum_{o \in O} \max_{s \in S} \Pr(o|s) \Pr(s)$}} for random variables $s$ and $o$~\cite{qifbook.2020}.
\looseness=-1

\bigskip

\begin{figure}[t!]

  \centering

  \includegraphics[
    width=\linewidth,
    clip,
    trim = 20mm 2mm 20mm 2mm
  ]{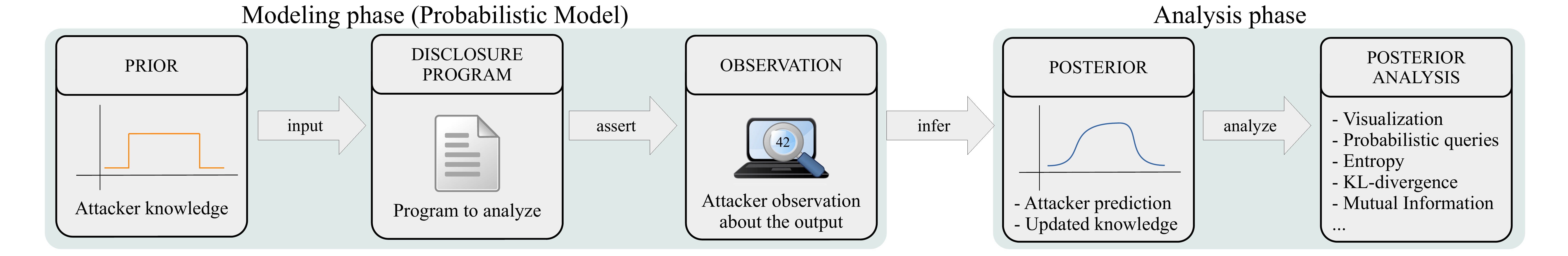}

  \caption{\label{fig:framework-overview} Overview of the steps in \privug method.}

  \vspace{-2.5mm plus 0.5mm minus 0.5mm}

\end{figure}

\noindent
This way we determine if specific attackers are capable of learning secrets, and assess whether disclosing the output of the program poses a privacy risk.  \Cref{fig:framework-overview} gives an overview of the steps in the \privug method.   \privug's intended users are data analysts with knowledge in statistics and probabilistic modeling.\label{required-expertise} These users are typically trained in probabilistic programming, an essential part of their toolbox~(\eg,\,\cite{BDA.gelman.2013}). This makes it easy to perform steps 1, 3, and 4.  Step 2 typically requires simply updating the datatypes of the input (as in \code{agg} and \code{ano}).  The analyst may, however, need to change the program to ensure differentiability, or replace some operators with their probability counterparts.  These are the same techniques that data analysts use to create advanced probabilistic models and analyses.  Step 5's probability queries, visualizations, and distribution statistics such as expected value or variance, are likewise familiar to data analysts.  The interpretation of leakage does, however, require privacy-specific expertise (information theory, quantitative information flow).
\looseness=-1

The results and conclusions drawn do depend on the to choice of prior.  The prior models what an attacker knows about the input of the program, the secondary knowledge that \emph{linked} with the observed output can lead to privacy violations.  The analysis may report no leakage if priors do not reflect the real information that an attacker has access to.  Ideally, priors should be informed from real world data.  For instance, if the program takes as input a set of records of US citizens, then it is advisable to inform priors from publicly available sources, \eg, the US census.  Alternatively, probabilistic programming frameworks can be used to automatically learn underlying distributions from data with better accuracy than simply using the empirical distributions~\cite{BDA.gelman.2013}.  For mutual information and multiplicative Bayes capacity (a derived measure from Bayes risk, it has been shown that running the analysis with uniform priors uncovers leakage if it exists, see\,\cite[Theorem 4]{uniform.leakage.2013} and \cite[Theorem 7.2]{qifbook.2020}.  This result can be used with good effect to detect leakage, but not to estimate its magnitude, which can be estimated using \privug.
\looseness=-1

\begin{figure}[p]%
%
%
  \begin{minipage}{\PLOTTW}
    \includegraphics[
      height=\PLOTTH,
      clip,
      trim = -2.05mm 4.5mm 0 -0.7mm,
    ]{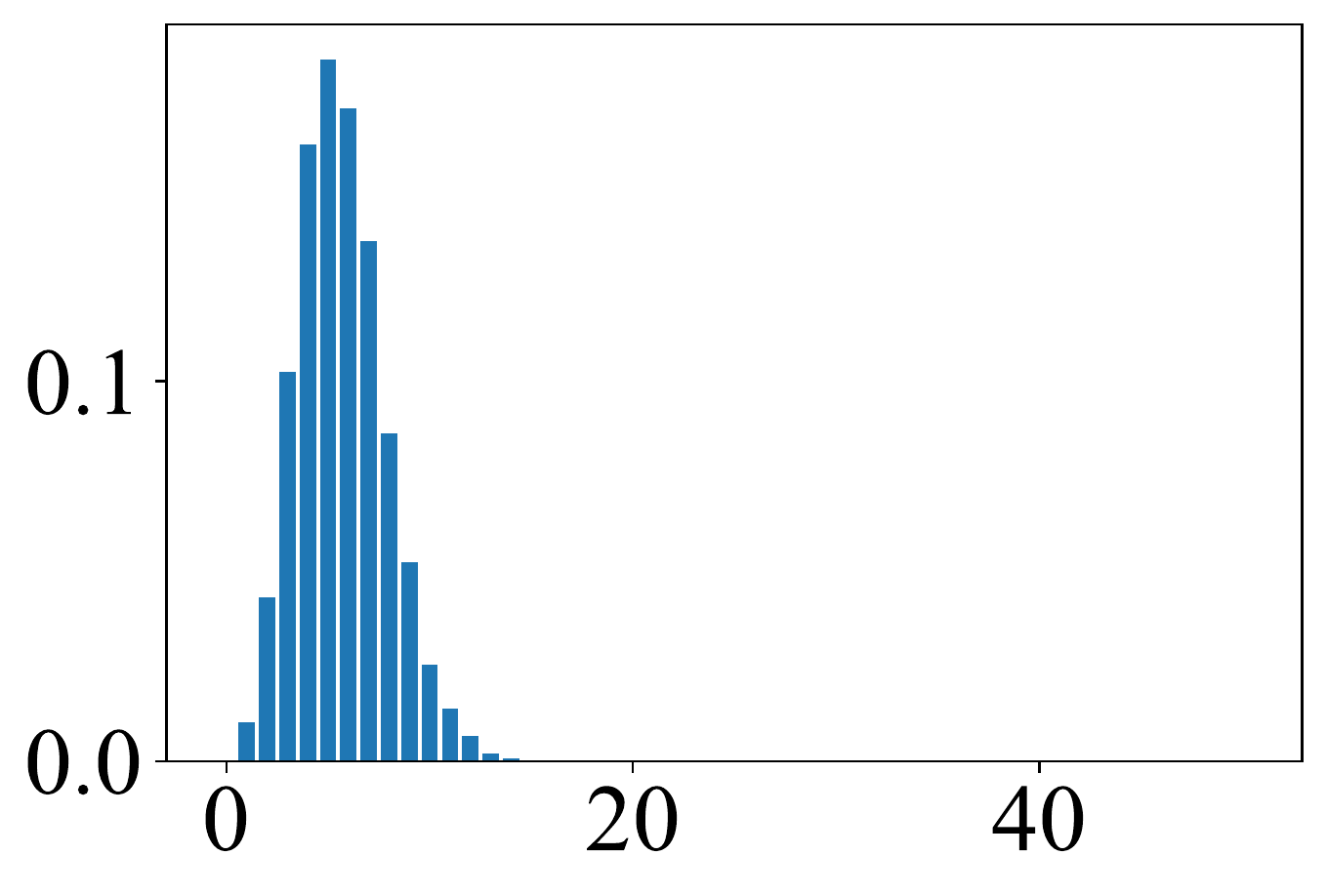}%
  \end{minipage}{\labelT{fig:numzip}}%
  \begin{minipage}{\PLOTTW}
    \includegraphics[
      height=\PLOTTH,
      clip,
      trim = 2.9mm 4.5mm 0 -0.7mm,
    ]{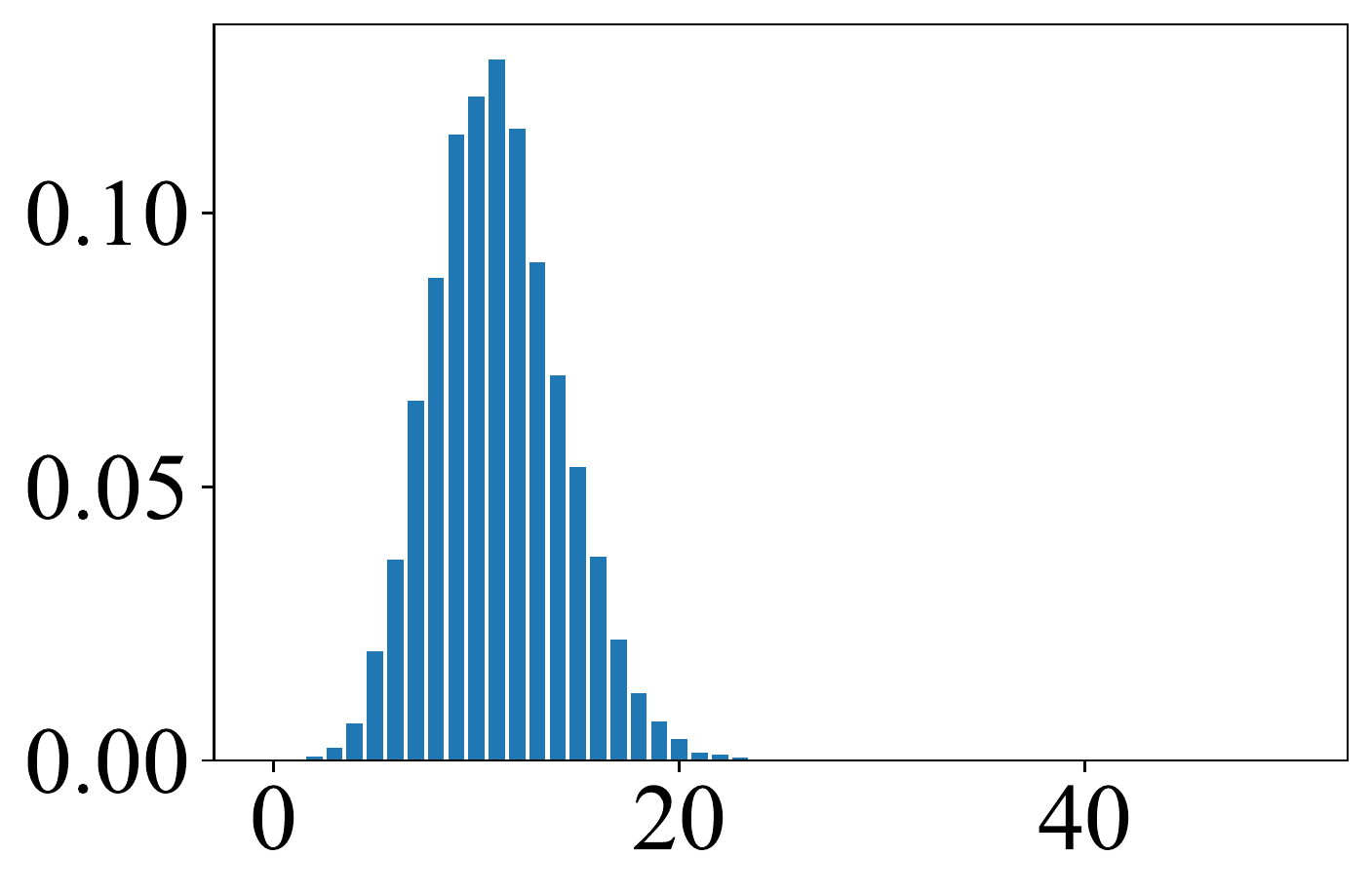}%
  \end{minipage}{\labelT{fig:numbday}}%
  \begin{minipage}{\PLOTTW}
    \includegraphics[
      height =\PLOTTH,
      clip,
      trim = 2.9mm 4.5mm 0 -0.7mm,
    ]{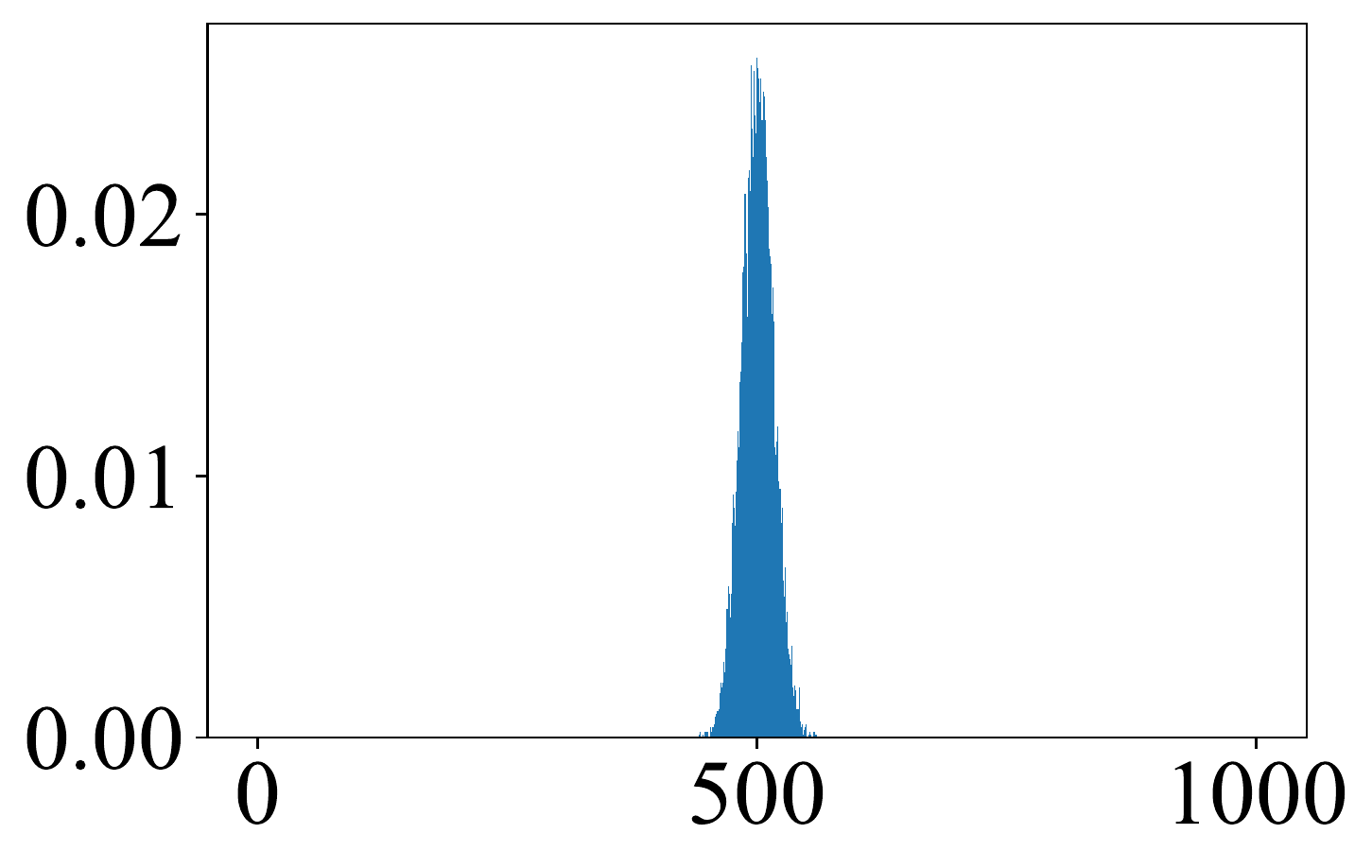}%
  \end{minipage}{\labelT{fig:numsex}}%
  \begin{minipage}{\PLOTTW}
    \includegraphics[%
      height = \PLOTTH,
      clip,
      trim = 8.35mm 14.9mm 0 -0.7mm,
    ]{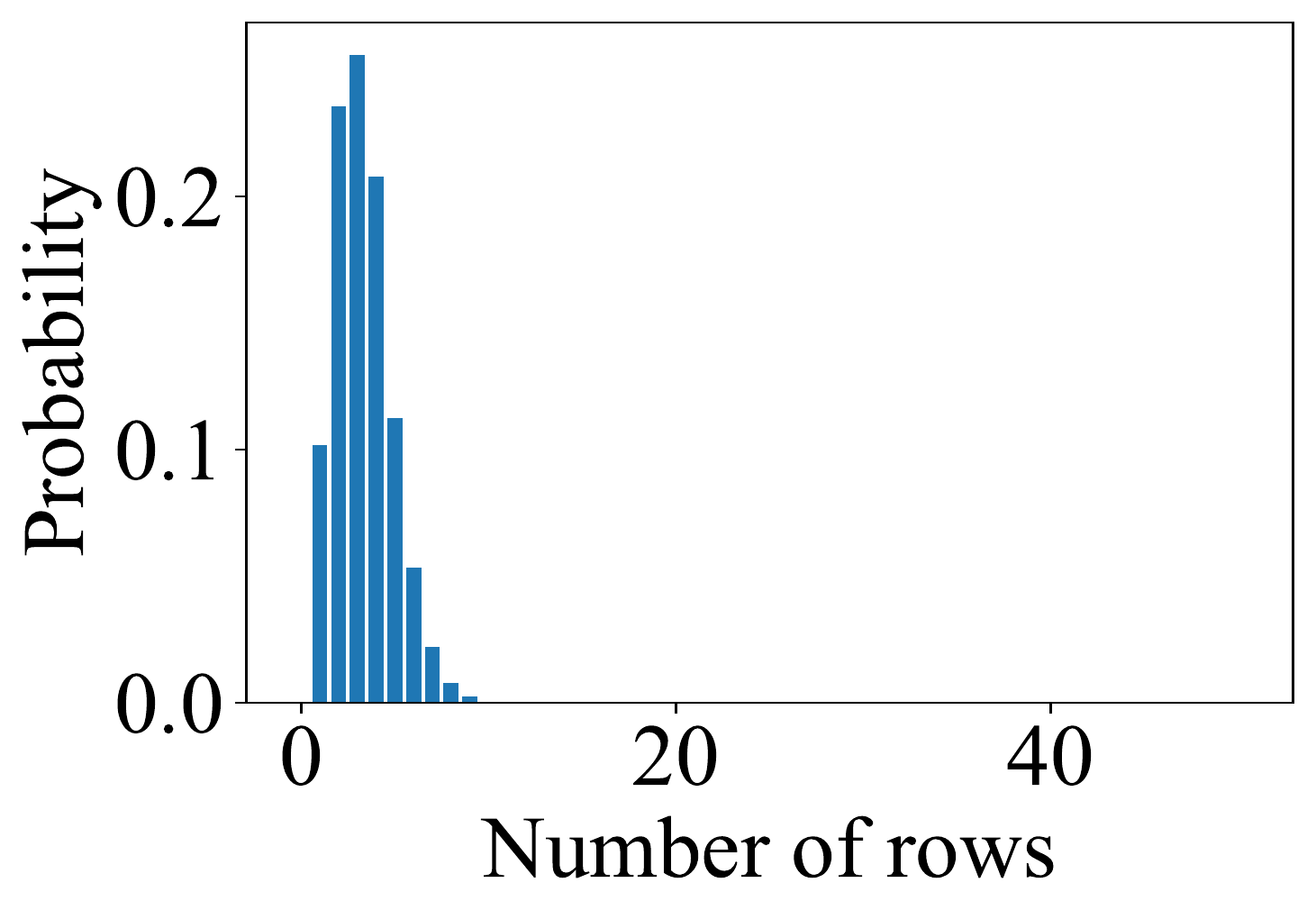}%
  \end{minipage}{\labelT{fig:numzipsex}}%
  \begin{minipage}{\PLOTTW}
    \includegraphics[
      height=\PLOTTH,
      clip,
      trim = -2.05mm 4.5mm -1.55mm -0.7mm,
    ]{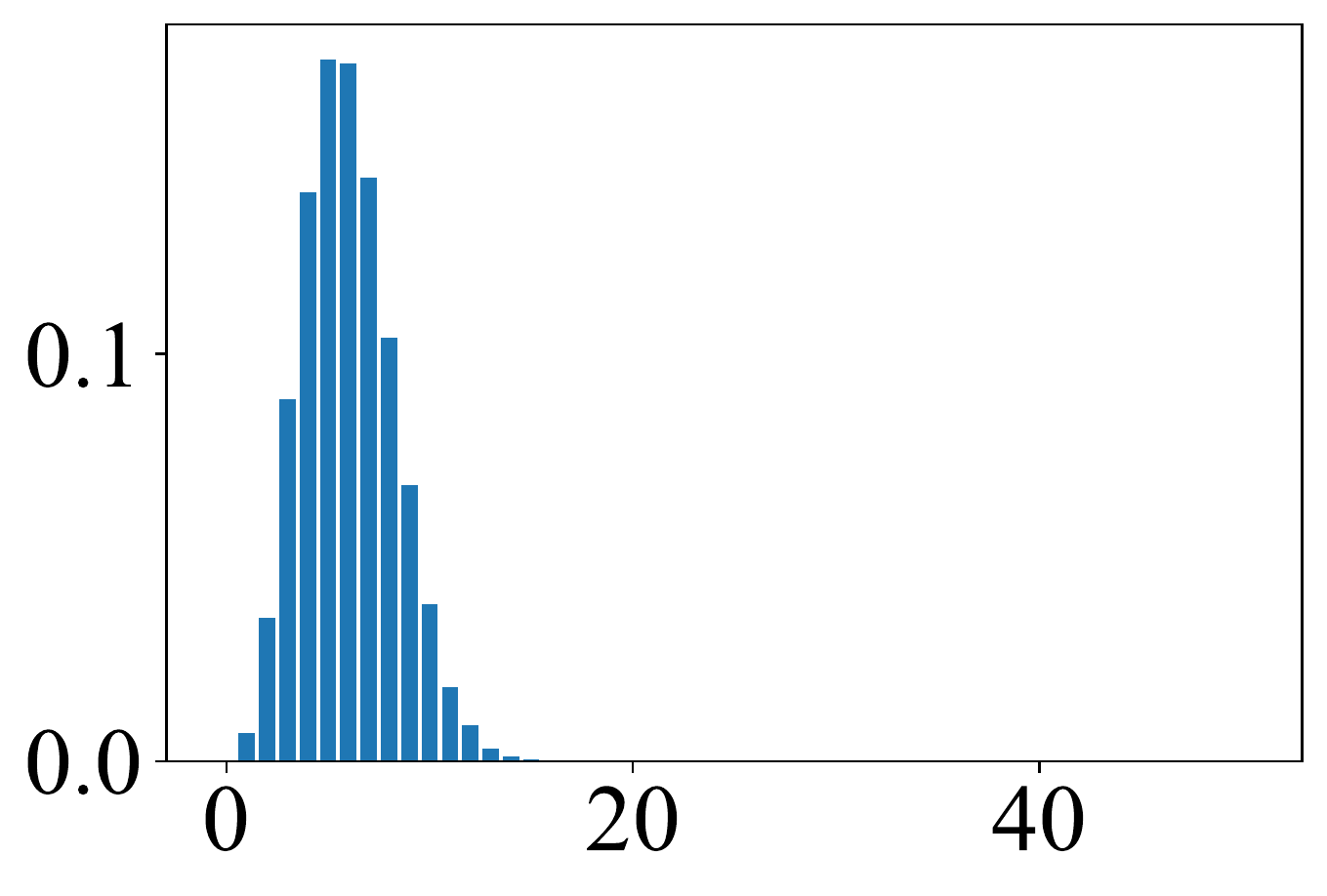}%
  \end{minipage}{\labelT{fig:numbdaysex}}%
  \begin{minipage}{\PLOTTW}
    \includegraphics[
      height=\PLOTTH,
      clip,
      trim = -2.05mm 4.5mm 0 2.65mm,
    ]{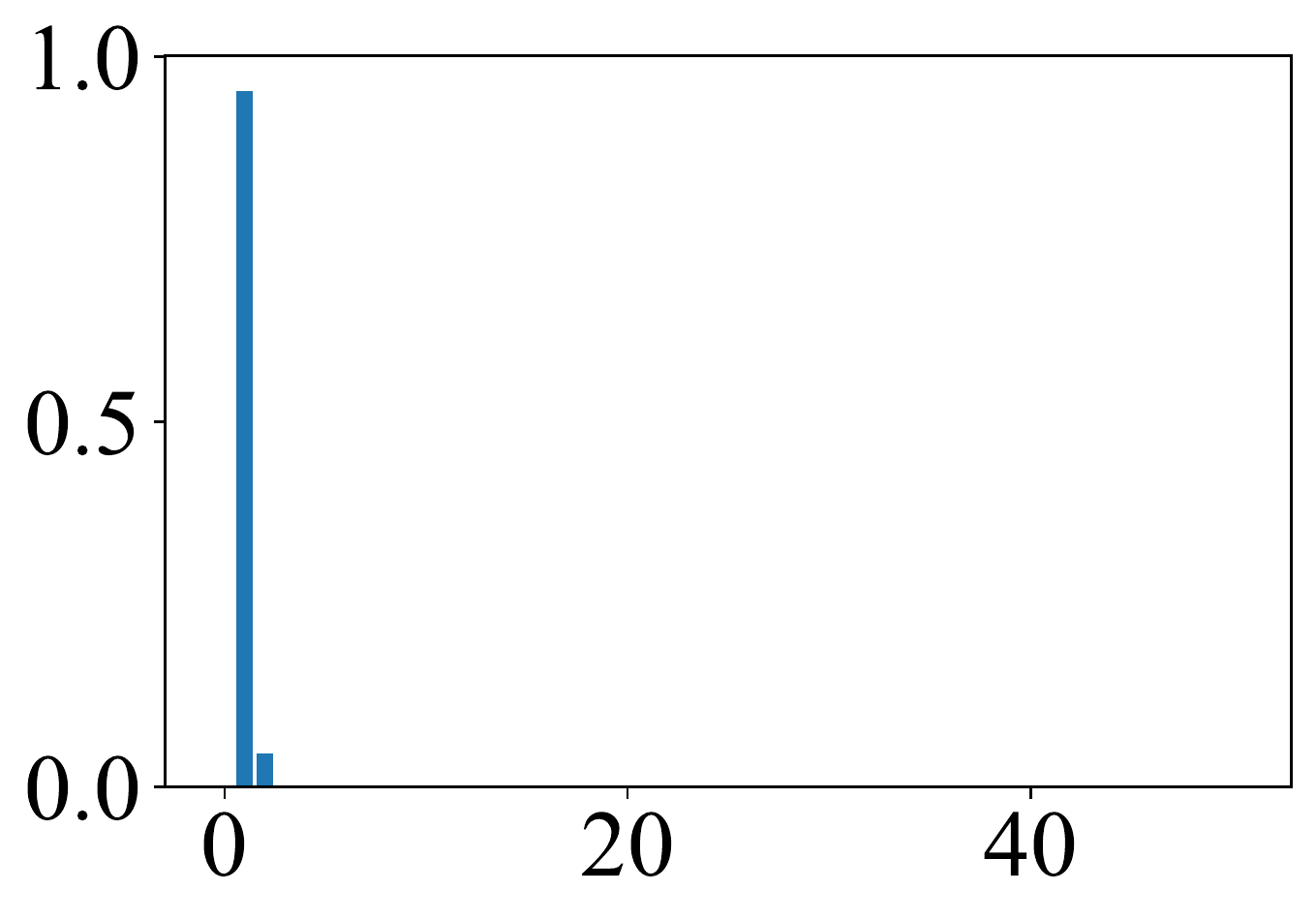}%
  \end{minipage}{\labelT{fig:numzipbday}}%

  \begin{minipage}{\PLOTTW}
    \includegraphics[
      height=\PLOTTH,
      clip,
      trim = -2.05mm 4.5mm 0 0.75mm,
    ]{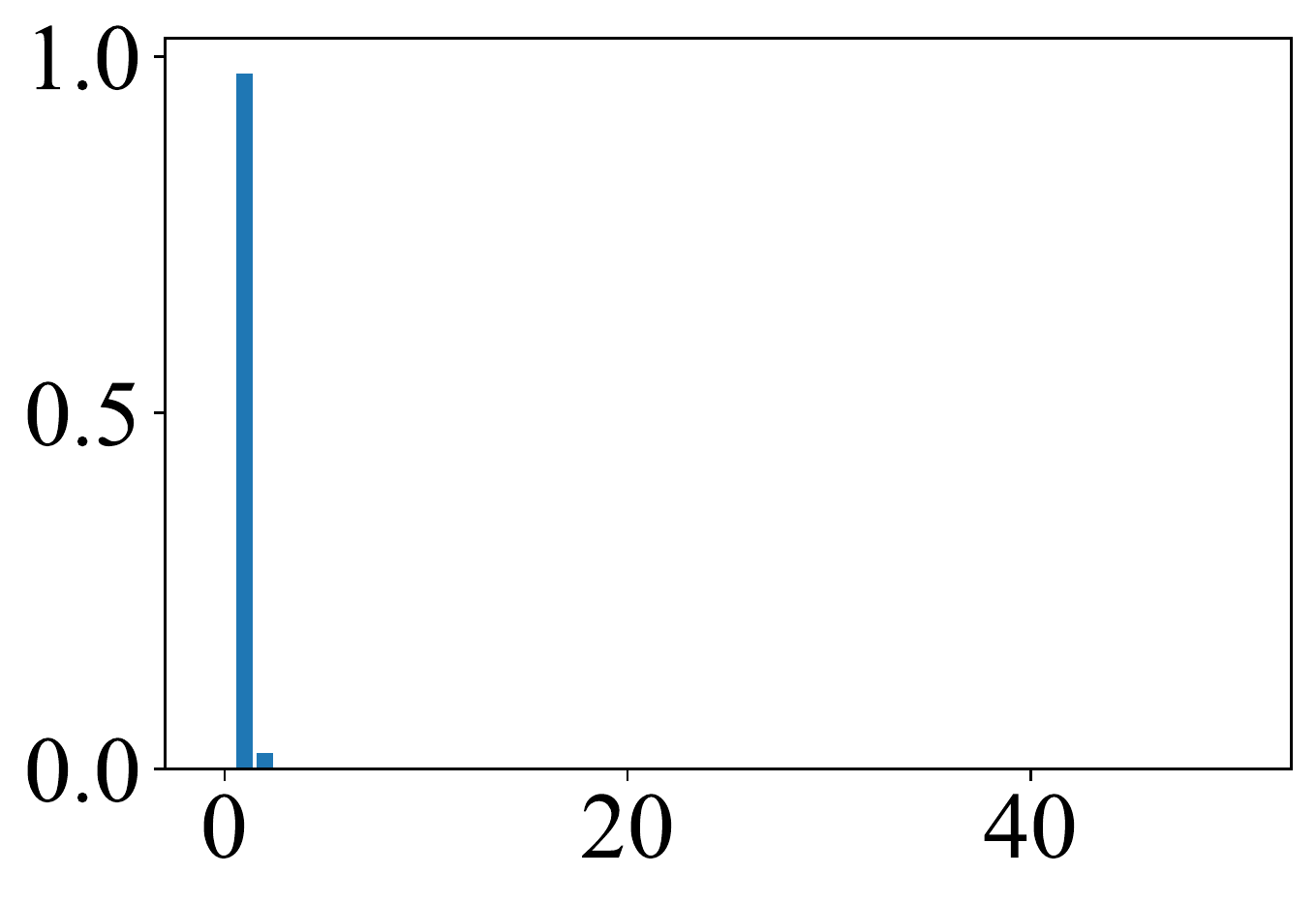}%
  \end{minipage}{\labelT{fig:numzipbdaysex}}%
  \begin{minipage}{\PLOTTW}
    \includegraphics[
      height=\PLOTTH,
      clip,
      trim = 8.35mm 14.9mm 0 -0.7mm,
    ]{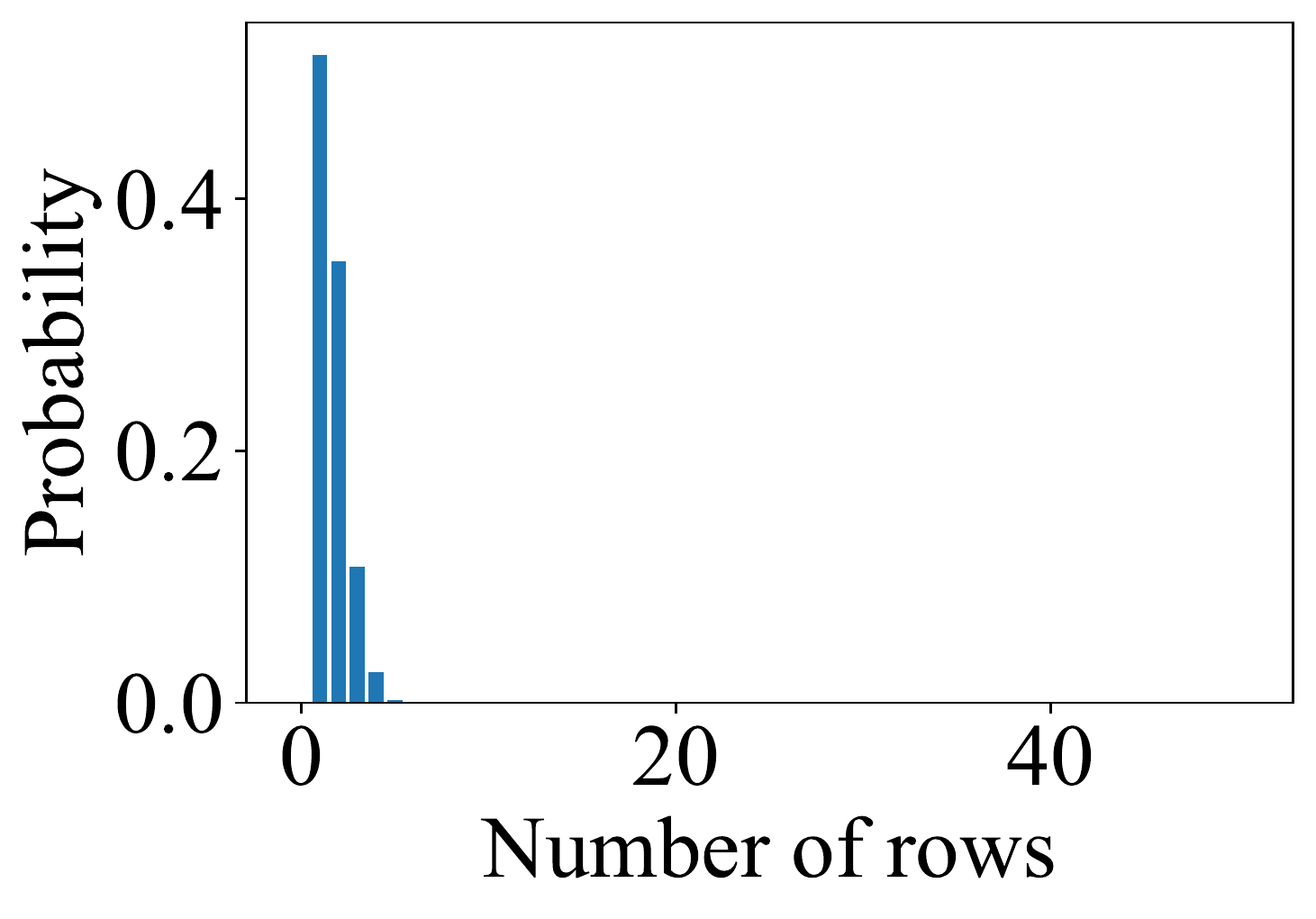}%
  \end{minipage}{\labelT{fig:uniqueness-14000-zipbday}}%
  \begin{minipage}{\PLOTTW}
    \includegraphics[
      height=\PLOTTH,
      clip,
      trim = -2.05mm 4.5mm 0 -0.7mm,
    ]{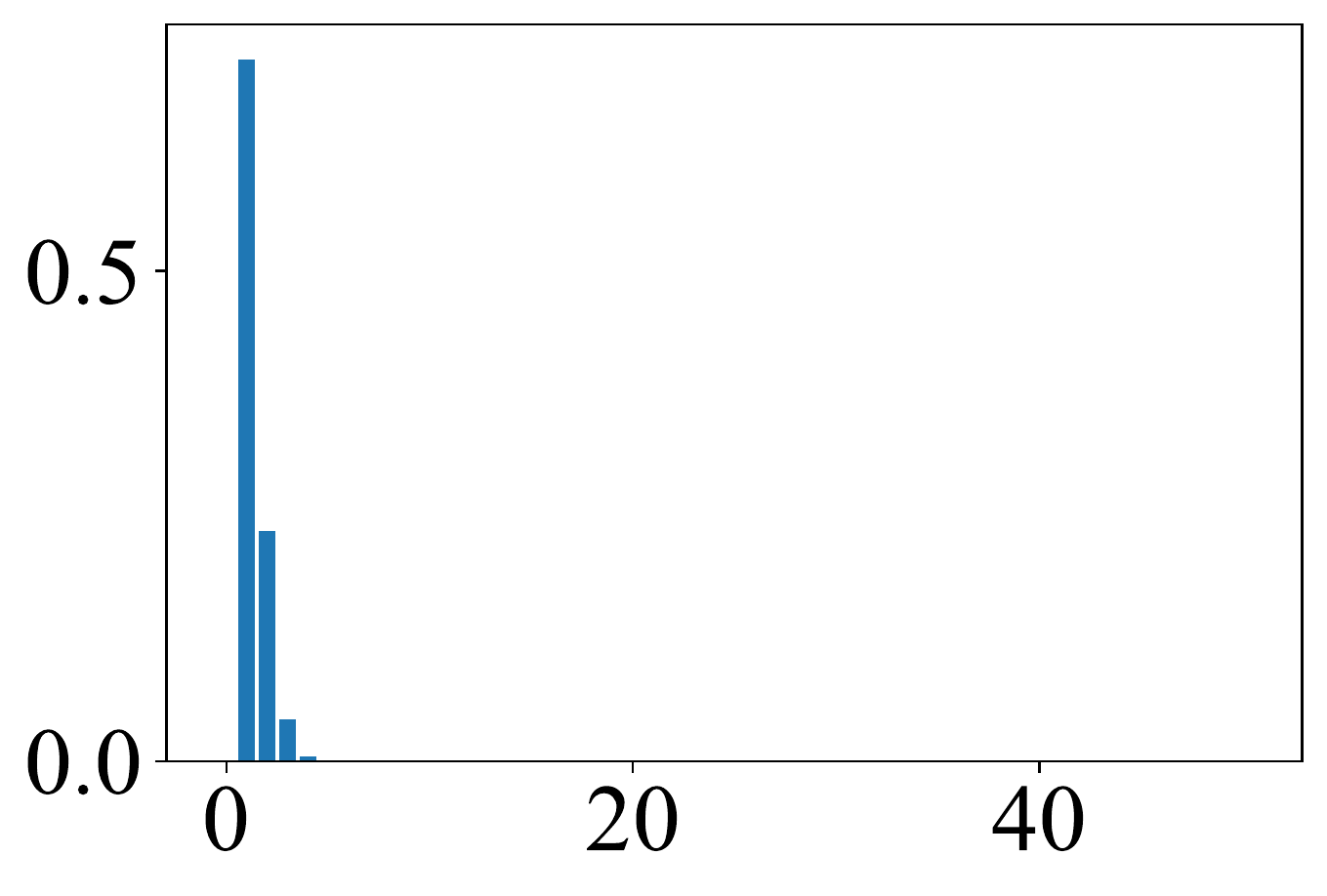}%
  \end{minipage}{\labelT{fig:uniqueness-14000-zipbdaysex}}%
  \begin{minipage}{\PLOTTW}%
    \includegraphics[
      height=\PLOTTH,
      clip,
      trim = 8.35mm 14.9mm 0 -0.7mm,
    ]{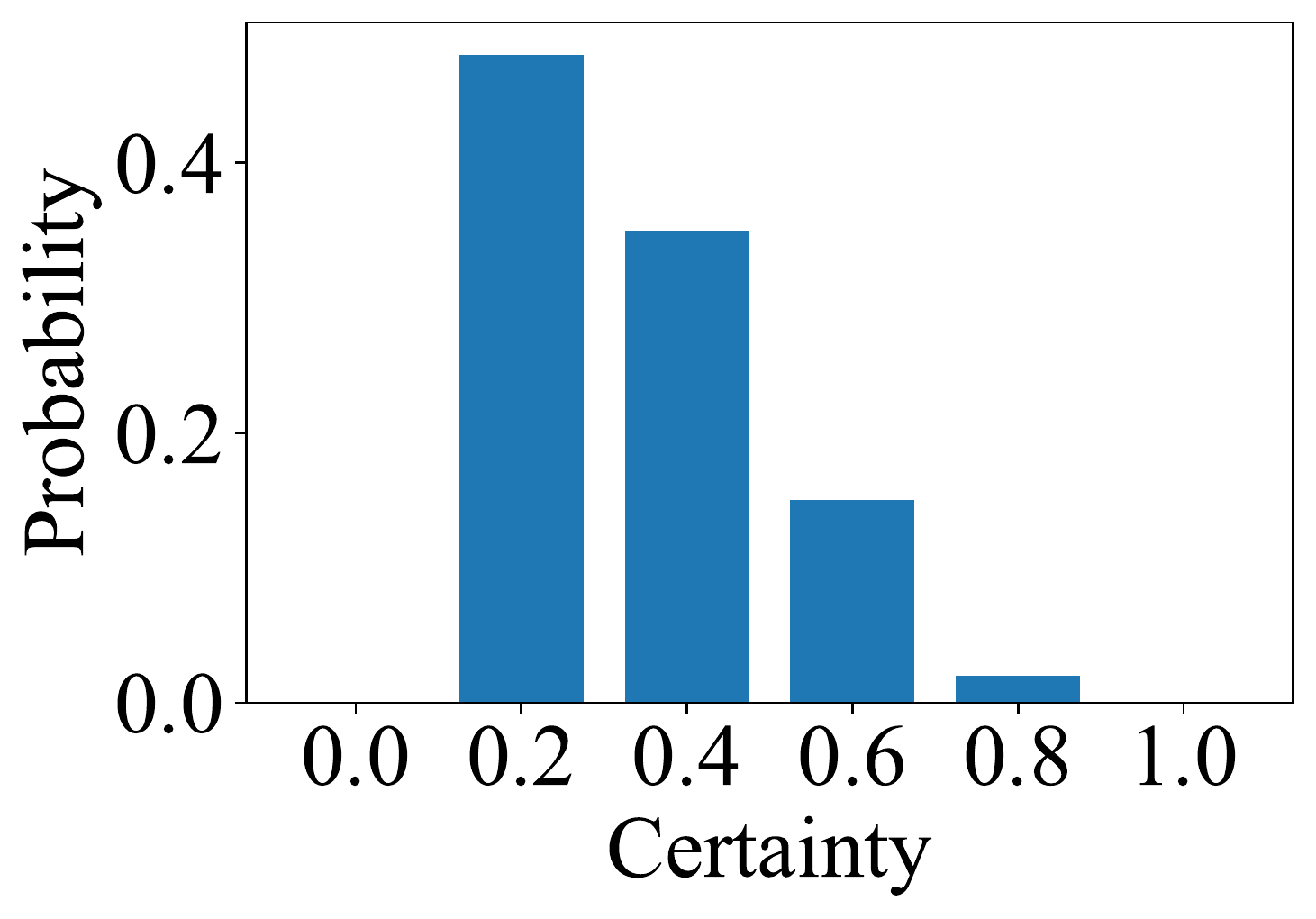}%
  \end{minipage}{\labelT{fig:illzip}}%
  \begin{minipage}{\PLOTTW}
    \includegraphics[
      height=\PLOTTH,
      clip,
      trim = 13.25mm 14.9mm 0 -0.7mm,
    ]{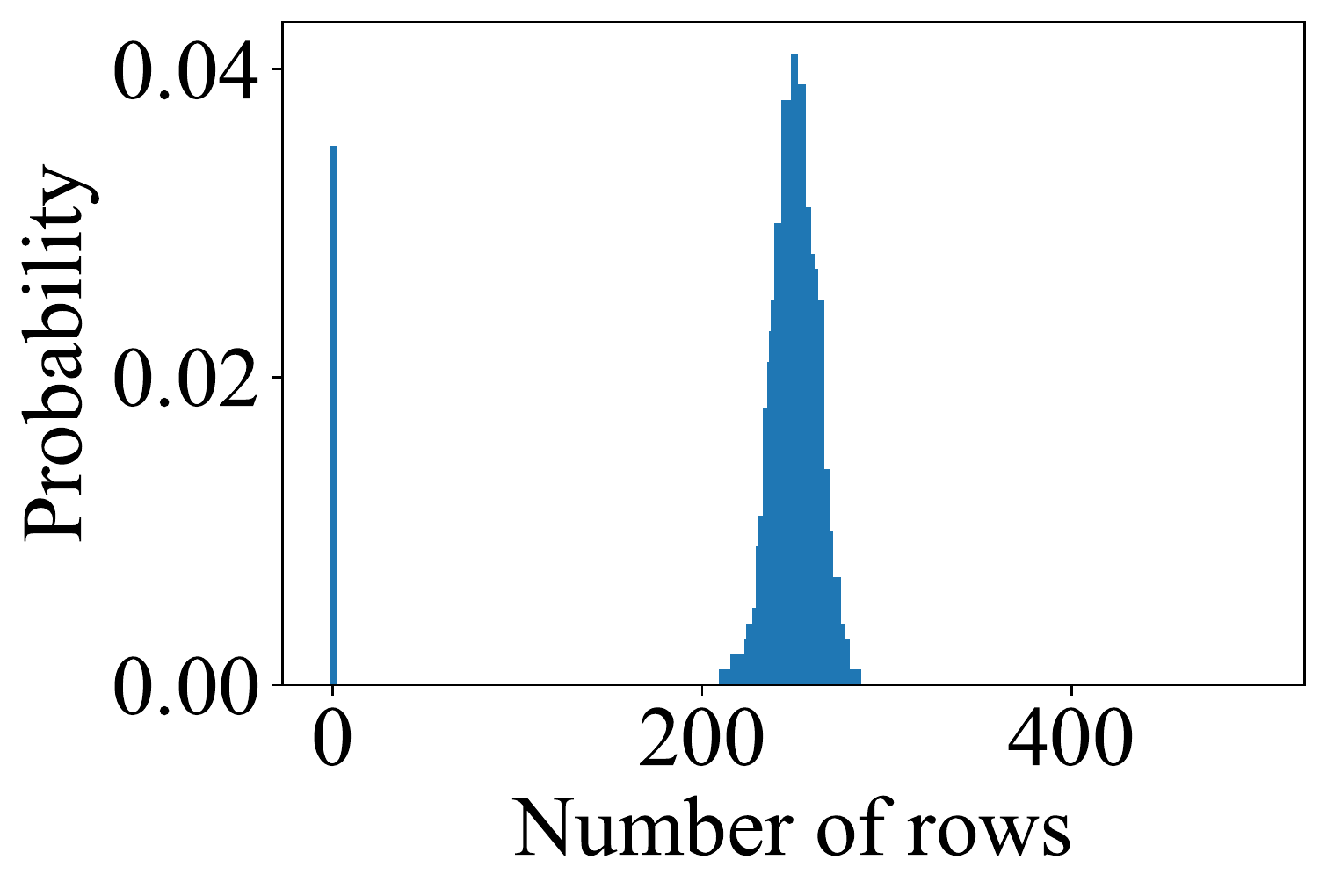}%
  \end{minipage}{\labelT{fig:uniqueness-k-anonymity-sex}}%
  \begin{minipage}{\PLOTTW}%
    \includegraphics[%
      height = \PLOTTH,
      clip,
      trim = -2.05mm 4.5mm 0 -0.7mm,
    ]{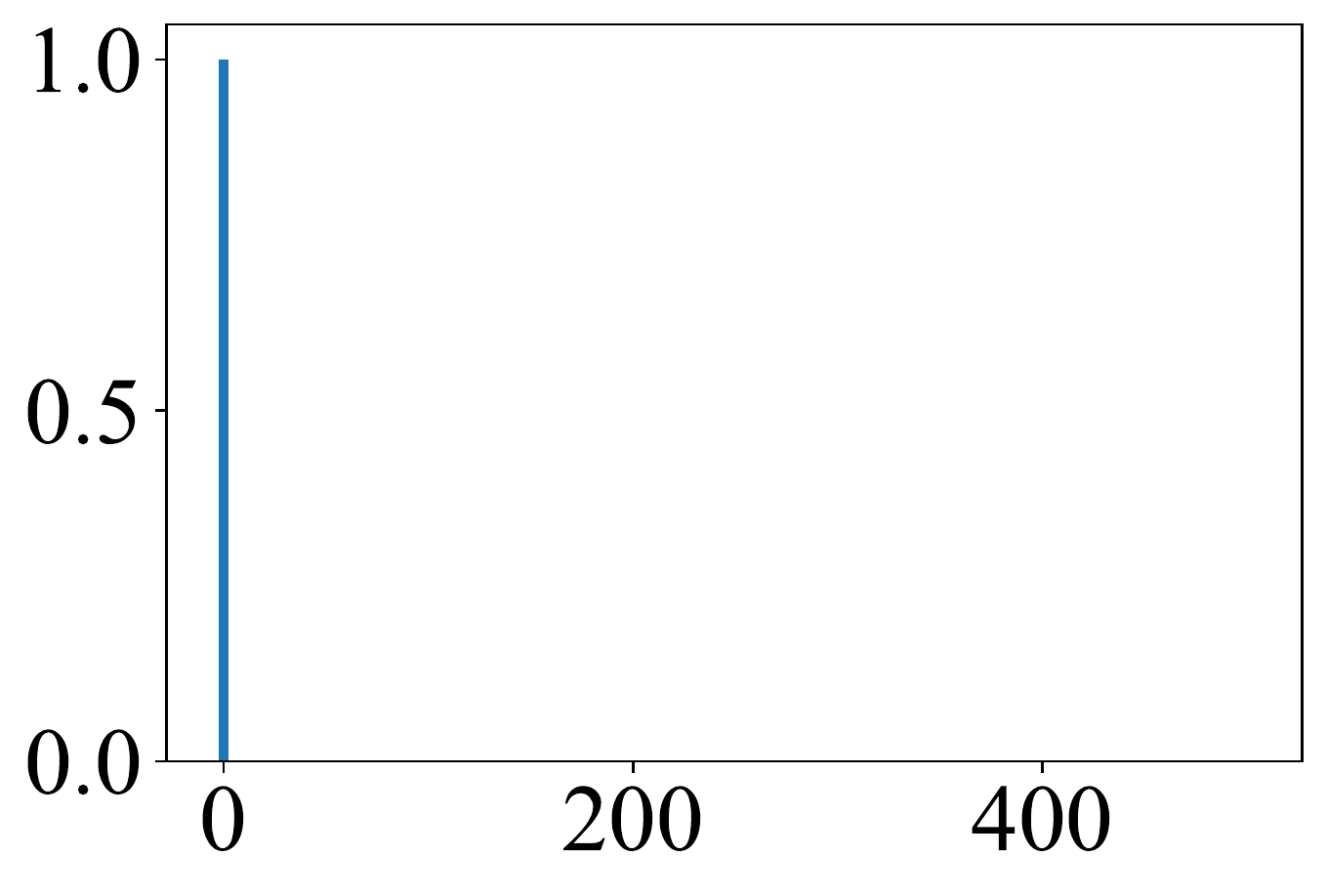}%
  \end{minipage}{\labelT{fig:uniqueness-k-anonymity-others}}%

  \begin{minipage}{\PLOTSW}%
    \includegraphics[
      height=\PLOTSH,
      clip,
      trim = -9.8mm 0.95mm 0 -5.2mm
    ]{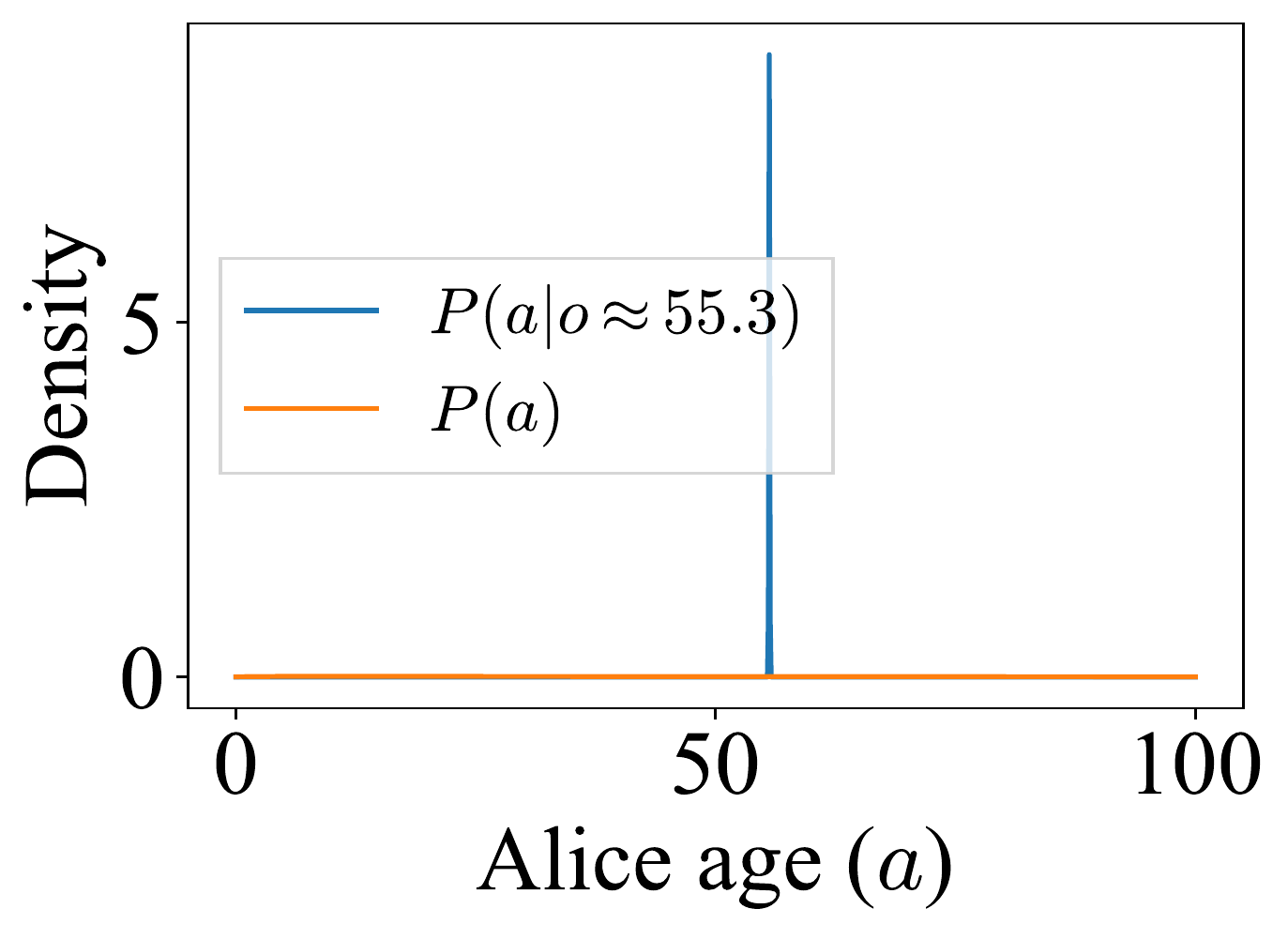}%
  \end{minipage}{\labelS{fig:alice-age_CONSTANT}}%
  \begin{minipage}{\PLOTSW}%
    \includegraphics[
      height=\PLOTSH,
      clip,
      trim = 2.55mm 0.95mm 0 -5.2mm
    ]{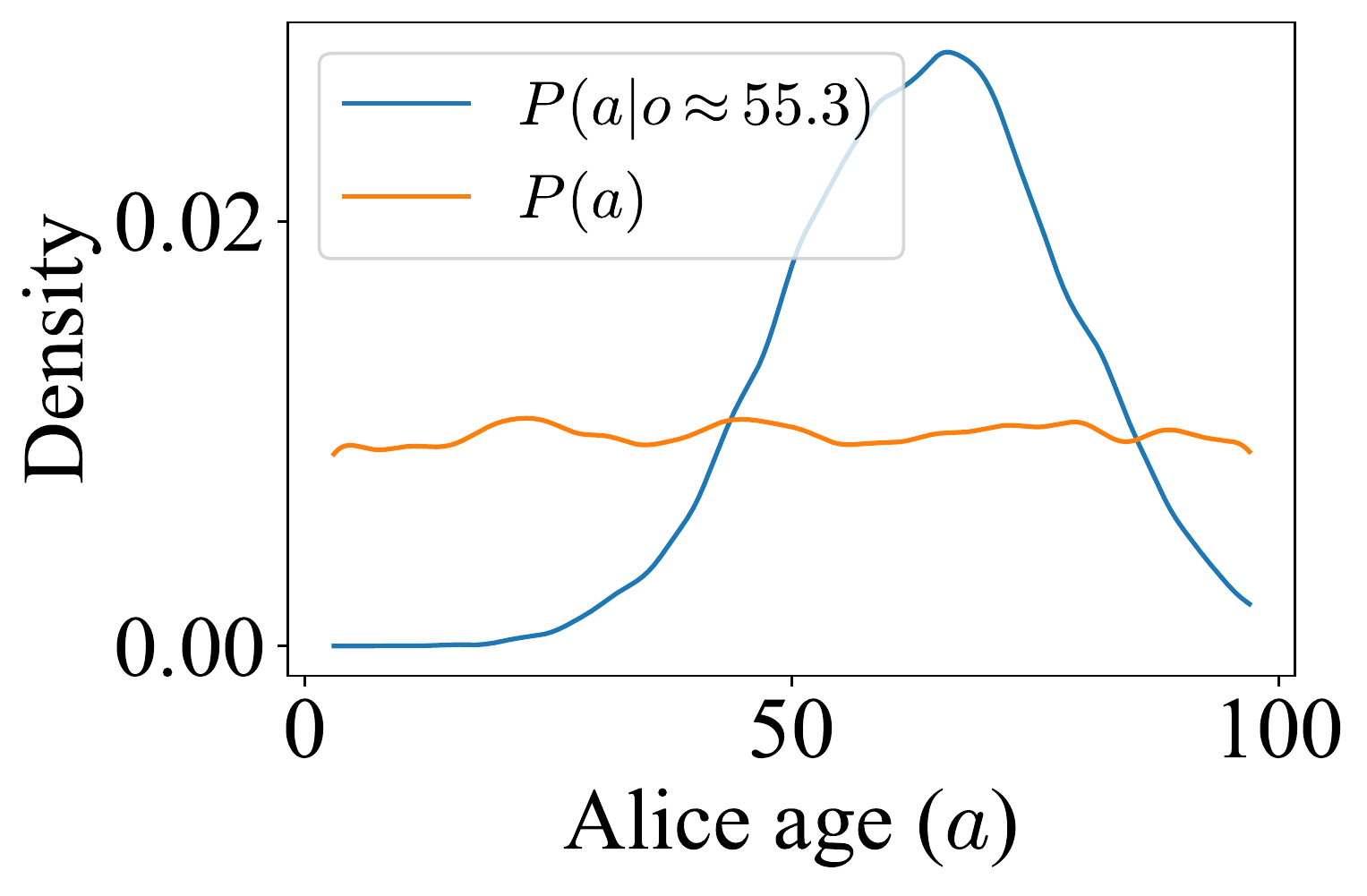}%
  \end{minipage}{\labelS{fig:alice-age-with-observation}}%
  \begin{minipage}{\PLOTSW}%
    \includegraphics[%
      height=\PLOTSH,
      clip,
      trim = -17.2mm 0.45mm 0 -5.2mm
    ]{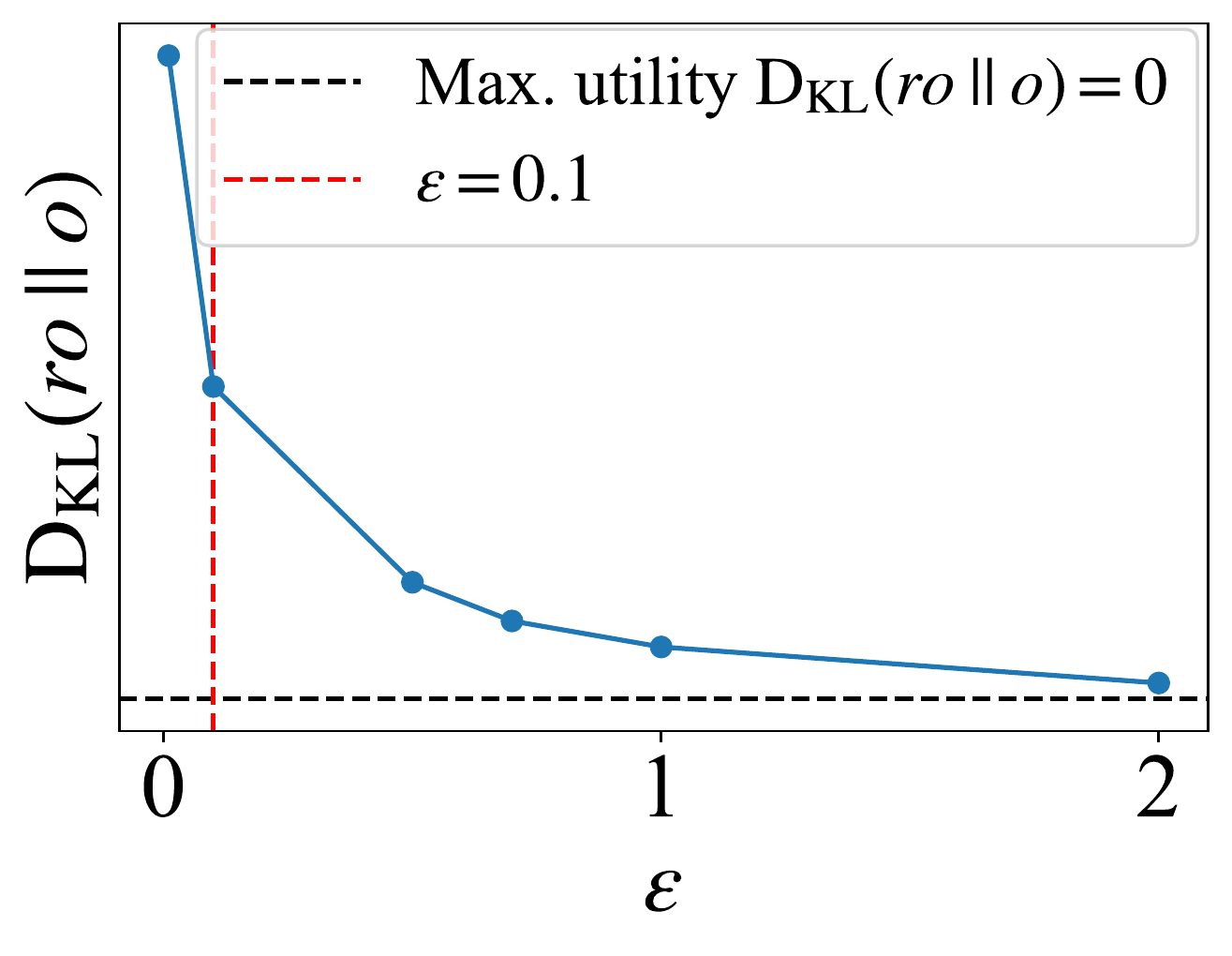}%
  \end{minipage}{\labelS[1.6mm]{fig:utility-dp}}%
  \begin{minipage}{\PLOTSW}%
    \includegraphics[%
      height=\PLOTSH,
      clip,
      trim = -17.2mm 0.45mm 0 -5.2mm
    ]{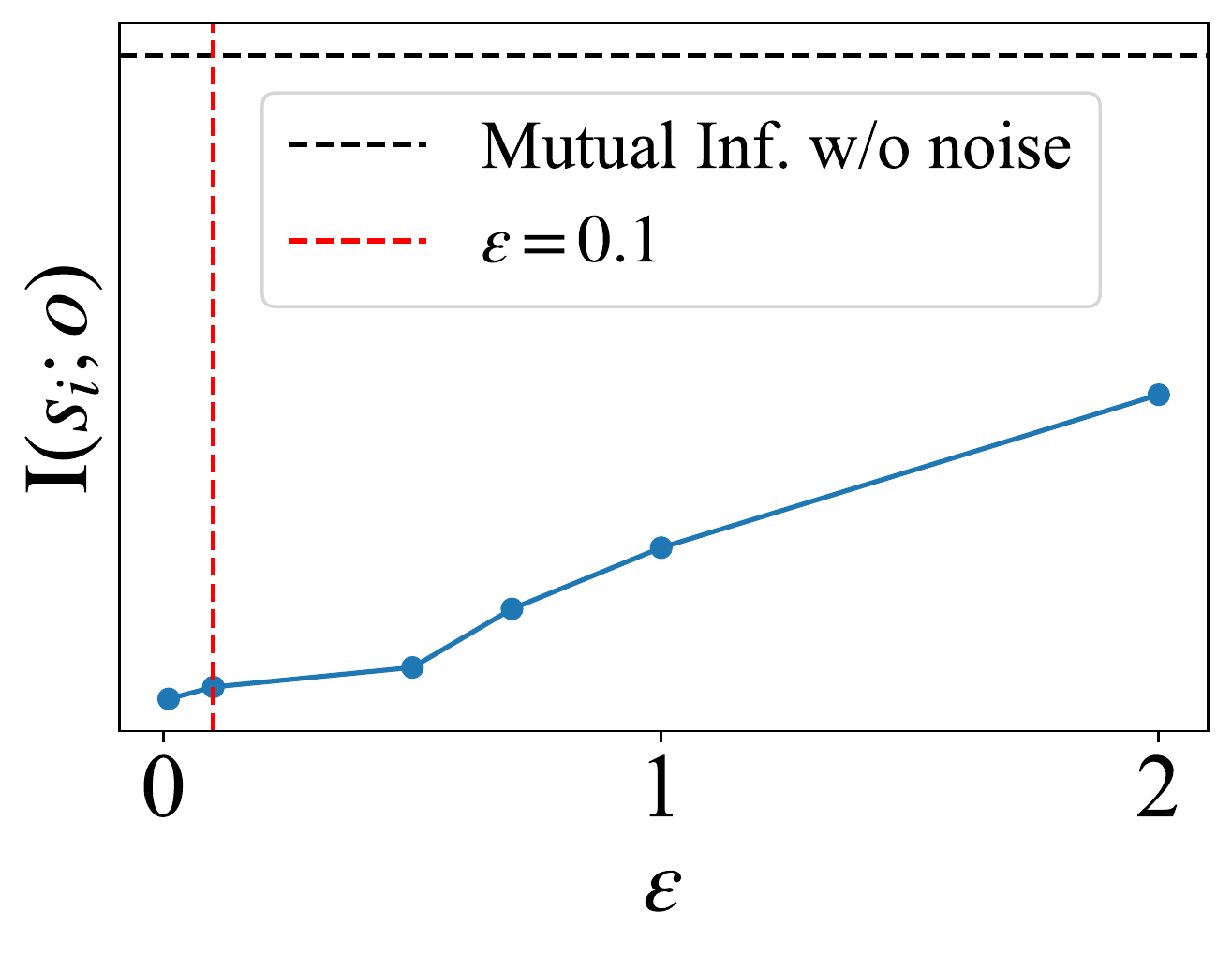}%
  \end{minipage}{\labelS[-4mm]{fig:mi-dp}}%
  \vspace{-.3em}
  
  \begin{minipage}{\PLOTSW}%
    \includegraphics[%
      height=\PLOTSH,
      clip,
      trim = -17mm 1.1mm 0 -5.2mm
    ]{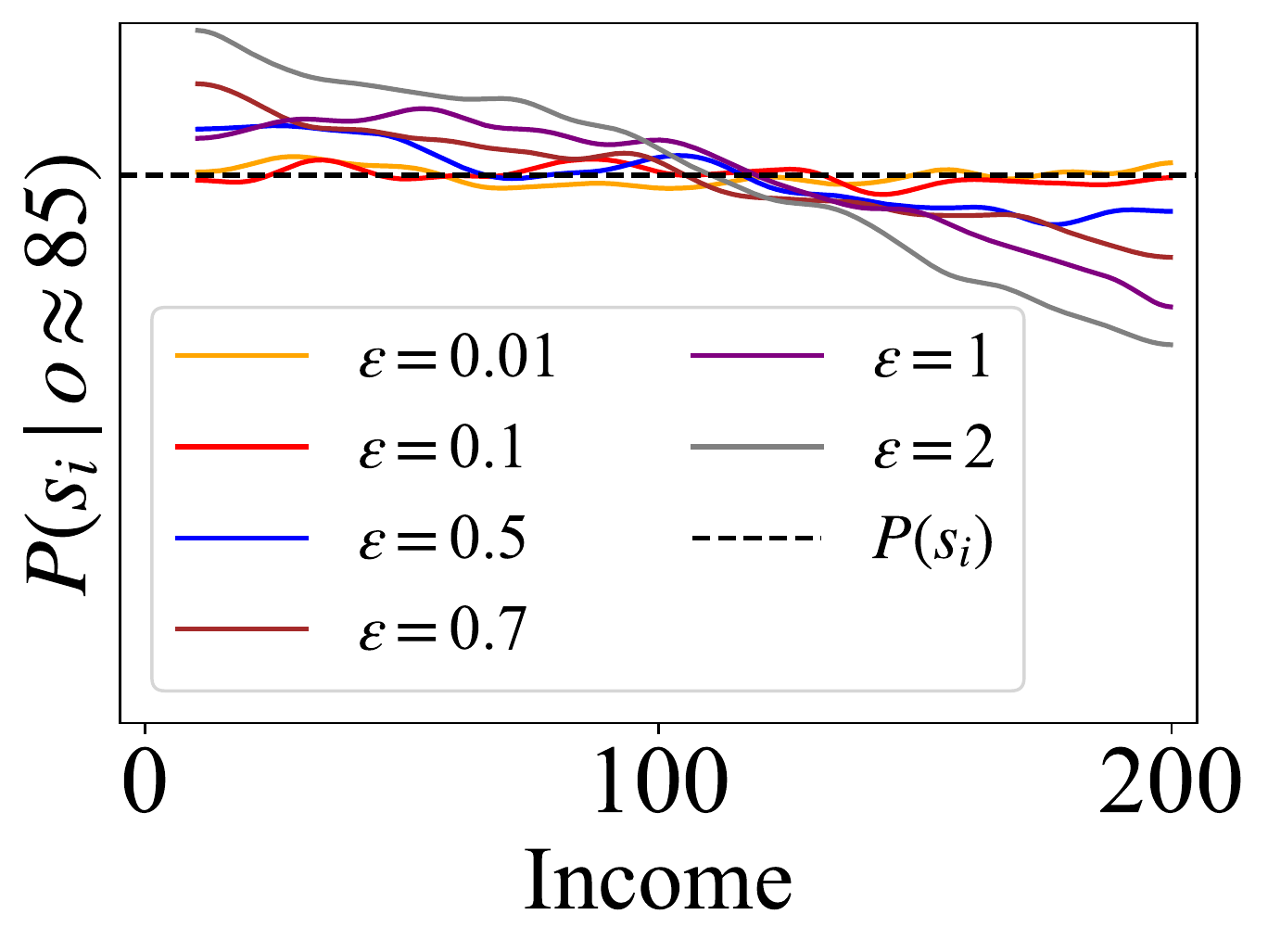}%
  \end{minipage}{\labelS[-4mm]{fig:kbp-dp}}%
  \begin{minipage}{\PLOTSW}%
    \includegraphics[
      height=\PLOTSH,
      clip,
      trim = -2.4mm 0.45mm 0 -4.15mm
    ]{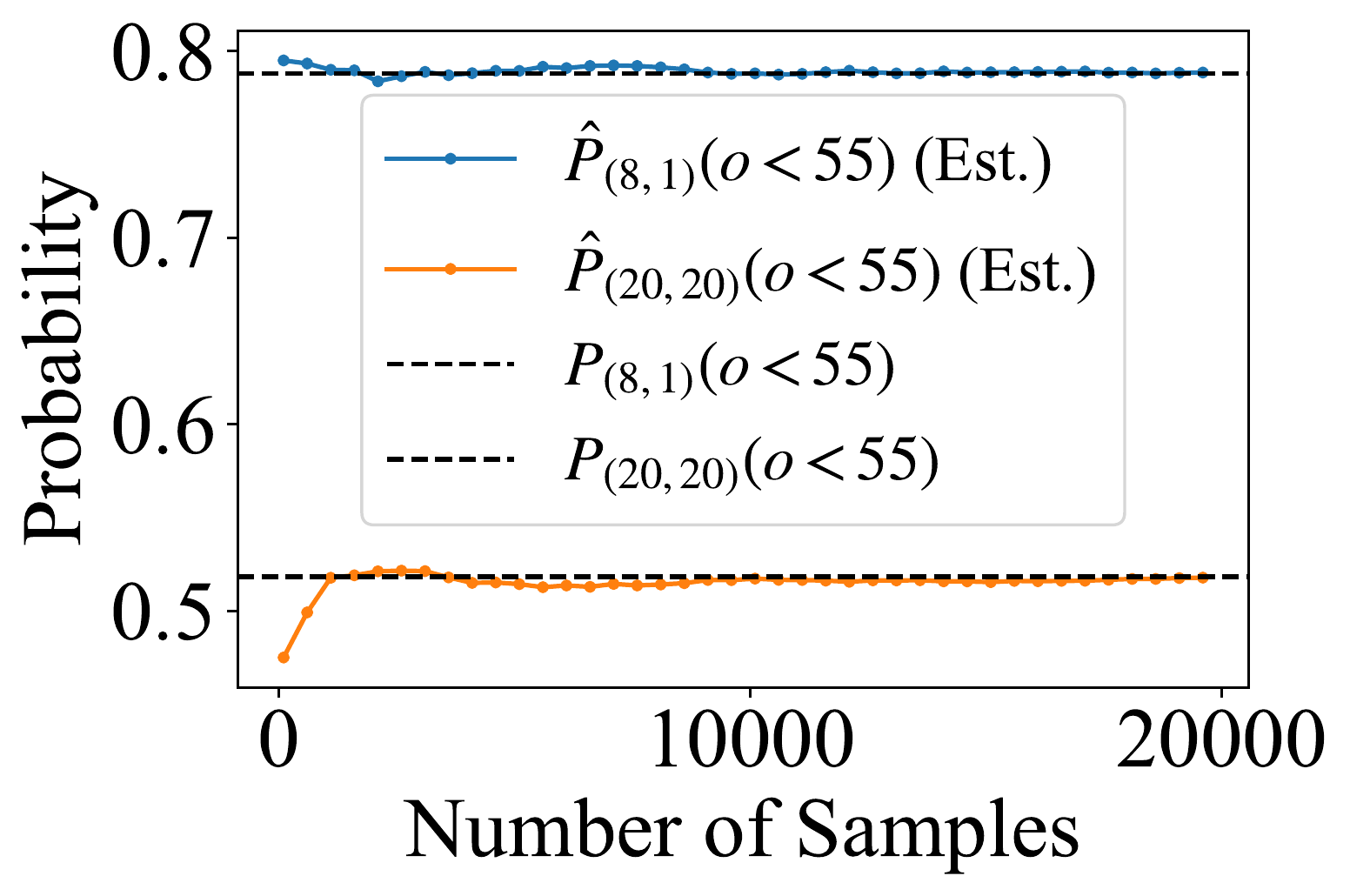}%
  \end{minipage}{\labelS[1.6mm]{fig:convergence-probability-query}}%
  \begin{minipage}{\PLOTSW}%
    \includegraphics[
      height=\PLOTSH,
      clip,
      trim = -2.4mm 0.45mm 0 -4.6mm
    ]{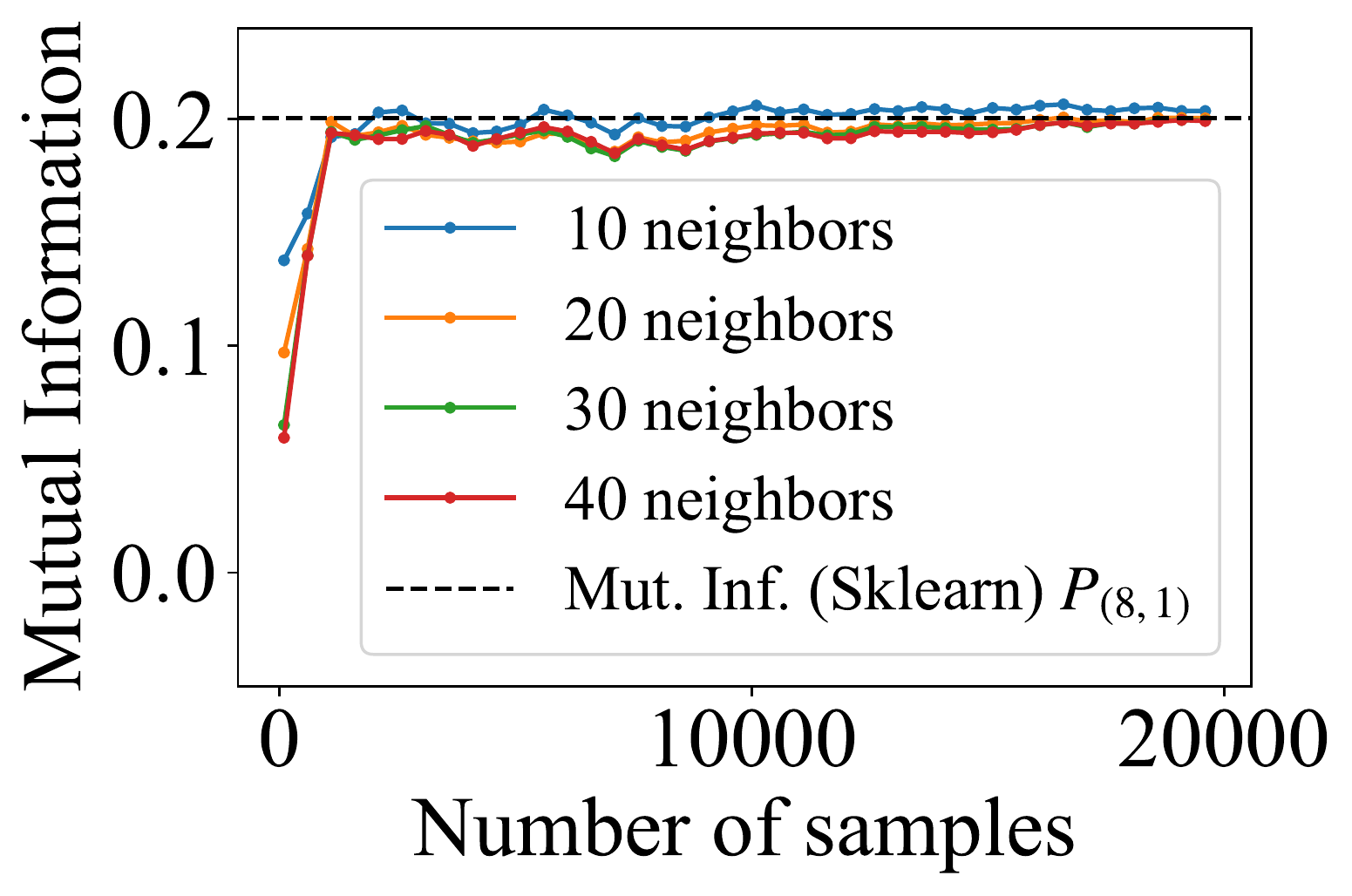}%
  \end{minipage}{\labelS[1.6mm]{fig:convergence-mutual-information-sklearn}}%
  \begin{minipage}{\PLOTSW}%
    \includegraphics[
      height=\PLOTSH,
      clip,
      trim = 2.55mm 0.45mm 0 -5.2mm
    ]{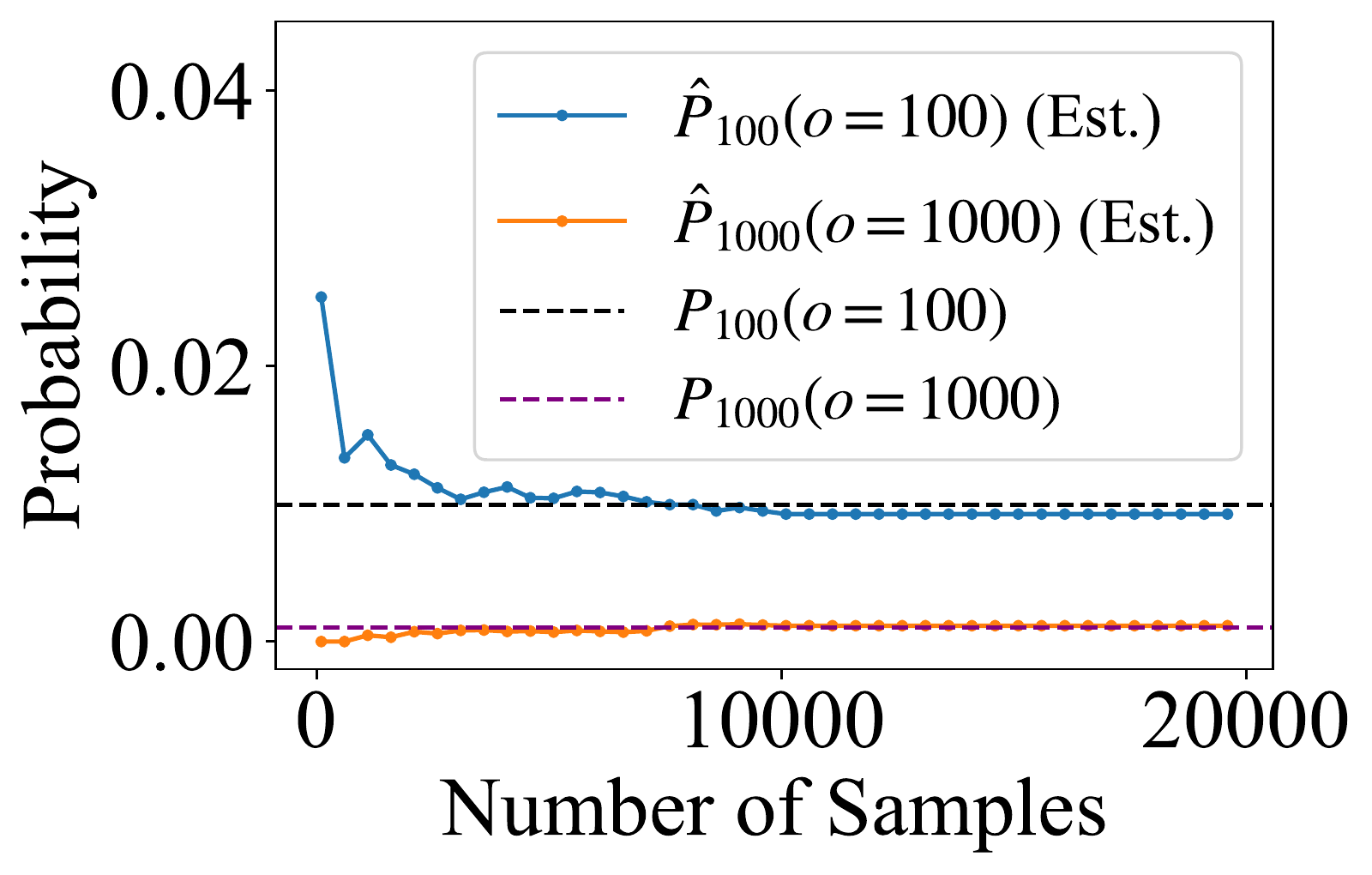}%
  \end{minipage}{\labelS[0.7mm]{fig:convergence-prob-query-discrete}}%
  \vspace{-.3em}
  
  \begin{minipage}{\PLOTSW}%
    \includegraphics[
      height=\PLOTSH,
      clip,
      trim = -9.85mm 0.45mm 0 2.7mm
    ]{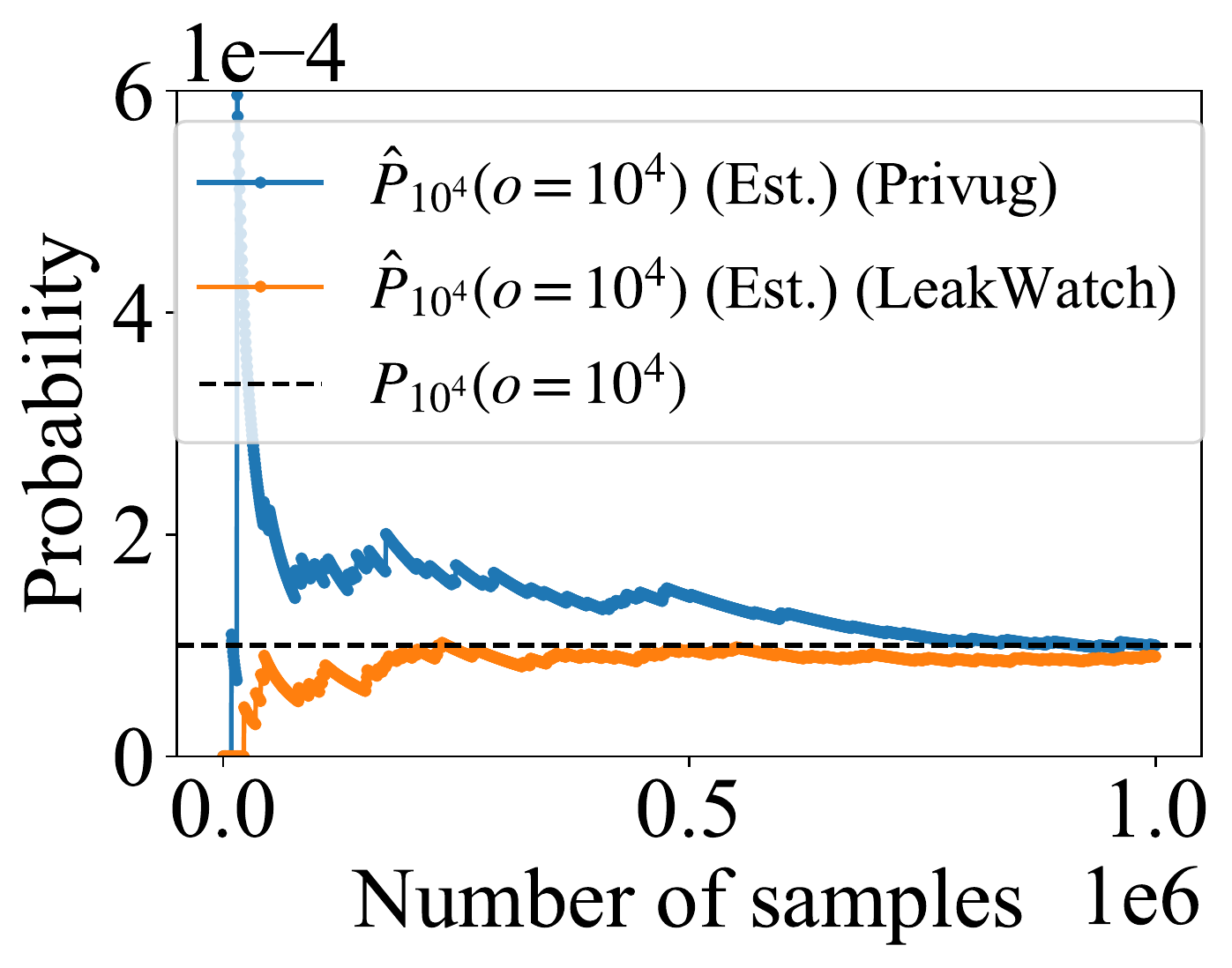}%
  \end{minipage}{\labelS[-0.2mm]{fig:convergence-prob-query-discrete-leakwatch-10000}}%
  \begin{minipage}{\PLOTSW}%
    \includegraphics[
      height=\PLOTSH,
      clip,
      trim = 2.5mm 0.4mm 0 -1.75mm
    ]{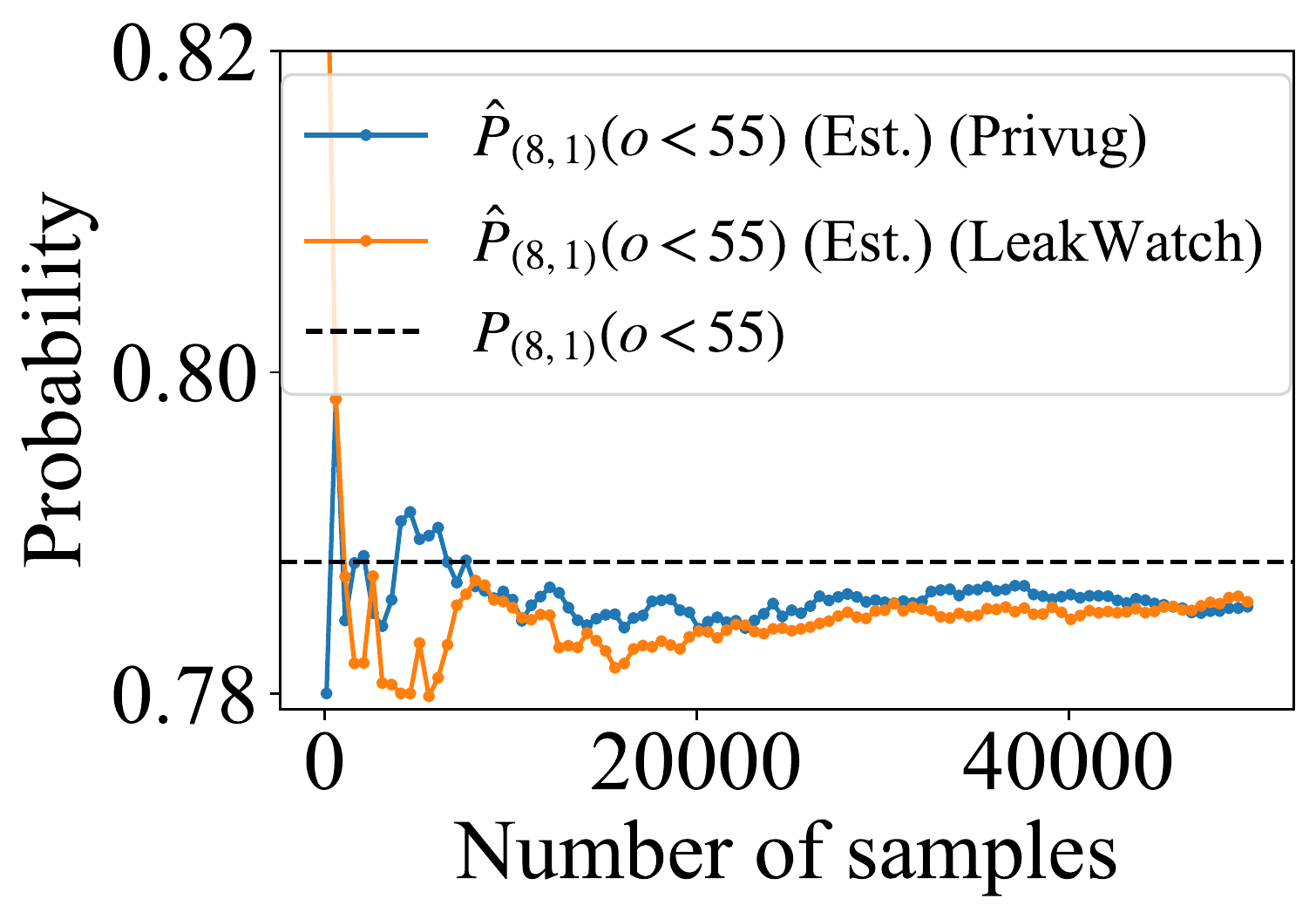}%
  \end{minipage}{\labelS[0mm]{fig:convergence-prob-query-discrete-leakwatch-200-vars}}%
  \begin{minipage}{\PLOTSW}%
    \includegraphics[
      height=\PLOTSH,
      clip,
      trim = 2.5mm 0.4mm 0 -4.45mm,
    ]{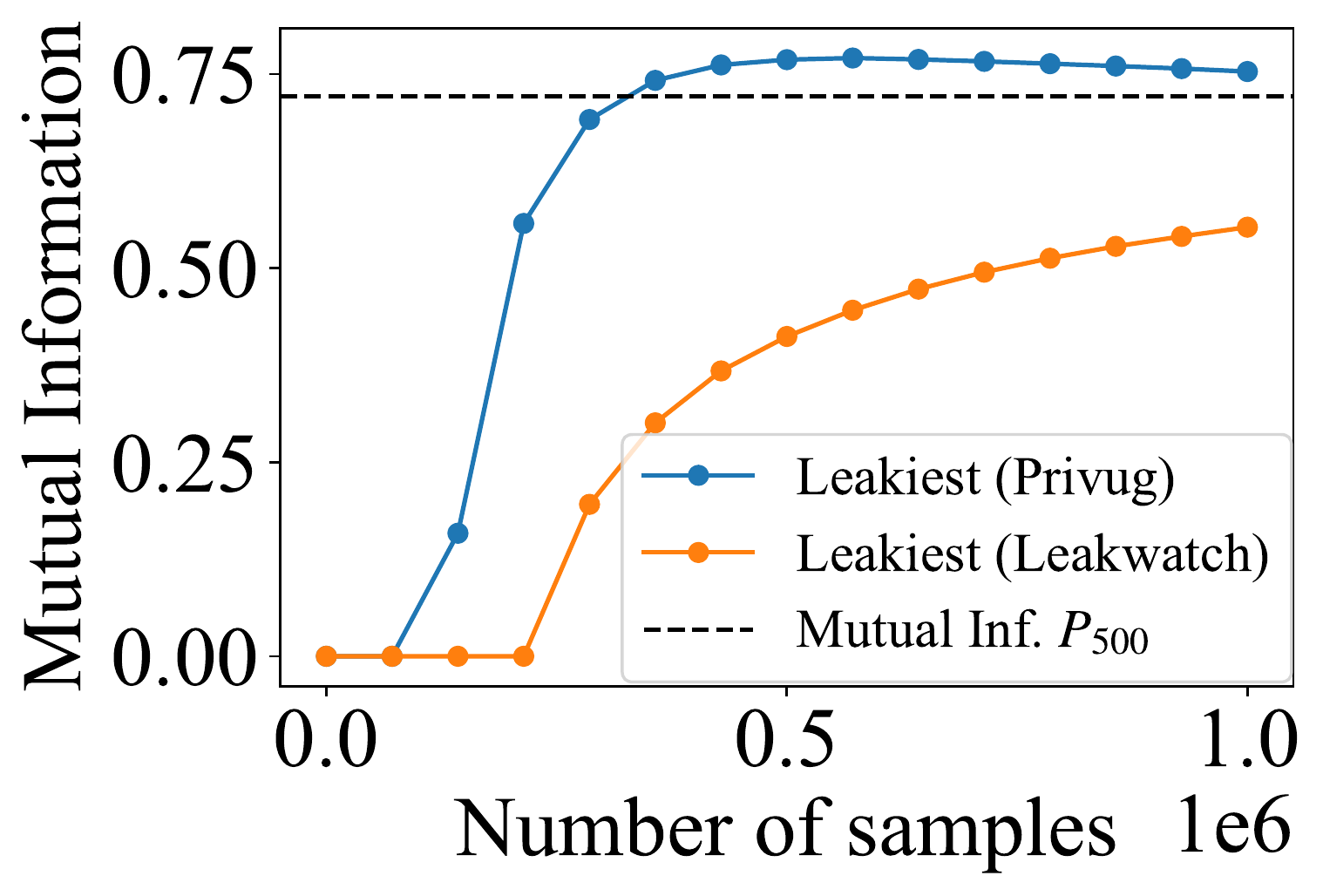}%
  \end{minipage}{\labelS[1.6mm]{fig:mi-leakiest-leakwatch-privug-discrete-500}}%
  \begin{minipage}{\PLOTSW}%
    \includegraphics[
      height=\PLOTSH,
      clip,
      trim = -2.4mm 0.4mm 0 -4.45mm,
    ]{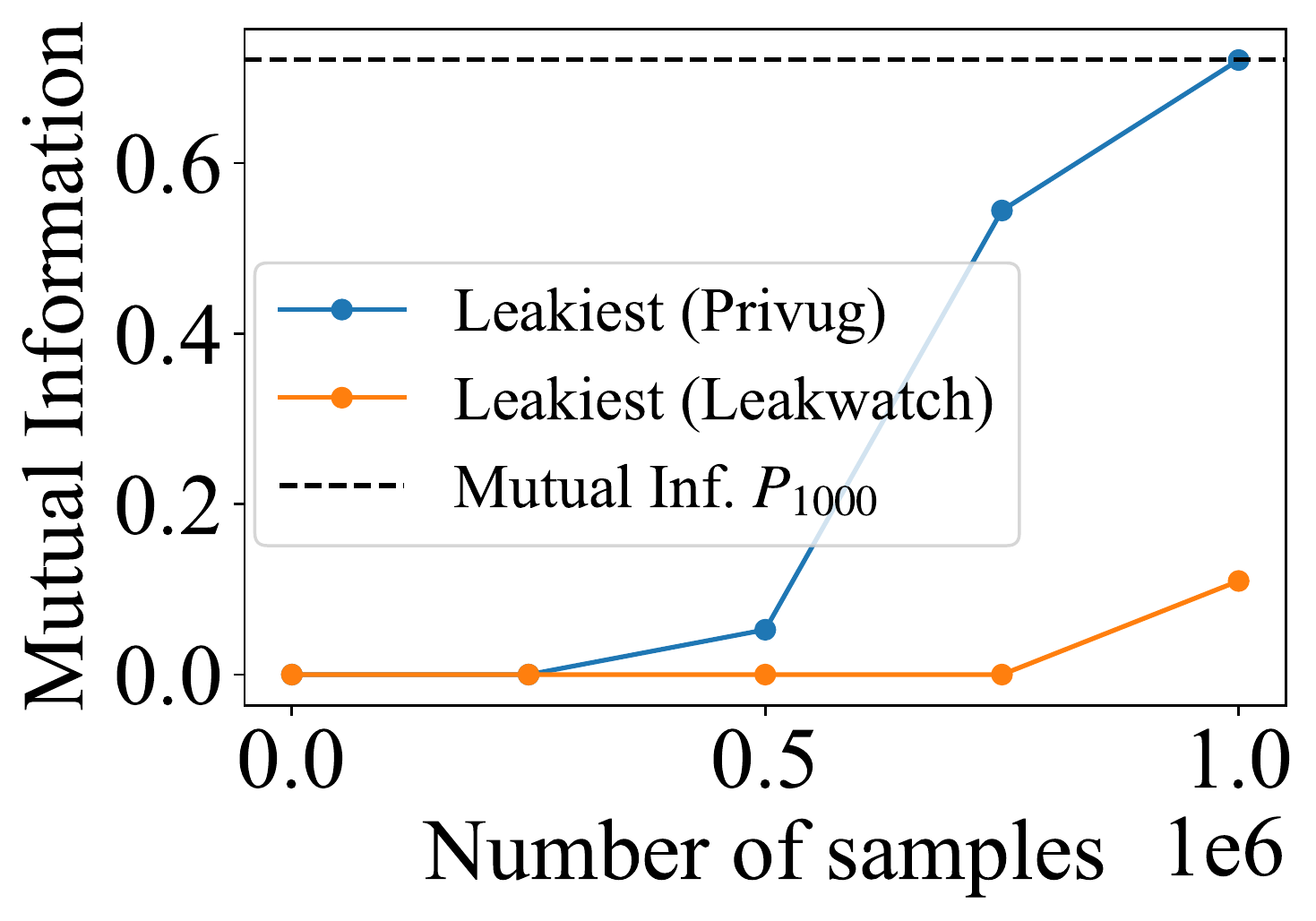}%
  \end{minipage}{\labelS[1.6mm]{fig:mi-leakiest-leakwatch-privug-discrete-1000}}%
  \vspace{-.3em}
  
  \begin{minipage}{\PLOTSW}%
    \includegraphics[
      height=\PLOTSH,
      clip,
      trim = -2.4mm 0.45mm 0 -4.45mm
    ]{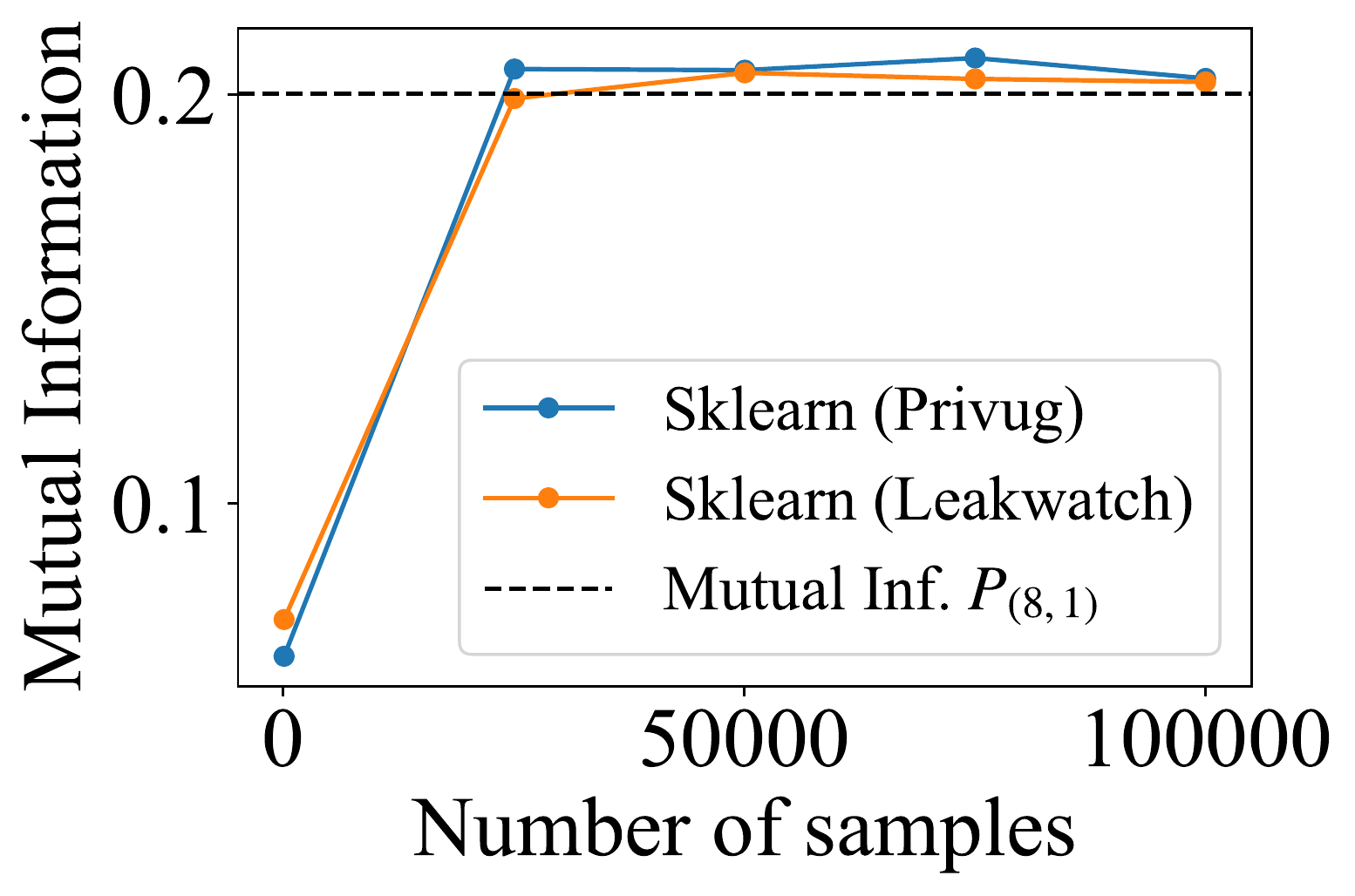}%
  \end{minipage}{\labelS[3.4mm]{fig:mi-sklearn-leakwatch-privug-continuous}}%
  \begin{minipage}{\PLOTSW}%
    \includegraphics[%
      height=\PLOTSH,
      clip,
      trim = 2.5mm 0.4mm 0 -1.8mm
    ]{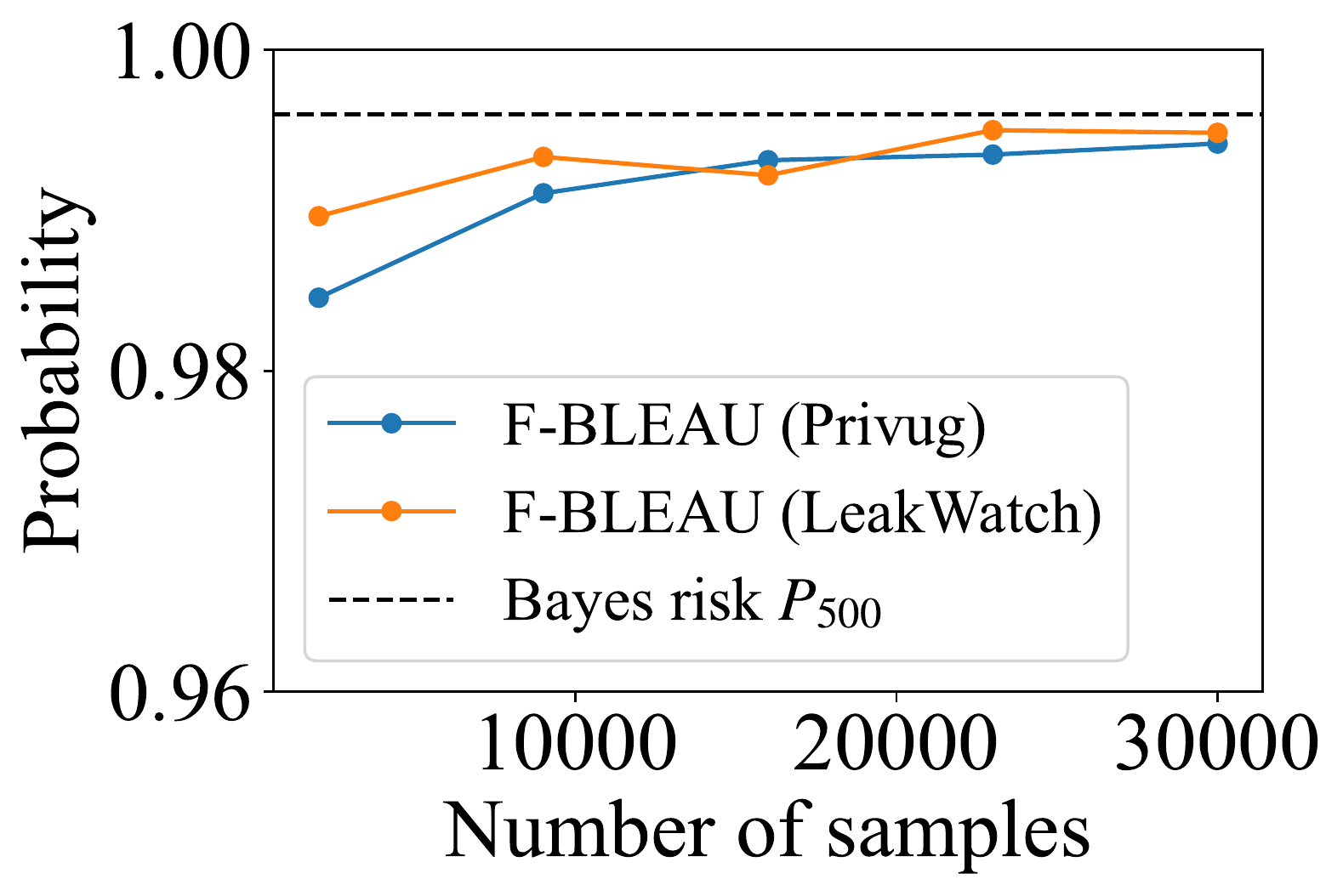}%
  \end{minipage}{\labelS[3.2mm]{fig:bayes-risk-discrete-leakwatch-500}}%
  \begin{minipage}{\PLOTSW}%
    \includegraphics[%
      height=\PLOTSH,
      clip,
      trim = 2.5mm 0.45mm 0 -1.8mm
    ]{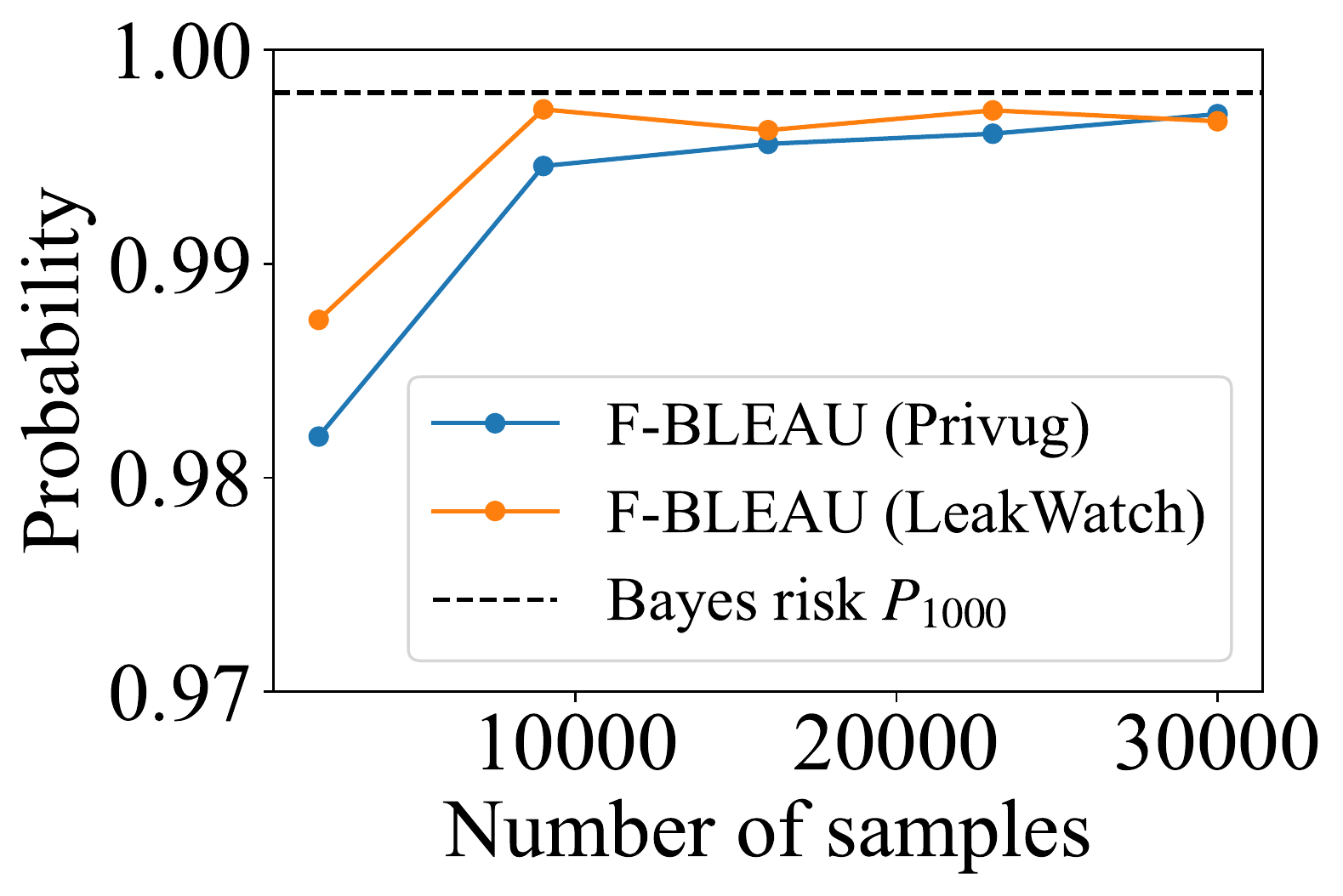}%
  \end{minipage}{\labelS[3.2mm]{fig:bayes-risk-discrete-leakwatch-1000}}%
%
%
  \begin{minipage}{\PLOTSW}%
    \includegraphics[
      height = \PLOTSH,
      clip,
      trim =-9.85mm 2.8mm 0 2.65mm
    ]{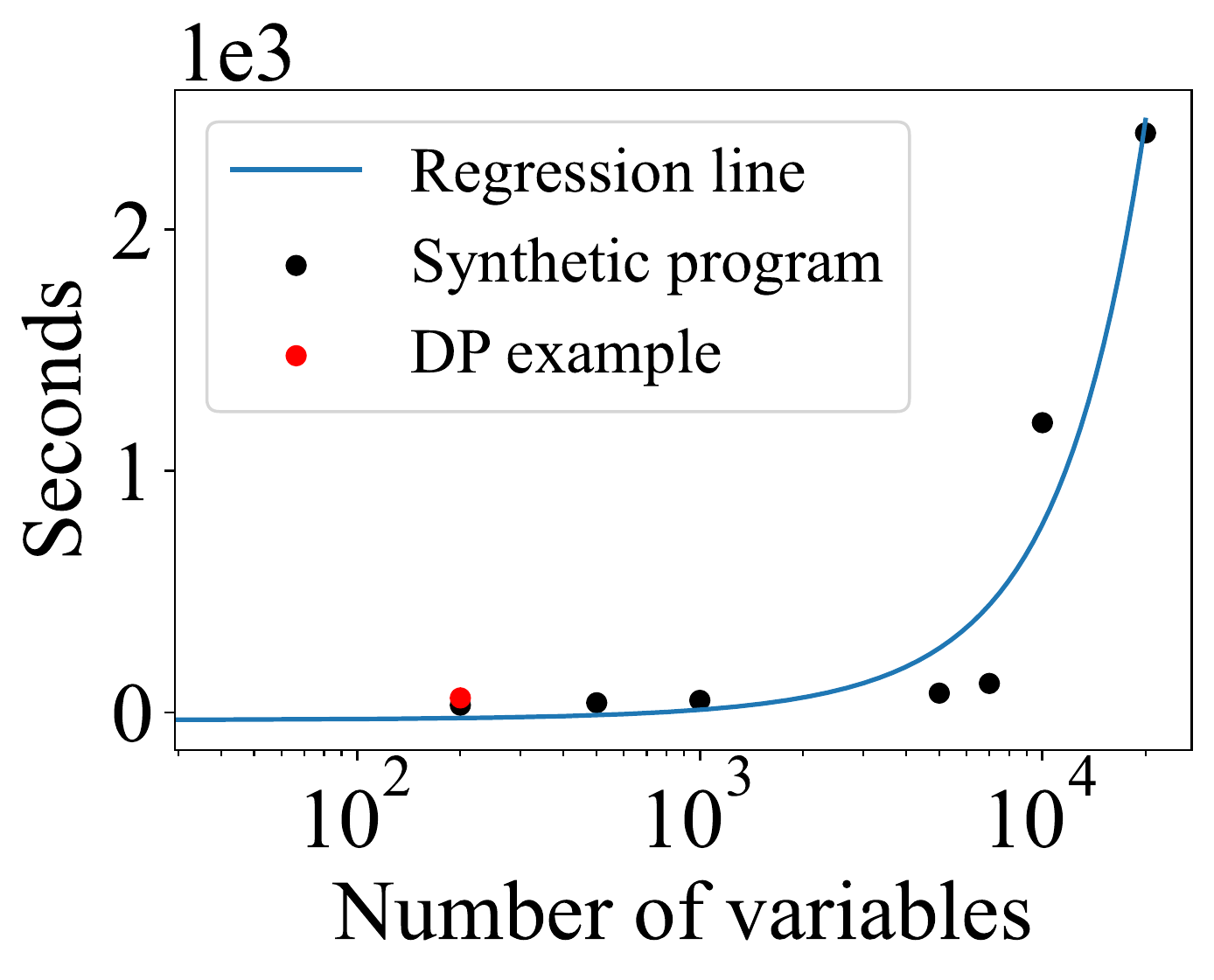}%
  \end{minipage}{\labelS[-4mm]{fig:scalability-continuous}}%
  \vspace{-.3em}%
  
  \begin{minipage}{\PLOTSW}%
    \includegraphics[ 
      height=\PLOTSH,
      clip,
      trim =-9.85mm 2.8mm 0 2.65mm
    ]{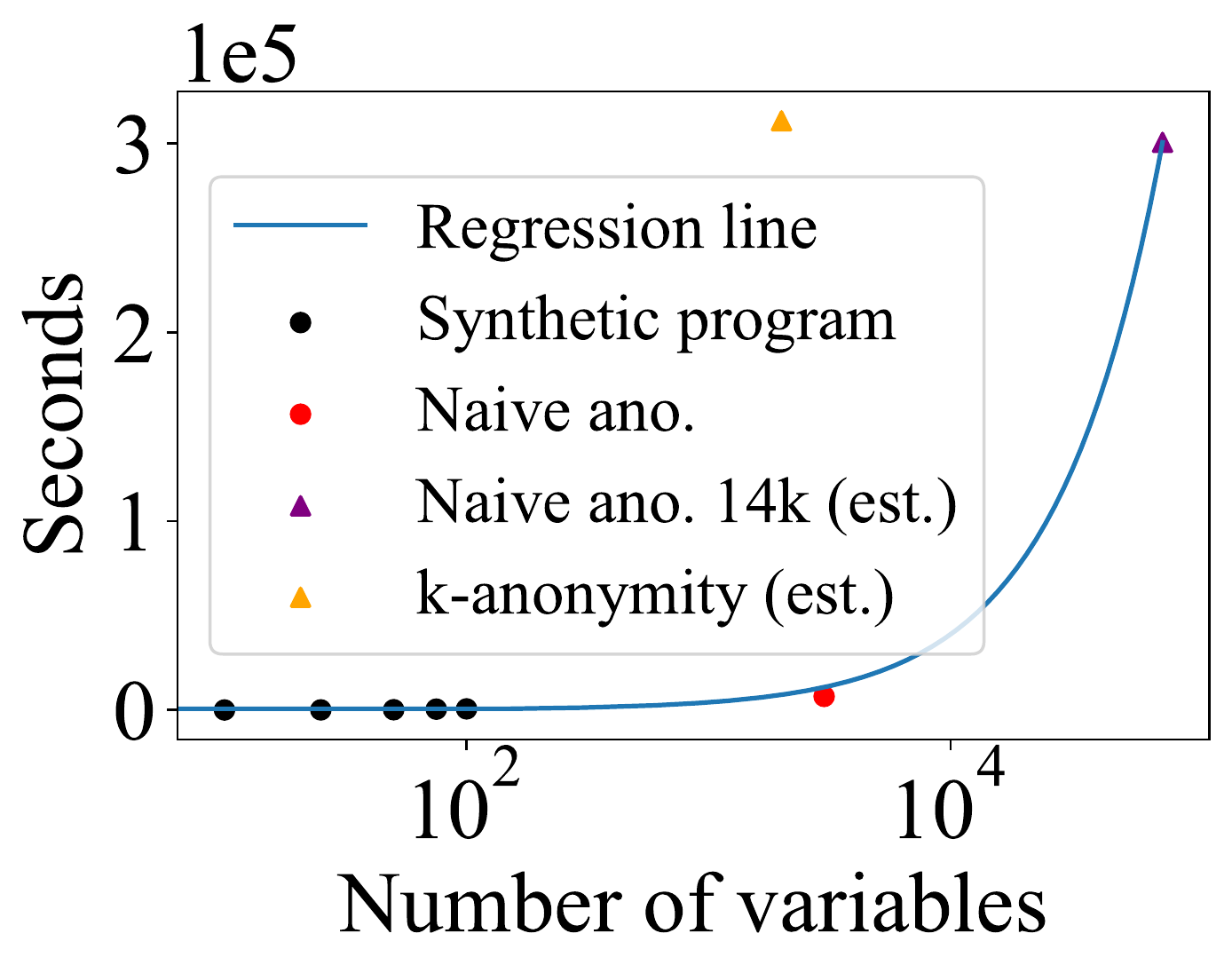}%
  \end{minipage}{\labelS[-4mm]{fig:scalability-discrete}}%
  \begin{minipage}{\PLOTSW}%
    \includegraphics[
      height=\PLOTSH,
      clip,
      trim=-2.45mm -9.85mm 0 7.5
    ]{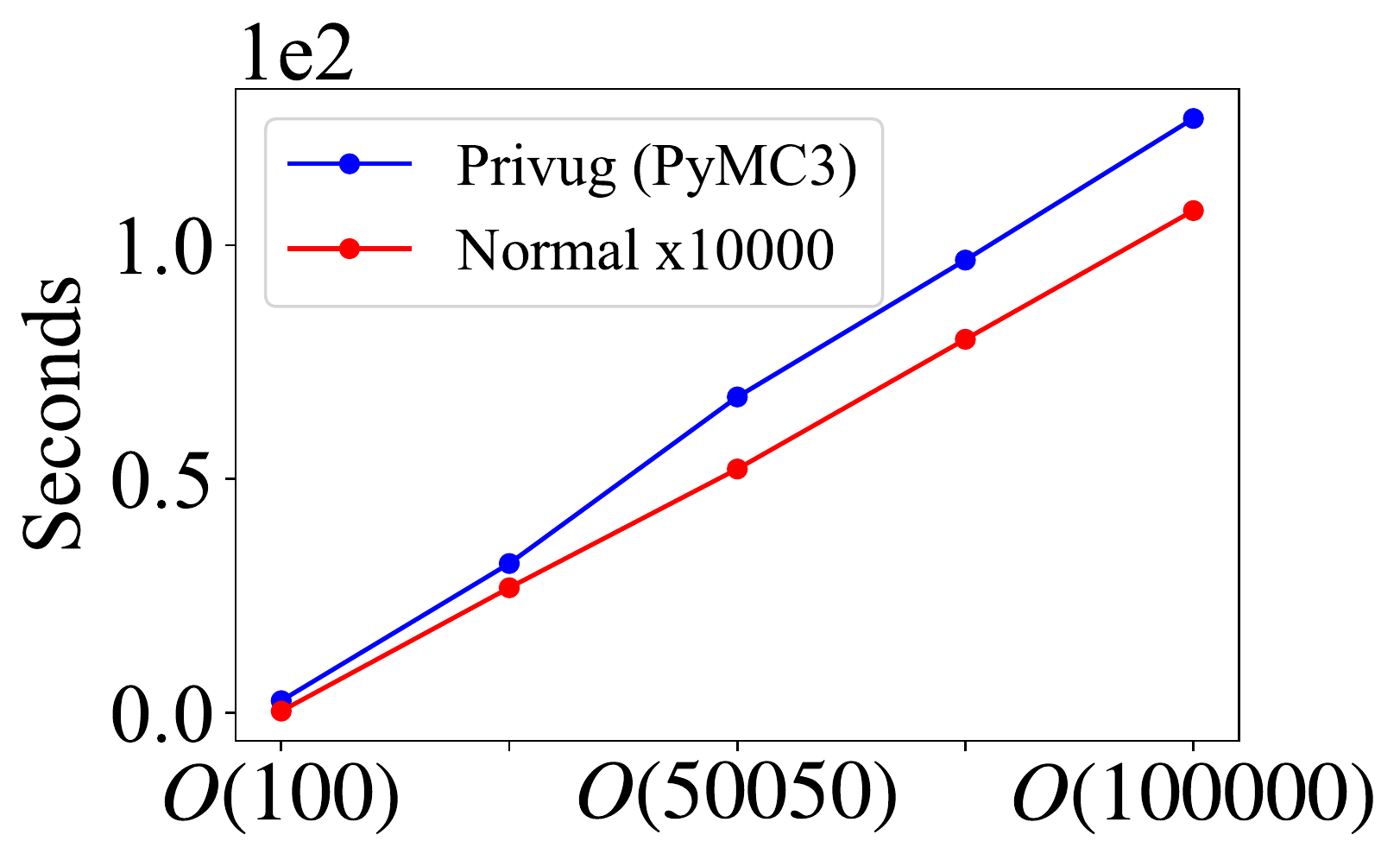}%
  \end{minipage}{\labelS[-4mm]{fig:scalability-complexity-n}}%
  \begin{minipage}{\PLOTSW}%
    \includegraphics[
      height=\PLOTSH,
      clip,
      trim=-0.25mm -9.85mm 0 -5.2mm
    ]{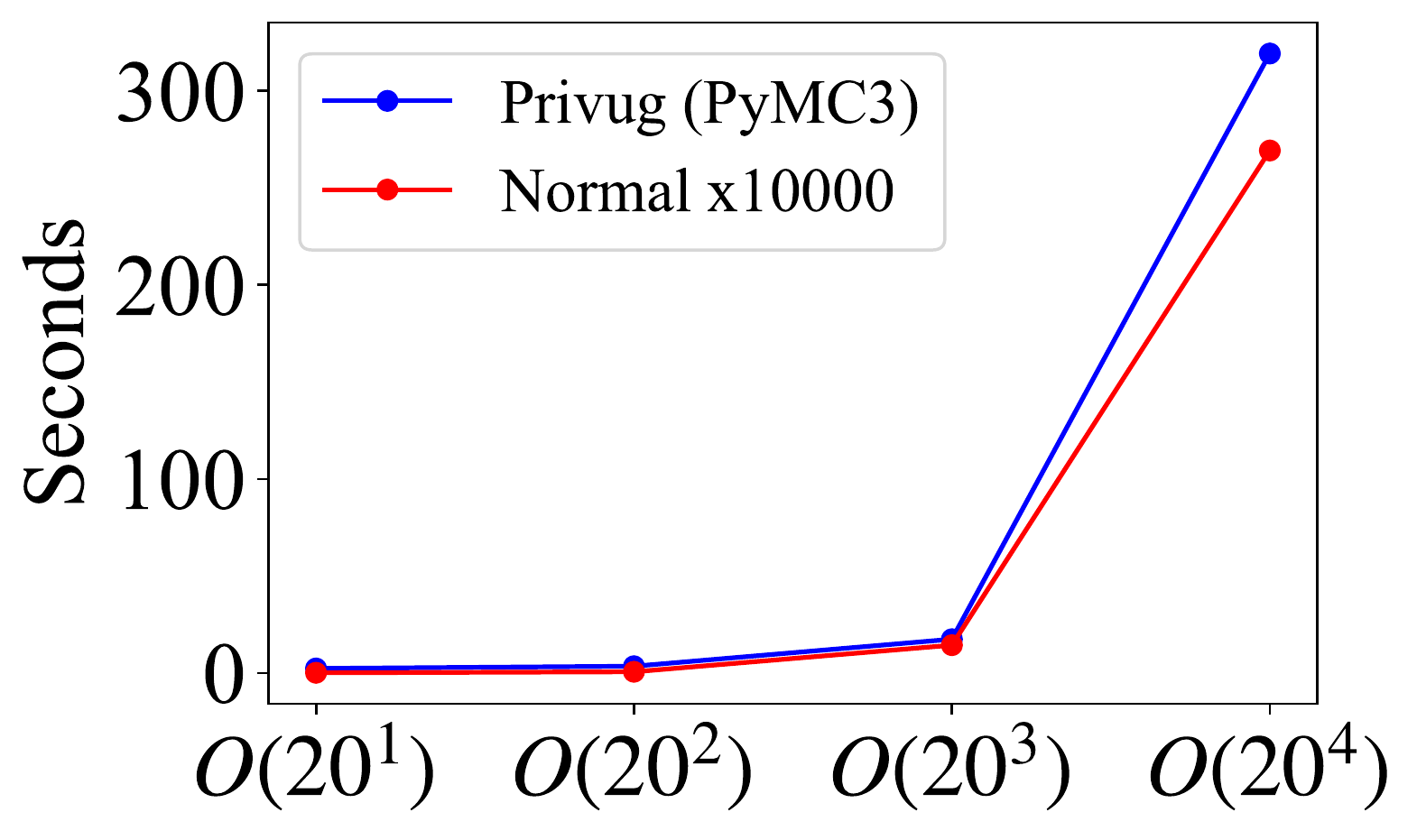}%
  \end{minipage}{\labelS[-4mm]{fig:scalability-complexity-c}}
  \begin{minipage}{\PLOTSW}%
    \includegraphics[
      height=\PLOTSH,
      clip,
      trim = -2.4mm 0.45mm 0 2.7mm,
    ]{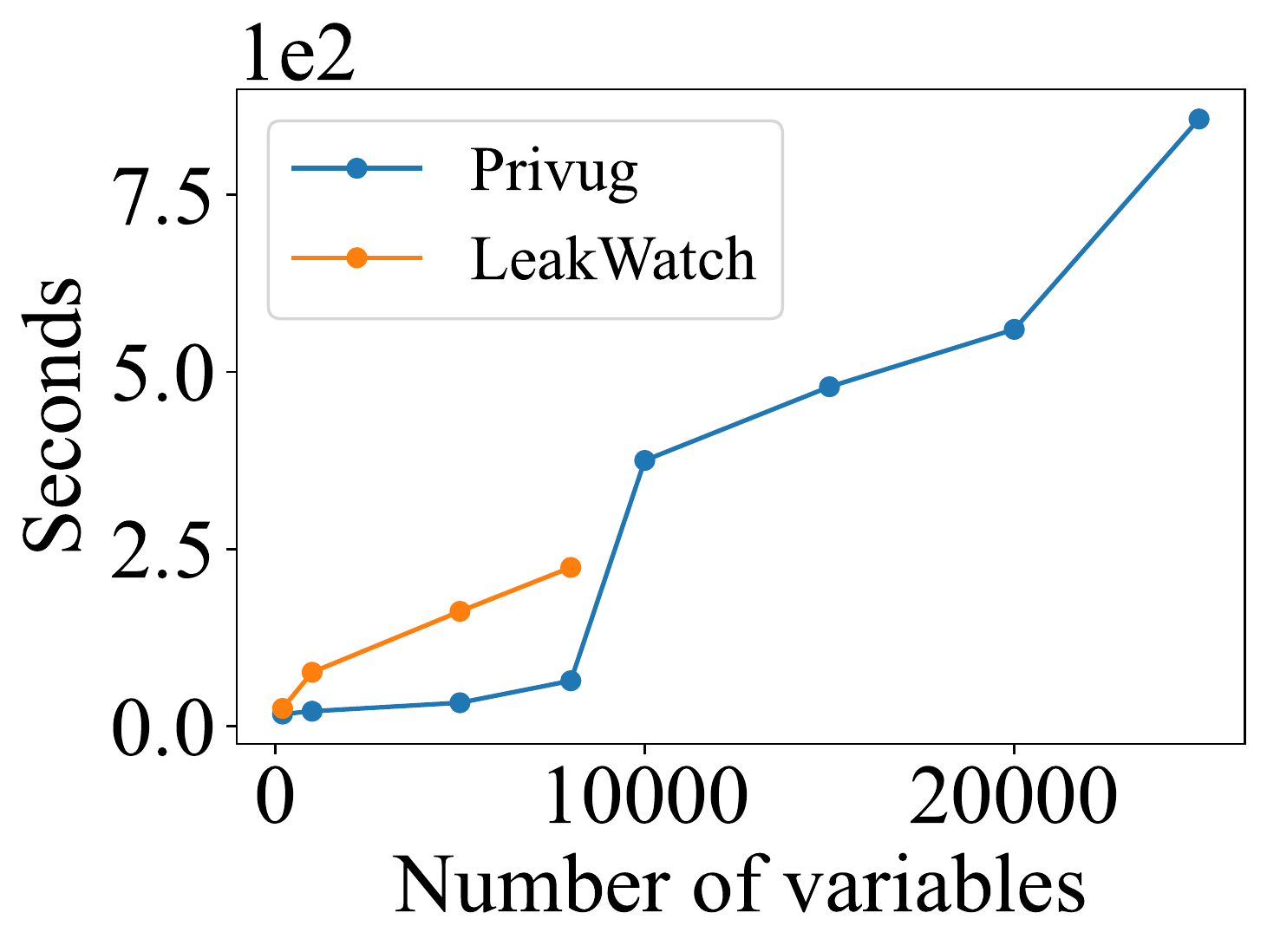}%
  \end{minipage}{\labelS[-4mm]{fig:scalability-privug-vs-leakwatch}}%

  \caption{%
    \mbox{\textbf{Analysis results.}}
    \textsl{Naive Anonymization:}
    Quasi-identifier analysis
    (\subref*{fig:numzip})~zip,
    (\subref*{fig:numbday})~day,
    (\subref*{fig:numsex})~sex,
    (\subref*{fig:numzipsex})~zip+sex,
    (\subref*{fig:numbdaysex})~day+sex,
    (\subref*{fig:numzipbday})~zip+day,
    (\subref*{fig:numzipbdaysex})~zip+day+sex.
    Large datasets:
    (\subref*{fig:uniqueness-14000-zipbday})~zip+day,
    (\subref*{fig:uniqueness-14000-zipbdaysex})~zip+day+sex.
    Sensitive attribute analysis:
    (\subref*{fig:illzip})~Chance of learning that governor is ill if 5 share his zip.
    \textsl{$k$-anonymity:}
    Number of rows matching governor's attributes ($k\!=\!2$):
    (\subref*{fig:uniqueness-k-anonymity-sex})~sex,
    (\subref*{fig:uniqueness-k-anonymity-others})~any other attribute combination.
    \textsl{Aggregate example}:
    \textcolor{darkgoldenrod}{Prior} and \textcolor{darkblue}{posterior} knowledge of age of Alice (distributions):
    (\subref*{fig:alice-age_CONSTANT}) \kal,
    (\subref*{fig:alice-age-with-observation}) \kab.
    \textsl{Differential privacy}:
    (\subref*{fig:utility-dp})~Utility,
    (\subref*{fig:mi-dp})~Mutual Information,
    (\subref*{fig:kbp-dp})~Probability Queries.
    \mbox{\textbf{Convergence of \privugbold.}}
    %
    (\subref*{fig:convergence-probability-query})~Probability query (Continuous),
    (\subref*{fig:convergence-mutual-information-sklearn})~Mutual Information (Continuous),
    %
    (\subref*{fig:convergence-prob-query-discrete})~Probability query (Discrete).
    %
    \textsl{Comparison w/ LeakWatch:}
    Probability queries 
    (\subref*{fig:convergence-prob-query-discrete-leakwatch-10000})~Discrete $P_{10000}(o=10000)$,
    (\subref*{fig:convergence-prob-query-discrete-leakwatch-200-vars})~Continuous $P_{(8,1)}(o<55)$.
    Mutual Information 
    (\subref*{fig:mi-leakiest-leakwatch-privug-discrete-500})~Discrete $P_{500}$ in LeakiEst,
    (\subref*{fig:mi-leakiest-leakwatch-privug-discrete-1000})~Discrete $P_{1000}$ in LeakiEst,
    (\subref*{fig:mi-sklearn-leakwatch-privug-continuous})~Continuous $P_{(8,1)}$ in SKlearn.
    Bayes Risk:
    (\subref*{fig:bayes-risk-discrete-leakwatch-500})~Discrete $P_{500}$ in F-BLEAU,
    (\subref*{fig:bayes-risk-discrete-leakwatch-1000})~Discrete $P_{1000}$ in F-BLEAU.
    \mbox{\textbf{Scalability of \privugbold.}}
    Inference time: 
    (\subref*{fig:scalability-continuous})~Continuous random variables,
    (\subref*{fig:scalability-discrete})~Discrete random variables.
    Time complexity on \code{f(arr,c)}:
    (\subref*{fig:scalability-complexity-n})~Increasing $n \in (10^2,10^5)$ for $O(n)$,
    (\subref*{fig:scalability-complexity-c})~Increasing $c \in (1,4)$ for $O(20^c)$.
    \textsl{Comparison w/~LeakWatch:}
    (\subref*{fig:scalability-privug-vs-leakwatch}) $P_{(8,1)}$.
  }
  \label{fig:convergence-continuous}%
  \label{fig:convergence-discrete}%
  \label{fig:convergence-probability-queries}%
  \label{fig:convergence-mutual-information}%
  \label{fig:convergence-bayes-risk}%
  \label{fig:scalability}%
  \label{fig:scalability-complexity}
  \label{fig:alice-age-distributions}%
  \label{fig:dp-analysis-plots}%
  \label{fig:numsgovattributes}%
  \label{fig:convergence}
\end{figure}

\vspace{-1mm plus 1mm}

\section{Evaluation}%
\label{sec:evaluation}

\subsubsection{RQ1: Can \privugbold\ analyze common privacy mechanisms?}
\label{subsec:applicability}
We analyze three (modern and traditional) privacy mechanisms in \privug. The purpose is twofold:
\begin{inparaenum}[i)]
\item Demonstrate the applicability of \privug, and
\item Serve as templates for data analysts.
\end{inparaenum}

\paragraph{Differential Privacy.}
\label{sec:rq3}%
\label{subsec:dp}

Consider a company computing the mean income of employees with \code{agg} and releasing the output publicly.  To protect the anonymity of employees, they add Laplacian noise to the output; a popular mechanism to enforce differential privacy~\cite{DBLP:conf/tcc/DworkMNS06,dp.book.2014}.  We use \privug\ to explore trade-offs between privacy protection and data utility by varying the values of parameters.  We assume a dataset of $200$ incomes.  The company have previously released some data on 195 of the 200 incomes, so it is publicly known that they are between \$80k and \$90k ($\Uniform(80,90)$).  There are 5 new employees of which no income information is known ($\Uniform(10,200)$).  The program to analyze is an extension of $\code{agg}$ that adds Laplacian noise to the output: $o \sim \code{agg} + \Laplace(0,\Delta \code{agg}/\epsilon)$ where $\Delta\code{agg}$ denotes the \emph{sensitivity}. This is known to preserve $\epsilon$-differential privacy\,\cite[Thm.\,3.6]{dp.book.2014}. Sensitivity captures the magnitude by which a single entry can change the output. The program with the mechanism incorporated takes as input the \code{epsilon} ($\epsilon$)
and a set of \code{records}, returning the average income. The implementation in Figaro after lifting is:
\looseness=-1

\begin{lstlisting}[escapeinside={(*}{*)}]
def dp_agg (epsilon: Double, records: FixedSizeArrayElement[(String,Double)]) =
  val delta  = Constant(200.0)/(*records*).length (*\label{dp:delta}*)
  val lambda = Constant(epsilon)/delta (*\label{dp:lambda}*)
  val X = continuous.Exponential(lambda) (*\label{dp:exp}*)
  val Y = Flip(0.5) (*\color{gray}\em// <-- Bernoulli*) (*\label{dp:bernu}*)
  val laplaceNoise = If(Y, X*Constant(-1.0), X) (*\label{dp:lapnoise}*)
  agg((*records*)) ++ laplaceNoise (*\label{dp:noisyoutput}*)
\end{lstlisting}

\noindent
Since the maximum income is 200k, the sensitivity (\code{delta}) is $\nicefrac{200}{|\mathit{records}|}$.  We construct the Laplace distribution from an exponential and a Bernoulli distributions in lines \ref{dp:delta}--\ref{dp:lapnoise} using a standard construction. \Cref{dp:noisyoutput} adds the noise to the result of \code{agg}.
\looseness=-1

The Laplacian mechanism includes a notion of accuracy to quantify utility~(e.g., \cite[Thm 3.8]{dp.book.2014}). Unconventionally, we opt for measuring utility as KL-divergence between the output with noise ($o$) and without ($\mathit{ro}$). The reason is that KL-divergence can be applied to any method based on
perturbing the output of the program (demonstrating broad applicability of our currently-supported measures). High KL-divergence indicates low utility as it represents loss of information wrt. the noiseless output. Maximum utility is achieved when KL-divergence equals 0. We observe in\,\cref{fig:utility-dp} an exponential decay of
KL-divergence as $\epsilon$ increases, consistent with the intuition that small values of $\epsilon$ result in high noise and reduced utility. The graph suggests that decrements in $\epsilon$ for $\epsilon<0.5$ may impact utility strongly.
\looseness=-1

Now we evaluate how $\epsilon$ influences the flow of information from the income of new employees ($s_i$) to the output ($o$). Mutual information is $0$ if one is independent of the other (\ie\ no information flow). \Cref{fig:mi-dp} plots mutual information for different values of $\epsilon$. The dashed line shows the baseline, \ie\ mutual information when the output has no added noise. This shows mutual information increases linearly with $\epsilon$.
\looseness=-1

Finally, we use a \emph{probability query} to evaluate the effect of $\epsilon$ on statistical information that an attacker can learn about the new members in the dataset. Differential privacy does not focus on protecting this type of information---it focuses on protecting the presence of a record in the dataset. Yet it is useful to quantify the effect of the noise on attacker knowledge. Suppose running \code{dp_agg} with $\epsilon\! =\! 0.5$ yields $85$k as a result. Suppose privacy regulations disallow revealing income data without employee consent. The company would like to determine, whether revealing this result could breach the regulation. In \privug, we analyze the distribution $P(s_i \mid o \approx 85)$ to determine this. \Cref{fig:kbp-dp} shows the distributions for different values of $\epsilon$. The dashed line marks the baseline, \ie, the probability distribution before the observation. The blue line corresponds to the run with $\epsilon\!=\!0.5$. Since this line is not parallel with the baseline, there is evidence of an increase in knowledge. The other lines show that any $\epsilon > 0.1$ increases the knowledge about the salary of the new employees. Consequently, releasing the average 85k computed using $\epsilon=0.5$ will result in a violation of the regulation.
\looseness=-1

In summary, we have discovered that releasing the average income 85k with $\epsilon > 0.1$ reveals information about the salary of new employees. Thus, for $\epsilon=0.5$, the company must seek consent from employees. Mutual information increases linearly with $\epsilon$, and for $\epsilon<0.5$ is notably low. Utility exhibits an exponential decay as $\epsilon$ decreases. This decay is especially pronounced with $\epsilon<1$, showing the impact of the added noise on utility.
\looseness=-1


\paragraph{Naive Anonymization}
\label{ssec:ano-analysis}
\label{subsec:naive-anonymization}
We quantify how strongly an attacker can determine the diagnosis of an individual (the governor) by observing the output of \smath{\code{ano}} from \cref{sec:overview}. Though this mechanism has well-known privacy flaws, it is still commonly used. Thus, we illustrate how \privug\ is used to effectively find these
flaws. First, we define the prior. In Figaro:
\looseness=-1
\begin{lstlisting}[escapeinside={(*}{*)}]
def(*\hspace{.5em}*)p(*\hspace{.25em}*):(*\hspace{.25em}*)FixedSizeArrayElement[(*\hspace{-0.1em}*)(Name(*\hspace{-.1em}*),(*\hspace{-.1em}*)Zip(*\hspace{-.1em}*),(*\hspace{-.1em}*)Day(*\hspace{-.1em}*),Sex(*\hspace{-.1em}*),(*\hspace{-.1em}*)Diag)(*\hspace{-0.1em}*)](*\hspace{.25em}*)=(*\hspace{.25em}*)
   VariableSizeArray (Constant (1000), i => for
     n <- if i==0 then Constant (GNAME) else Uniform (names:_*)
     z <- if i==0 then Constant (GZIP)  else Uniform (zips:_*)
     b <- if i==0 then Constant (GDAY)  else Uniform (days:_*)
     s <- if i==0 then Constant (GSEX)  else Uniform (Male,Female)
     d <- if i==0 then Constant (GILL)  else If (Flip (.2), Ill, Healthy)
   yield (n,z,b,s,d))
\end{lstlisting}
\mbox{\smath{\code{Name}}} is an identifier for an individual. For the sake of clarity, we assume that \mbox{\smath{\code{Zip}}}, \mbox{\smath{\code{Day}}} and \mbox{\smath{\code{Sex}}} are non-sensitive attributes, and \mbox{\smath{\code{Diag}}} is sensitive. \mbox{\smath{\code{Name}}}, \mbox{\smath{\code{Zip}}}, \mbox{\smath{\code{Day}}} and \mbox{\smath{\code{Sex}}} are uniformly distributed, and \mbox{\smath{\code{Diag}}} is \mbox{\smath{\code{Ill}}} with probability 0.2. The first row in the dataset is fixed, containing the governor's record. The prior fixes a dataset size of $1000$ records.
\looseness=-1

The lifted version of \code{ano} follows. Note that, compared to \code{ano}, only the type changed.
\begin{lstlisting}[escapeinside = {(*}{*)}]
def ano_p (records : FixedSizeArrayElement[(Name,Zip,Day,Sex,Diag)]) =
  (*records*).map { (n, z, b, s, d) => (z, b, s, d) }
\end{lstlisting}

First, we assess re-identification risk. We check whether an attacker can uniquely identify an individual's row using \emph{quasi-identifiers}, which enables linking attacks. We  inspect subsets of attributes to determine how uniquely they identify subjects. We query for the probability of a certain number \mbox{\smath{$x$}} of rows in the output satisfying a predicate \mbox{\smath{$\phi$}}, where \mbox{\smath{$\phi$}} models which attributes we want to match with the governor. For example:
\mbox{%
  \smath{%
    \code{probability(}%
    $x$%
    \code{,(v:Int) => v == 5)}%
  }%
},
where \mbox{\smath{$x =\ $\code{output.count(}$\phi$\code{)}}}, yields the probability that there are 5 such rows ($x = 5$). \Cref{fig:numzip,fig:numbday,fig:numsex,fig:numzipsex,fig:numbdaysex,fig:numzipbday,fig:numzipbdaysex} show the results. The governor is most likely to share zip code with \tildecustom\smath{$5$} rows, and sex with \tildecustom\smath{$500$} rows. With more attributes (\eg{} zip+day, zip+day+sex) it becomes likely that only the governor's record has those values. Disclosing those together thus poses significant re-identification risk.
\looseness=-1

Next, we assess positive disclosure risk\,\cite{ldiversity.2007}: Can the attacker determine the diagnosis of an individual (w/o necessarily identifying its row)?
Consider the following property of datasets, \mbox{\smath{$\forall r \in D \cdot \fun{\psi}{r} \implies r.d = \text{\text{\code{Ill}}}$}}, which stipulates that all records satisfying $\psi$ are ill. We instantiate $\psi$ in various ways; with \mbox{\smath{$\fun{\psi}{r} = (r.z = \text{\code{GZIP}})$}}, the property stipulates that all records with the governor's zip code are ill. We compute the probability that this property holds for the anonymized dataset by issuing a \smath{\code{forall}} query on the posterior. Column 2 in \cref{tab:diagnosis-probability} displays the result. Like in the original case study~\cite{sweeney.kanonymity.2002}, we conclude that with access to the governor's zip code, birthday, and sex (last row), an attacker can determine the diagnosis of the governor with high probability (98\%). Unlike in the original study, we concluded this for all datasets satisfying our prior model.
\looseness=-1

\begin{table}[t!]
\renewcommand{\arraystretch}{1.09}
  \begin{tabularx}{\linewidth}{
    >{\small}X
    >{\hspace{0mm}\small}r
    >{\hspace{0mm}\small}r
    >{\hspace{0mm}\small}r
  }
    \textbf{Query}
    & \textbf{\llap{Prob.}\,naive}
    & \rotatebox{0}{\textbf{Prob.\,14k}}
    & \rotatebox{0}{\textbf{Prob.\,$k$-ano}}
    \\

    \midrule

    \smath{$\Pr(\forall r \!\in\! D \cdot r.z \!=\! \text{\codescriptsize{GZIP}} \implies r.d \!=\! \text{\codescriptsize{Ill}})$}
    & \smath{$.02000$}
    & \smath{$.00$}
    & \smath{$.0$}
    \\

    \smath{$\Pr(\forall r \!\in\! D \cdot r.b \!=\! \text{\codescriptsize{GDAY}}\implies r.d \!=\! \text{\codescriptsize{Ill}})$}
    & \smath{$.00006$}
    & \smath{$.00$}
    & \smath{$.0$}
    \\

    \smath{$\Pr(\forall r \!\in\! D \cdot r.s \!=\! \text{\codescriptsize{GSEX}}\implies r.d \!=\! \text{\codescriptsize{Ill}})$}
    & \smath{$.00000$}
    & \smath{$.00$}
    & \smath{$.0$}
    \\

    \smath{$\Pr(\forall r \!\in\! D \cdot r.z \!=\! \text{\codescriptsize{GZIP}} \!\land\! r.b \!=\! \text{\codescriptsize{GDAY}} \implies r.d \!=\! \text{\codescriptsize{Ill}})$}
    & \smath{$.96000$}
    & \smath{$.51$}
    & \smath{$.0$}
    \\

    \smath{$\Pr(\forall r \!\in\! D \cdot r.z \!=\! \text{\codescriptsize{GZIP}} \!\land\! r.s \!=\! \text{\codescriptsize{GSEX}} \implies r.d \!=\! \text{\codescriptsize{Ill}})$}
    & \smath{$.16000$}
    & \smath{$.00$}
    & \smath{$.0$}
    \\

    \smath{$\Pr(\forall r \!\in\! D \cdot r.b \!=\! \text{\codescriptsize{GDAY}} \!\land\! r.s \!=\! \text{\codescriptsize{GSEX}} \implies r.d \!=\! \text{\codescriptsize{Ill}})$}
    & \smath{$.01800$}
    & \smath{$.00$}
    & \smath{$.0$}
    \\

    \smath{$\Pr(\forall r \!\in\! D \cdot r.z \!=\! \text{\codescriptsize{GZIP}} \!\land\! r.b \!=\! \text{\codescriptsize{GDAY}} \!\land\! r.s \!=\! \text{\codescriptsize{GSEX}} \implies r.d \!=\! \text{\codescriptsize{Ill}})$}\hspace{-1em}
    & \smath{$.98000$}
    & \smath{$.71$}
    & \smath{$.0$}
\end{tabularx}

\bigskip

\caption{Probability of learning governor's diagnosis.}%
\label{tab:diagnosis-probability}

\vspace{-1.5ex plus 0.5mm minus 0.5mm}

\end{table}

We assess whether the dataset size affects our risk analyses. We re-run quasi-identifier and positive disclosure analyses for a dataset size of $14000$---closer to Sweeney's\,\cite{sweeney.kanonymity.2002}. This probabilistic model contains $70000$ random variables ($5$ variables per row, $14000$ rows). Our results (cf.~\cref{fig:uniqueness-14000-zipbdaysex}) are close to those originally reported \cite{sweeney.kanonymity.2002}: There is a \mbox{\smath{71\%}} probability that no other record shares the governor's zip code, birthday, and sex. For zip code and birthday (cf.~\cref{fig:uniqueness-14000-zipbday}) the probability is \mbox{\smath{51\%}}. Positive disclosure analysis shows a decrease in the probability of learning the diagnosis (column 3 in \cref{tab:diagnosis-probability}). These results indicate that, for this program and prior, increasing the size of the dataset does not uncover new privacy risks (in fact, smaller datasets are more vulnerable).
\looseness=-1

Finally, we assess how certain the attacker is about the governor's diagnosis. Say the dataset contains \smath{$5$} records with the governor's zip
code (cf. \cref{fig:numzip}). Suppose that out of those \smath{$5$} people, \smath{$k$} are ill. Then the probability of the governor being ill is \smath{$k/5$}. Notably, if all \smath{$5$} are ill, then the attacker is certain that the the governor is ill. This corresponds to the query
$
  \Pr(\,
  \text{\textcmtt{output.count}}(\phi \wedge \psi) = k
  \mid
  \text{\textcmtt{output.count}}(\phi) = 5
  \,),
$
where $\phi$ and $\psi$ are predicates; $\psi$ is true iff the record is ill, and $\phi$ iff it has the governor's zip code. We use \mbox{\smath{\code{setCondition}}} to observe that \mbox{\smath{$\text{\textcmtt{output.count}}(\phi) = 5$}}. \Cref{fig:illzip} shows the result. The first bar (0.2) reflects the prior probability, so there is 50\% chance that the attacker learns nothing from an actual data set. However, there is a 50\% chance that the belief of an attacker in a positive diagnosis grows: \smath{$0.4$} with 35\% probability, etc. This demonstrates that \privug{} can not only reason about the risk of an attacker learning something with certainty, but about decrease of uncertainty as well.
\looseness=-1


\paragraph{$k$-anonymity}
\label{subsec:kanonymity-governor}
We analyze an algorithm that produces a $k$-anonymous dataset of health records. That is, for any combination of attributes, at least $k$ rows in the dataset share those attribute values\,\cite{sweeney.kanonymity.2002}. This case study illustrates the use of \privug\ for a non-trivial program with quadratic complexity. In terms of privacy analysis, we compare the results of running the program with $k=2$ to those of naive anonymization above.
\looseness=-1

We start by presenting the prior and program. We use the same prior as \code{ano_p} above, but with a dataset size of $500$ records (due to sampling performance, see~RQ3). As for the program, we implemented \smath{\code{k_ano}}, which takes as parameter $k$ and a dataset, and outputs a $k$-anonymous dataset. The lifted version of \smath{\code{k_ano}} has type
$
(\Lift \code{k\_ano}) : \Dist(\code{Int},\, \code{List[(Name,Zip,Day,...)]}) \to \Dist(\code{List[(Zip,Day,...)]}).
$
Due to space contraints, we refer interested readers to our code repository for implementation details.
\looseness=-1

We analyze re-identification and positive disclosure risks. \Cref{fig:uniqueness-k-anonymity-sex} shows that the number of records in the output dataset matching the governor's sex in the input, is like we saw before (cf. \cref{fig:numsex}), save for the rare (\tildecustom3.5\%) occasion where sex was part of some quasi-identifier. In those instances, \smath{\code{k_ano}} \emph{masked} \smath{\code{Sex}}, replacing everyone's \smath{\code{Sex}} with \texttt{*} to enforce $2$-anonymity. \Cref{fig:uniqueness-k-anonymity-others} shows that for any other attribute combination, none of the records in the output share those attribute values with the governor's values from the input. Thus, \smath{\code{k_ano}} always mask \smath{\code{Zip}} and \smath{\code{Day}}. Regarding positive disclosure, the risk of learning the governor's
diagnosis is $0$ for any attribute combination (column 4 in~\cref{tab:diagnosis-probability}), since \smath{\code{k_ano}} always mask \smath{\code{Zip}} and \smath{\code{Day}}.
\looseness=-1

In summary, \smath{\code{k_ano}} eliminates disclosure risk compared to \smath{\code{ano}}. However, \smath{\code{k_ano}} destroys most (or all) utility; when \smath{\code{Sex}} also gets anonymized, then only the distribution on \smath{\code{Diag}} remains (which is public). With \privug, an analyst can thus investigate the privacy-utility tradeoff of changes made to a program, and compare programs for disclosure risk.
\looseness=-1



\subsubsection{RQ2: Does \privugbold\ produce accurate results? How fast does it converge?}

\label{subsec:convergence}

\noindent
We study the convergence and accuracy of \privug\ for continuous and discrete variables, as the type of variables affects convergence---different methods are used for the continuous and discrete case \cite{BDA.gelman.2013}.  The goal is to confirm that \privug's results are accurate, and check how effective the sampling methods are for the leakage estimation problem.   In total, we have successfully driven five different estimators with \privug samples derived from program code and priors: probability queries (continuous and discrete), mutual information (SKlearn, LeakiEst), Bayes Risk (F-BLEAU).   All the estimators behaved as expected, \privug converges to correct results (dashed black lines in the plots of \cref{fig:convergence} represent ground truth obtained in a pen-and-paper analysis).
Furthermore, \privug{} meets and exceeds performance of the main competing sampler for programs, LeakWatch\,\cite{chothia.leakwatch.2014}, without inheriting some of its disadvantages: It is not bound to a single execution environment (JVM), it is naturally extensible with probabilistic programming ecosystem, and it is much more lightweight (very little code is required).
\looseness=-1

In all these experiments, 5000 samples give accurate results (except for LeakiEst that requires >500k samples for large domain spaces).  This is reassuring regarding the validity of experiments executed for RQ1. We generated 10k samples for \code{dp_agg} and \code{ano}; sufficient to obtain accurate results.  For 14k dataset size with \code{ano} and \code{k_ano}, we only generated 1000 samples, due to the long running time.  Still, since we only used discrete probability queries there, 1000 samples shall approximate the correct result well (\cref{fig:convergence-prob-query-discrete}). Below we provide key details on the experiments leading to the above conclusions.
\looseness=-1

\smallskip

\noindent
We start with continuous problems and the most popular sampler for such (NUTS\,\cite{NUTS}, Hamiltonian).  We use a program that computes the average $o$ of random variables $s, p_1, ..., p_{200}$ distributed as \mbox{$s \!\sim\! \Normal(42,\sigma_s)$} and $p_i \!\sim\!  \Normal(55,\sigma_p)$.  We vary $\sigma_s$ and $\sigma_p$ to control sample dispersion.  We check how many samples are needed to accurately answer probability and mutual information queries.  \Cref{fig:convergence-probability-query} shows the accuracy for the probability query $P(o\!<\!55)$ for $\sigma_s\!=\!8, \sigma_p\!=\!1$ and $\sigma_s\!=\!\sigma_p\!=\!20$, labeled as $P_{(8,1)}$ and $P_{(20,20)}$ in the graph.  The error is below $0.01$ after 5000 samples in both cases.  Increasing the dispersion does not impact convergence.  We also estimate mutual information for $P_{(8,1)}$.  After 5000 samples the estimation error drops below $0.02$ (\cref{fig:convergence-mutual-information-sklearn}). The mutual information estimator of SKlearn uses $k$-nearest neighbour distance, but we observe no significant impact when varying $k$.  For discrete variables, we use a program that adds two input variables $x, y \sim \Uniform(0,n)$ giving the output $o \sim x\!+\!y$, and sample with  Metropolis algorithm, the method of choice for discrete problems (Importance sampling performs comparably).   \Cref{fig:convergence-prob-query-discrete} shows accuracy of the probability query $P(o\!=\!n)$ with $n \!=\! 100, 1000$.  After 5000 samples the estimation error drops below $0.01$ for both values of $n$, indicating that the support of $x$ and $y$ does not significantly impact convergence.  We evaluate convergence of \emph{mutual information} for this case using LeakiEst.  Less than 5000 samples suffice for LeakiEst to converge.  Finally, we also check the convergence of \emph{Bayes risk} estimation using the state-of-the-art F-BLEAU estimator\,\cite{cherubin.fbleau.2019} driven by Metropolis sampling in \privug.  As few as 1000 samples suffice for F-BLEAU to converge.
\looseness=-1

We make LeakWatch, the most similar work to \privug, drive the same estimators as above and compare with \privug.  LeakWatch does not directly support continuous inputs or Bayes risk.  We have extracted the sample sets generated by LeakWatch and manually implemented the queries.  We test the same estimators with LeakWatch as above.  \Cref{fig:convergence-prob-query-discrete-leakwatch-10000,fig:convergence-prob-query-discrete-leakwatch-200-vars} show convergence of \emph{probability queries} for the discrete system with an input domain of size 10000, $P_{10000}$, and for the continuous system, $P_{(8,1)}$. \Cref{fig:bayes-risk-discrete-leakwatch-500,fig:bayes-risk-discrete-leakwatch-1000} show convergence using F-BLEAU to estimate \emph{Bayes risk}. \Cref{fig:mi-leakiest-leakwatch-privug-discrete-500,fig:mi-leakiest-leakwatch-privug-discrete-1000} show the convergence of using LeakiEst to estimate \emph{mutual information}.   For continuous random variables, we use the SKlearn estimator (\cref{fig:mi-sklearn-leakwatch-privug-continuous}). In all these cases except for mutual information queries (\cref{fig:mi-leakiest-leakwatch-privug-discrete-500,fig:mi-leakiest-leakwatch-privug-discrete-1000}), the two samplers perform comparably.  Strikingly, in~\cref{fig:mi-leakiest-leakwatch-privug-discrete-1000}, \privug\ needs 300k fewer samples to start converging; much less than LeakWatch which has been specifically designed to work with LeakiEst!
\looseness=-1


\subsubsection{RQ3: Does \privugbold\ scale? Does program complexity impact running time?}
\label{subsec:scalability}

We evaluate how long it takes for NUTS (continuous) and Metropolis (discrete) samplers to produce two chains of 10000 samples for synthetic programs of increasing size.  As the efficiency of MCMC sampling depends on the dimensionality of the domain, we use the example from RQ2, but scaled up to 20000 variables (continuous:
$(
s+\linebreak[0]p_1+\linebreak[0]p_2+\linebreak[0]\ldots\linebreak[0]+\linebreak[0]p_{20000}
)/20001$, and discrete:
$x+\linebreak[0]y_1+y\linebreak[0]_2+\linebreak[0]\ldots\linebreak[0]+y_{20000}$).
This number permits modeling large and complex systems.  We include several realistic programs in the scalability experiment:  naive anonymization, $k$-anonymity and differential privacy, see RQ1 details.  \Cref{fig:scalability-continuous,fig:scalability-discrete} show the data points measured. The blue line overlays the main tendency of the measurements, black points correspond to the above synthetic programs, and the remaining symbols refer to  the realistic programs.  We run the experiments on a machine with 8x1.70GHz cores, 16GB RAM, except for the two experiments with naive anonymization, which have been run on 8x3.60GHz machine with 32 GB RAM.
\looseness=-1

Execution time of synthetic programs in \cref{fig:scalability-discrete} follows a linear trend.  The red point corresponds to the naive anonymization case with 5000 variables.  This data point follows the linear trend of synthetic programs, with run-time exceeding 2h.  Interestingly, inference for continuous variables is more efficient, as  Hamiltonian samplers can leverage continuity to generalize faster~\cite{NUTS}.  We can generate samplers for a model with 20000 random variables in around 40 minutes (\cref{fig:scalability-continuous}).  Notably, the differential privacy case exhibits particularly low execution time (red in \cref{fig:scalability-continuous}), consistent with the trend of the synthetic examples.  The purple and orange triangles correspond to the naive anonymization with a dataset of size 14k, and to the $k$-anonymity case, respectively.  To account for low sampling performance, we generated only 1000 samples for each and scaled the time linearly to place it in the graph.  Both cases took over 5h (80h after scaling).
\looseness=-1

The $k$-anonymity program is an interesting outlier: even with a small database of 500 entries.  The exponential $k$-anonymity algorithm used to produce each sample dominates the cost of inference.  This leads us to ask  how the subject program impacts the execution time of \privug.  We use the Metropolis sampler in this experiment, since it performed slower above.  To this end, we use a program \mbox{$f(\mathit{arr},c) \triangleq \sum_{0..c} \sum_{i \in \mathit{arr}} i$} with running time $O(n^c)$ for $n=|\mathit{arr}|$ (see our code repository for the implementation).  Increasing $n$ and $c$ induces linear and exponential growth respectively.  We compare the running time of generating 10000 samples with $f$ in \privug against 10000 executions of $f$ without \privug.  \Cref{fig:scalability-complexity-n} and \ref{fig:scalability-complexity-c} show similar execution times for both.  Thus the execution time of \privug\ is dominated by the number of samples requested and the cost of running the subject program, but the Metropolis sampler itself incurs no significant overhead.
\looseness=-1

Finally, we compare the scalability of \privug\ and LeakWatch, by measuring the execution time to generate 20000 samples for $P_{(8,1)}$ with increasing number of variables.  \Cref{fig:scalability-privug-vs-leakwatch} shows that up to 9000 variables, both perform comparably well---with \privug\ slightly faster.  However, LeakWatch crashes from out-of-memory errors on cases with more than 10000 variables.  In contrast, \privug\ exhibits much better scalability; it runs out of memory after 30000 variables.
\looseness=-1

In summary, \privug\ can handle complex programs without introducing major overhead over the subject program's running time.  \privug\ scales better than LeakWatch, making it better fit for larger systems and more complex priors.  This is largely due to probabilistic programming frameworks being heavily optimized by the data science community.   We thus advocate use of these framework in information leakage research.
\looseness=-1



\section{Related Work and Concluding Remarks}%
\label{sec:related}

\begin{table*}[t]
  \fontsize{6}{12}\selectfont
  \scalebox{1}{
    \setlength{\tabcolsep}{-1mm}
    \renewcommand{\arraystretch}{0.9}
  \begin{tabularx}{\textwidth}{
      >{\scriptsize} l
      >{\hspace{5.5mm} \scriptsize} l
      >{\hspace{1mm} \scriptsize} l
      >{\hspace{2mm} \scriptsize} l
      >{\hspace{1mm} \scriptsize} l
      >{\hspace{-0.5mm} \scriptsize} c
      >{\hspace{2.5mm} \scriptsize} l
      >{\hspace{0mm} \scriptsize} l
      >{\hspace{-3mm} \scriptsize} l
      >{\hspace{0mm} \scriptsize} l
      >{\hspace{2mm} \scriptsize} l
      >{\hspace{-1mm} \scriptsize} l
      >{\hspace{-3mm} \scriptsize} l
      >{\hspace{-1mm} \scriptsize} l
      >{\hspace{-2mm} \scriptsize} l
  }

    & \multicolumn{4}{c}{\scriptsize \hspace{2mm}\textbf{\makecell{Random\\Variable Type}}}
    & \multicolumn{1}{c}{\scriptsize \hspace{1mm}\textbf{Input Type}}
    & \multicolumn{3}{c}{\scriptsize \hspace{2mm}\textbf{\makecell{Tool\\Capabilities}}}
    & \multicolumn{6}{c}{\scriptsize \hspace{2mm}\textbf{\makecell{Supported Quantitative\\Information Flow Measures}}}
    \\[2mm]

    & \multicolumn{2}{c}{\scriptsize \hspace{4.2mm} {\it input}}
    & \multicolumn{2}{c}{\scriptsize \hspace{0mm} {\it output}}
    &
    &
    &
    &
    &
    &
    &
    &
    &
    &
    \\[-3.5mm]

    & \rotatebox{32}{\textls[-5]{\llap{d}isc.}}
    & \rotatebox{32}{\textls[-5]{\llap{c}ont.}}
    & \rotatebox{32}{\textls[-5]{\llap{d}isc.}}
    & \rotatebox{32}{\textls[-5]{\llap{c}ont.}}
    &
    & \rotatebox{40}{\textls[-5]{\llap{e}xact}}
    & \rotatebox{40}{\textls[-5]{\llap{s}ampling}}
    & \rotatebox{40}{\textls[-5]{\llap{e}stimation}}
    & \rotatebox{40}{\textls[-5]{\llap{K}SP}}
    & \rotatebox{40}{\textls[-5]{\llap{e}ntropy}}
    & \rotatebox{40}{\textls[-5]{\llap{m}in-entropy}}
    & \rotatebox{40}{\textls[-5]{\llap{m}utual-inf}}
    & \rotatebox{40}{\textls[-5]{\llap{B}ayes-risk}}
    & \rotatebox{40}{\textls[-5]{\llap{K}L-diverg}}
    \\[1mm] \midrule

    LeakiEst\,\cite{chothia.leakest.2013}
    & \yes
    & \no
    & \yes
    & \yes
    & set of samples
    & \no
    & \no
    & \yes
    & \no
    & \no
    & \yes
    & \yes
    & \no
    & \no
    \\

    F-BLEAU\,\cite{cherubin.fbleau.2019}
    & \yes
    & \llap{(}\yes\rlap{)$^{a}$}
    & \yes
    & \yes
    & set of samples
    & \no
    & \no
    & \yes
    & \no
    & \no
    & \yes
    & \no
    & \yes
    & \no
    \\

    SPIRE\,\cite{spire}
    & \yes
    & \no
    & \yes
    & \no
    & custom:\,PSI
    & \yes
    & \no
    & \no
    & \yes
    & \no
    & \no
    & \no
    & \no
    & \no
    \\

    DKBP\,\cite{DKBSP}
    & \yes
    & \no
    & \yes
    & \no
    & custom:\,Polyhedra
    & \yes
    & \no
    & \no
    & \yes
    & \no
    & \no
    & \no
    & \no
    & \no
    \\

    QUAIL\,\cite{QUAIL}
    & \yes
    & \no
    & \yes
    & \no
    & custom:\,QUAIL
    & \yes
    & \no
    & \no
    & \no
    & \no
    & \no
    & \yes
    & \no
    & \no
    \\

    HyLeak\,\cite{HyLeak}
    & \yes
    & \no
    & \yes
    & \no
    & \textls[-5]{custom:\,QUAIL\,2.0}
    & \no
    & \yes
    & \yes
    & \no
    & \yes
    & \no
    & \yes
    & \no
    & \no
    \\

    LeakWatch\,\cite{chothia.leakwatch.2014}
    & \yes
    & \llap{(}\yes\rlap{)$^{b}$}
    & \yes
    & \yes
    & Java
    & \no
    & \yes
    & \yes
    & \llap(\yes\rlap{)$^{b}$}
    & \no
    & \llap(\yes\rlap{)$^c$}
    & \llap(\yes\rlap{)$^{b,c}$}
    & \no
    & \no
    \\

    \privug\hspace*{-.5mm}(this\,wo\rlap{rk)}
    & \yes
    & \yes
    & \yes
    & \yes
    & Java/Scala/Python
    & ?
    & \yes
    & \yes
    & \yes
    & \yes
    & \llap(\yes\rlap{)$^{c}$}
    & \llap(\yes\rlap{)$^{c}$}
    & \llap(\yes\rlap{)$^{c}$}
    & \yes
    \\

  \end{tabularx}}

  \medskip

  \caption{%
      Overview of leakage quantification tools.  Legend: \textbf{KSP} = knowledge-based security policy; \textbf{custom} = custom input language; $^{a}$Cherubin~\cite{cherubin.phd.thesis} lays the foundation to handle continuous input but this has not been implemented; $^{b}$we  show how to handle continuous and discrete KSP and mutual information with LeakWatch in \cref{subsec:convergence}---this was not demonstrated originally\,\cite{chothia.leakwatch.2014};  $^{c}$via integrated 3rd party tools (F-BLEAU/LeakiEst/SKlearn) and pmf estimation for discrete input/output. ? = not studied.
  }%
  \label{tab:related-works}

  \vspace{-1.5ex plus 0.5mm minus 0.5mm}
\end{table*}

We have shown that probabilistic programming with Monte-Carlo Bayesian inference is a promising basis for implementing  privacy risk and data leakage analyses. \privug analyses follow a well-defined architecture: modeling attackers, extracting models by lifting programs, and using a state-of-the-art sampler to drive an estimator. We know of no similarly broad competing framework to compare against.  Several tools exist to quantify leakage using probabilistic reasoning. \Cref{tab:related-works} provides a detailed comparison. The first 4 columns specify the type of input/output variables; \privug fully supports discrete and continuous distributions (unlike existing tools that mostly focus on discrete variables). The fifth column indicates whether the tool works on a (externally generated) set of samples, a custom specification language or a general purpose programming language; \privug works directly on general purpose programming languages. Columns 6-8 indicate whether the tool can perform exact analytical inference, sample from distributions (e.g., via naive sampling or MCMC), or can estimate leakage measures; in \privug we can perform all of these, but we have not studied exact inference in this paper. The last 6 columns show whether the tools support the corresponding measures; all of them are supported by \privug (unlike any other existing tool). In the following, we discuss the existing tools in two groups, white- and black-box.  These tools are highly-specific; they feature a design and architecture of samplers and estimators highly optimized for a single purpose. In contrast, the idea of \privug\ is to build on a broad platform of probabilistic programming, which has not been used for this purpose before, and to reuse as many components as possible to provide a comprehensive assessment of a program.
\looseness=-1

\emph{Black-box methods}
estimate leakage by analyzing a set of input/output pairs of the system.  LeakiEst\,\cite{chothia.leakest.2013} estimates min-entropy\,\cite{renyi.minentropy.1961} and mutual information\,\cite{elementsofinformationtheory.2006} using frequentist statistics, \ie, counting the relative frequency of the outputs given inputs. F-BLEAU\,\cite{cherubin.fbleau.2019} and its generalization \cite{romanelli.leaves.2020} use nearest neighbor classifiers to estimate Bayes risk\,\cite{chatzikokolakis.catuscia.bayesrisk.2008} and g-leakage \cite{qifbook.2020}.  Classifiers can exploit patterns in the data and scale better than LeakiEst for large output spaces.  Black-box tools require a set of independent and identically distributed samples over inputs. Obtaining such a sample is not easy as discussed by Chothia\,\etal\,\cite{chothia.leakwatch.2014}. \privug\ automates this process, obtaining synergy with black-box
methods in two ways:
\begin{enumerate*}[label=(\roman*)]
\item black-box methods can be used easily within \privug (\cref{subsec:convergence});
\item black-box methods can leverage the well-studied sampling mechanisms~\cite{mcmc} used in \privug\ to produce the set of samples they work on. \Cref{subsec:convergence} shows that LeakiEst converges faster using \privug\ than with LeakWatch for mutual information queries.
\end{enumerate*}
\looseness=-1

\emph{White-box methods}
exploit the source code of the program to compute leakage analytically or via sampling. We distinguish white-box methods working on \emph{custom specification languages} from those working on \emph{general purpose programming languages}. Custom specification languages are languages designed for program analysis and are typically not directly executable. Mardziel \etal\ introduce abstract probabilistic polyhedra to capture attacker beliefs, and define transformations over the polyhedra to analytically obtain the revised belief of the attacker after observing an output of the program\,\cite{DKBSP}.  They are able to check whether queries to a database violate a knowledge-based security policy.  SPIRE\,\cite{spire} uses the symbolic inference engine PSI\,\cite{PSI} to analytically compute the updated beliefs of an attacker given an observation.  Then, it uses Z3\,\cite{Z3} to verify whether a knowledge-based security policy holds.  QUAIL performs forward state exploration of a program to construct a Markov chain capturing its semantics, which is then used to compute mutual information\,\cite{QUAIL}.  HyLeak is an evolution of QUAIL to use hybrid statistical estimation\,\cite{HyLeak}.  The method works on the control flow graph of the program.  It first uses several symbolic reductions to simplify the program, then applies standard statistical reasoning via sampling. These works support programs with discrete inputs and outputs.  In contrast, \privug\ handles \emph{discrete and continuous inputs and outputs}.  In principle, it also allows obtaining analytical solutions, \eg, using variable elimination (in Figaro)\,\cite{pfeffer2016practical} but we have not explored this.  Unlike HyLeak, we do not reduce the program graph, but Hamiltonian samplers compute gradients of the model (probabilistic program) to improve sampling effectiveness.  QUAIL computes mutual information; HyLeak computes mutual information and Shannon entropy.  Others  support only analysis of knowledge-based security policies\,\cite{DKBSP,spire}.
\looseness=-1

\privug\ is, perhaps, the first work whose goal is supporting estimation of many measures for programs written not in custom specification languages, but in \emph{general purpose programming languages} (Python, Scala, and Java via the Scala interface).  LeakWatch samples a Java program and uses LeakiEst to estimate mutual information and min-entropy leakage\,\cite{chothia.leakwatch.2014}. There are several differences between \privug\ and LeakWatch. First, \privug\ uses efficient and scalable Bayesian inference methods as opposed to LeakWatch that relies on direct sampling from target distributions. We found that the Bayesian methods used in \privug\ scale better (\cref{subsec:scalability}). We also found that LeakiEst, the estimator LeakWatch was designed for, converges faster when using \privug's samples
(\cref{subsec:convergence}). Bayesian inference is proven to be very effective in the presence of conditions~\cite{mcmc}, which are not directly available in LeakWatch. Second, LeakWatch relies on its users to select appropriate Pseudo-Random Number Generators (PRNGs). The authors recommend \code{java.security.SecureRandom}~\cite{leakwatch.prng.example}, which only support sampling from uniform and normal distributions. In contrast, probabilistic programming frameworks (used in \privug) support a wide range of probability distributions with high quality PRNGs. This emphasizes another key contribution of this work for leakage research: It is beneficial to build on top of strong statistical and probabilistic platforms over custom solutions, with Bayesian probabilistic programming being one such platform.
\looseness=-1


\bibliographystyle{plain}
\bibliography{references}

\appendix
\section{Programs as Models}
\label{sec:language}


%
%
%

%
This section is aimed at readers who wish to gain insight into the
semantic underpinning of \privug.
The main idea is that probability distributions form a monad
\cite{DBLP:conf/popl/RamseyP02}; we interpret a program as a
probabilistic model by mapping it into said monad.

\paragraph{Language}
\label{subsec:syntax}
%
%
Our language is the untyped lambda calculus, extended with data
constructors (à la ML and Haskell), and case expressions to eliminate
them. We define expressions, ranged over by \smath{$\xpr \in \Xpr$}, inductively.
%
$
\xpr ::= \var \mid \lam{\var}{\xpr} \mid \app{\xpr}{\xpr} \mid \con{K}[\lstof{\xpr}] \mid \cse{\con{K}}{\xpr}{\xpr}
$
%
where \smath{$x \in \Var$} ranges over variables, \smath{$\lam{\var}{\xpr}$}
denotes abstraction, and \smath{$\app{\xpr}{\xpr}$} application.
Data constructors are ranged over by \smath{$\con{K} \in \Con$},
\smath{$\lstof{\xpr}$} is a (possibly empty) list of expressions,
and
\smath{$\con{K}[\xpr_1 \cdots{} \xpr_k]$} denotes an expression constructed by
\smath{$\con{K}$} and containing \smath{$\xpr_1$} through \smath{$\xpr_k$}.
Finally, \smath{$\cse{\con{K}}{\xpr}{\xpr_\texttt{F}}$} performs
pattern-matching, matching for expressions constructed by \smath{$\con{K}$}.
%



\paragraph{Semantics: Computation}
%
An expression can be seen as defining a computation.
Each computation step of an expression involves a reduction defined as follows,
%
%
\senv{%
  \vspace{-2mm}
\begin{align}
&\step{\env}{\var}{\fun{\env}{\var}}\label{eq:sem:var}\\
&\step{\env}{\app{\xpr}{\xpr''}}{\app{\xpr'}{\xpr''}}\hspace{11em}\text{, if }\step{\env}{\xpr}{\xpr'}\label{eq:sem:appl}\\
  &\step{\env}{\app{\val}{\xpr}}{\app{\val}{\xpr'}}\hspace{11.96em}\text{, if }\step{\env}{\xpr}{\xpr'}\label{eq:sem:appr}\\
  &\step{\env}{\app{(\lam{\var}{\xpr})}{\val}}{\sub{\var}{\val}{\xpr}}\label{eq:sem:lam}\\
&\step{\env}{\app{(\cse{\con{K}}{\xpr}{\xpr_\texttt{F}})}{(\con{K}[\xpr_1 \cdots{} \xpr_k])}}{\app{\xpr}{\xpr_1\cdots{}\xpr_k}}\label{eq:sem:cset}\\
&\step{\env}{\app{(\cse{\con{K}}{\xpr}{\xpr_\texttt{F}})}{\val}}{\xpr_\texttt{F}}\hspace{7.6em}\text{, if }\val \neq \con{K}[\lstof{\xpr}]\label{eq:sem:csef}
\end{align}%
}%
\noindent%
This transition relation
\smath{$(\step{\env}{}{}) : \Xpr \times \Xpr$},
specifies how an expression reduces, small-step, towards an
irreducible expression, \ie\ a value \smath{$\val \in \Val = \{\xpr \mid \step{}{\xpr \arrownot}{}\} \subseteq \Xpr$}.
The relation is parameterized by an environment
\smath{$\env\in \Env = \Var \to \Val$},
which assigns free variables to values \eqref{eq:sem:var}.
Before an application is performed, the operator and its operand are
reduced to a value, in that order \eqref{eq:sem:appl}
\eqref{eq:sem:appr}.
If the operator is an abstraction, then the application yields
its body,
with (free) occurrences of the variable it
binds replaced with the operand \eqref{eq:sem:lam}.
If the operator is
of the form
\smath{$\cse{\con{K}}{\xpr}{\xpr_\texttt{F}}$}, then application
pattern-matches the operand.
If the operand is \smath{$\con{K}[\lstof{\xpr}]$}, then the application yields
\smath{$\xpr$} applied to the expressions in \smath{$\lstof{\xpr}$}~\eqref{eq:sem:cset}.
Otherwise, application yields \smath{$\xpr_\mathtt{F}$} (pattern-matching
failure)~\eqref{eq:sem:csef}.
%

\paragraph{Semantics: Probabilistic Model}
%
In probabilistic programming, a program defines a probabilistic
model. Likewise, an expression in our language can  be viewed as defining a probabilistic model.

First, we present \emph{Monad-Lift}. We define the probabilistic model that an expression describes in
terms of the computation semantics.
This definition relies on the observation that \smath{$\Dist$} is a
monad, a fact described in detail by Ramsay and Pfeffer
\cite{DBLP:conf/popl/RamseyP02}.
By virtue of being a monad, the following two functions are defined for
\smath{$\Dist$}:
$
\Return : \boldsymbol{A} \to \Dst{\boldsymbol{A}} \text{ and }
(\Bind) : \Dst{\boldsymbol{A}} \to (\boldsymbol{A} \to \Dst{\boldsymbol{B}}) \to \Dst{\boldsymbol{B}}
$
%
Here, \smath{$\boldsymbol{A}$} and \smath{$\boldsymbol{B}$} are arbitrary sets. Concretely, the functions are polymorphic in $\boldsymbol{A}$ and $\boldsymbol{B}$. So, \smath{$\Return$} maps each element
of \smath{$\boldsymbol{A}$}, to a distribution on \smath{$\boldsymbol{A}$}. Concretely, \smath{$\Return\ a$} is simply the
so-called Dirac-measure concentrated at \smath{$a$}:
\senv{%
  \begin{equation*}
  (\Return\ a)\ A = \left\{
  \begin{array}{l@{\ }l}
    1 & \text{, if }a \in A\\
    0 & \text{, otherwise.}
  \end{array}\right.
\end{equation*}%
}
%
%
Suppose
you have random variables
\smath{$a$} and \smath{$b$}, and that we know their distributions
\smath{$\Pr(a)$} and \smath{$\Pr(b|a)$}.
%
Then you
can compute \smath{$\Pr(b)$} by marginalizing out \smath{$a$} in
%
\mbox{\smath{$\Pr(b|a)*\Pr(a)$}}.
This is what \smath{$(\Bind)$} does;
in \mbox{\smath{$\Pr(b) = (\Pr(a) \Bind \Pr(b|a))$}} with
%
%
$
(\Pr(a) \Bind \Pr(b|a))\ B = \int_{\boldsymbol{A}} (\,\lambda a\,.\, \Pr(B|a)*\Pr(a)) \mathrm{d}a.
$
%
%
%
These functions have been used to implement a whole host of functions
for monads. One of these functions is the standard monad lift
operation.
%
$
{\Lift{}} : \monad{M} \Rightarrow (\boldsymbol{A} \to \boldsymbol{B}) \to M\ \boldsymbol{A} \to M\ \boldsymbol{B} \text{ and }
{\Lift{}} f\ m = m {\Bind} (\ \lambda\ x\ .\ {\Return}\ (f\ x)\ ).
$
%
With types
\smath{${\Lift{}} : (\Env \to \Val) \to \Dst{\Env} \to \Dst{\Val}$},
we see that \smath{${\Lift{}}$} looks very much like \smath{$\mathit{dist}$} from
before.
The main advantage of using \smath{${\Lift{}}$} is that we can use
monad laws, and other results proven for monads, to reason about it,
and thus, about distributions.
%
%
So, like before, the probabilistic model that \smath{$\xpr$} describes is
simply
\mbox{\smath{$({\Lift{}}\ \eaf{{}}{\xpr}) : \Dst{\Env} \to \Dst{\Val}$}}.

\begin{example}\label{ex:monad-lift}
  Let \smath{$f = \lambda\ xy\ .\ (x+y)/2$} be a simplified version of
  the program $\code{agg}$ which computes the average of two numbers
  $x$ and $y$.
  Let \smath{$\{0,1,2\}^2$} be a set of environments (\Env) where the
  first element of the pair defines the value of $x$ and the second
  the value of $y$ in $f$.
  Let \smath{$P(x,y) = \Uniform(\{0,1,2\}^2)$} be a discrete
  distribution over environments which allocates the same probability
  to all environments---this distribution corresponds to the prior.
  Then, \smath{${\Lift{}}\ \eaf{{}}{f}\ \Pr(x,y)$} defines the prediction $\Pr(o)$.
  Here we show the steps to compute $\Pr(o=2)$ via monad-lift.
  {\small
  \begin{align*}
    \Pr(o=2) & = \Pr(x,y) {\Bind{}} (\ \lambda\ xy\ .\ {\Return}\ (f\ xy)\ ) \; \{2\} \\
             & = \sum\limits_{(x,y)\in\{\mathrlap{0,1,2\}^2}} (\lambda\ xy\ .\ \Return\ (f\ xy)\ \{2\})\ \Pr(x,y) \\
             & = \sum\limits_{(x,y)\in\{\mathrlap{(0,2),(1,1),(2,0)\}}} (\Return\ f\ x\ y)\ \{2\} \cdot \Pr(x,y)
             = 1 \cdot 1/9 + 1 \cdot 1/9 + 1 \cdot 1/9
             = 3/9
  \end{align*}}
  %
  %
  We replace the $\int$ to $\sum$ because the prior is discrete.
\end{example}

%
%
%

%
One drawback of this semantics is that it ignores the structure of the probabilistic model.
%
Alternatively, one can build the structure of the probabilistic model
embedding the expressions of our programming language in the \Dist\ monad:
%
\senv{%
\begin{align*}
  \wsem{\env}{\var}
  &=
  \fun{\env}{\var}
  \\
  \wsem{\env}{\lam{\var}{\xpr}}
  &=
  \Return\ (\ \lam{\val}{\wsem{\sub{\var}{\Return\ \val}{\env}}{\xpr}}\ )
  \\
  \wsem{\env}{\app{\xpr}{\xpr'}}
  &=
  \wsem{\env}{\xpr}\hspace{.36em} \Bind\
  \lam{f}{\wsem{\env}{\xpr'} \Bind\ \lam{\val}{\app{f}{\val}}}
  \\
  \wsem{\env}{\con{K}[\xpr_1 \cdots{} \xpr_k]}
  &=
  \wsem{\env}{\xpr_1} \Bind\ \lam{\val_1}{ \cdots{}
    \\ & \phantom{=}\ \,
    \wsem{\env}{\xpr_k} \Bind\ \lam{\val_k}{\Return\ (\ \con{K}[\val_1 \cdots{} \val_k]\ )}}
  \\
  \wsem{\env}{\cse{\con{K}}{\xpr}{\xpr_\texttt{F}}}
  &=
  \Return\ (\ \lam{\val}{}\texttt{case}\ \val\ \texttt{of}
  \\
  &
  \ \ \ \ \ \ \con{K}[\var_1 \cdots \var_k]\ \texttt{=>}\ \wsem{\env}{\xpr}\,\Bind\,\lam{f\hspace{-.1em}}{\hspace{-.1em}f\,\var_1\cdots{}\var_k}
  \\
  &
  \ \ \ \ \ \ \_\hspace{4em}\texttt{=>}\ \wsem{\env}{\xpr_\texttt{F}}\ )
\end{align*}%
}%
Here, $\env : \Var \to \Dst{\Val}$.
It is important to note that, whereas \eg{} \mbox{$\wsem{\env}{\var} :
  \Dst{\Val}$}, the type for denotated abstractions is different;
\mbox{$\wsem{\env}{\lam{\var}{\xpr}} : \Dst{\Var \to \Dst{\Val}}$}.
Also note the $\mapsto$ in the denotation of abstractions; $\var$
stores a \emph{distribution}.
With this, \smath{$\wsem{}{\xpr} : \Dst{\Env} \to \Dst{\Val}$}
is the probabilistic model that \smath{$\xpr$} describes.
%
%


\end{document}